\documentclass[twocolumn,prb,amsmath,amssymb,floatfix]{revtex4}

\usepackage{bbold}
\usepackage{bm}
\usepackage{color}
\usepackage{dcolumn}
\usepackage{feynmf} 
\usepackage{graphics}
\usepackage{graphicx}
\usepackage{subfigure}
\usepackage{slashed}
\usepackage{tensor}
\usepackage{placeins}
\usepackage{natbib}

\def\Tr{\mbox{Tr}}

\def\al{\alpha}
\def\eps{\epsilon}

\def\veps{\varepsilon}

\def\be{\begin{equation}}
\def\ee{\end{equation}}
\def\bea{\begin{eqnarray}}
\def\eea{\end{eqnarray}}
\def\bse{\begin{subequations}}
\def\ese{\end{subequations}}
\def\bc{\begin{center}}
\def\ec{\end{center}}

\def\ra{\rightarrow}

\def\nonum{\nonumber}

\def\E{{\rm e}}

\def\D{{\rm d}}
\def\Ord{{\rm O}}

\def\Amu{{A_\mu}}
\def\Anu{{A_\nu}}
\def\Flmn{{F_{\mu \nu}}}

\def\Bpsi{{{\bar{\psi}}}}

\def\Sp{{\slashed p}}
\def\Sq{{\slashed q}}
\def\Sk{{\slashed k}}
\def\Dmu{{\partial_\mu}}
\def\Dnu{{\partial_\nu}}
\def\DDmu{{D_\mu}}

\def\Lag{{\mathcal{L}}}

\newcommand{\comment}[1]{}

\catcode`,\active

\catcode`\,12

\begin{document}

\title{Two-loop fermion self-energy in reduced quantum electrodynamics \\ and application to the ultra-relativistic limit of graphene}

\author{A.~V.~Kotikov$^1$ and S.~Teber$^{2,3}$}
\affiliation{
$^1$Bogoliubov Laboratory of Theoretical Physics, Joint Institute for Nuclear Research, 141980 Dubna, Russia.\\
$^2$Sorbonne Universit\'es, UPMC Univ Paris 06, UMR 7589, LPTHE, F-75005, Paris, France.\\
$^3$CNRS, UMR 7589, LPTHE, F-75005, Paris, France.}

\date{\today}

\begin{abstract}
We compute the two-loop fermion self-energy in massless reduced quantum electrodynamics for an arbitrary gauge using the method of integration by parts.
Focusing on the limit where the photon field is four-dimensional, our formula involves only recursively one-loop integrals and can therefore be evaluated exactly.
From this formula, we deduce the anomalous scaling dimension of the fermion field as well as the renormalized fermion propagator up to two loops.
The results are then applied to the ultra-relativistic limit of graphene and compared with similar results obtained for four-dimensional and three-dimensional 
quantum electrodynamics.
\end{abstract}

\maketitle

\begin{fmffile}{fmfsigma}

\section{Introduction}

In condensed matter physics, an emergent relativity at low energies appears
for systems with two stable Fermi points, see, {\it e.g.}, the textbook Ref.~[\onlinecite{Volovik09}].
This is the case of undoped graphene, a one-atom thick layer of graphite, see, {\it e.g.}, Ref.~[\onlinecite{KotovUPGC12}] for a review,
where the quasiparticle spectrum is Dirac-like and massless at low-energies~\cite{Wallace47,Semenoff84}.
In Ref.~[\onlinecite{GonzalezGV93}] a renormalization group approach indeed revealed the existence of an
infrared Lorentz invariant fixed point for graphene. Approaching this fixed point, the Fermi velocity, $v_F$, flows to the velocity of light, $c \approx 300 v_F$,
while the fine structure constant of graphene, $\al_g \approx e^2/4\pi \veps \hbar v_F$, which is of order one, flows to the usual fine structure constant, $\al \approx 1/137$.
Moreover, while electrons in graphene are confined to a three-dimensional space-time, $d_e = 2+1$, interactions between them are mediated by four-dimensional photons, $d_\gamma = 3+1$.
The Lorentz invariant fixed point may therefore be effectively described by a massless relativistic quantum field theory (QFT) 
model whereby $2+1$-dimensional electrons interact via a long-range fully retarded potential.
Such a model belongs to the class of reduced quantum electrodynamics (RQED) [\onlinecite{GorbarGM01}], or RQED$_{d_\gamma,d_e}$, and corresponds to RQED$_{4,3}$ in the case of graphene.
Following Ref.~[\onlinecite{Marino93}], the latter is also sometimes referred to as pseudo-QED, see, {\it e.g.}, Ref.~[\onlinecite{Alves13}].
In the general case, RQED$_{d_\gamma,d_e}$ is a relativistic QFT describing the interaction of an abelian $U(1)$ gauge field
living in $d_\gamma$ space-time dimensions with a fermion field living in a space-time of $d_e$ dimensions.
The case where $d_e =d_\gamma$ corresponds to usual QEDs, see the textbooks Refs.~[\onlinecite{ItzyksonZ05,PeskinS95,Grozin07}].
The reduced case corresponds to $d_e < d_\gamma$ where the fermion field is localized on a $(d_e-1)$-brane.
Motivated by potential applications to condensed matter physics as well as interest in branes we focus on
the computation of radiative corrections in a general model of RQED$_{d_\gamma,d_e}$. 
The computations are based on sophisticated methods devoted to the exact evaluation of multi-loop Feynman diagrams in relativistic
QFTs; these methods include, {\it e.g.}, the Gegenbauer polynomial technique~\cite{ChetyrkinKT80,Kotikov96}, integration by parts~\cite{VasilievPK81,ChetyrkinT81} (IBP), and
the method of uniqueness.~\cite{VasilievPK81,Usyukina83,Kazakov84,Kazakov85,Kazakov-lecture}

In Refs.~[\onlinecite{TeberRQED12,KotikovT13}] multi-loop corrections in a general theory of RQED$_{d_\gamma,d_e}$ were computed with a
special emphasis on electromagnetic current correlations and the link between RQED$_{4,3}$ and the ultrarelativistic limit of graphene was made.
In particular, the two-loop interaction correction coefficient to the polarization operator of graphene in the ultrarelativistic limit
was derived and found to be small in qualitative agreement with theoretical~\cite{C-non-relat}
as well as experimental~\cite{C-experiments} results in the non-relativistic regime.
These results were used in Ref.~[\onlinecite{HerbutM13}] to show that the optical conductivity of graphene in the ultrarelativistic limit
has a relative deviation which is within experimental uncertainty~\cite{C-experiments} and of the order of one percent with respect to the non-interacting value.
This striking qualitative agreement with the non relativistic limit together with the efficient tools available to tackle the relativistic limit suggest
the crucial importance of fully exploring the properties of the Lorentz invariant fixed point.

We pursue this task in the present paper by computing multi-loop corrections to the fermion propagator and the corresponding anomalous scaling dimension of the fermion field
in RQED$_{4,d_e}$. A peculiarity of graphene  is that the fermion field gets renormalized but not the electric charge, see
Ref.~[\onlinecite{GonzalezGV93}].
A similar feature was found for RQED$_{4,3}$ in Ref.~[\onlinecite{TeberRQED12}] where the anomalous dimension of the fermion field was computed at one-loop.
In the following we extend these results to two loops. From the field theory point of view, the main technical difficulty involves the computation of some peculiar
two-loop massless propagator diagrams with two non-integer indices on non-adjacent lines (see Eq.~(\ref{complicatedG}) below). Presently, no explicit analytical expression
for such diagrams is known. It turns out that, as will be shown below, repeated use of some well chosen IBP identities
allows us to express each complicated, and eventually divergent, diagram as a sum of primitive (or recursively one-loop) diagrams plus
a complicated but convergent diagram which is multiplied by a factor $\eps_\gamma = 2-d_\gamma/2$. In the limit,
$d_\gamma \ra 4$, the complicated part does not contribute which solves the problem in the case of RQED$_{4,d_e}$.

The paper is organized as follows. In Sec.~\ref{Sec:RQED}, we review the basics of massless RQED at one-loop.
In Sec.~\ref{Sec:Sigma}, we compute the two-loop fermion self-energy in RQED$_{d_\gamma,d_e}$ using IBP identities and
show that, for $d_\gamma \ra 4$, it can be expressed only as a function of recursively one-loop diagrams.
In Sec.~\ref{Sec:FermiDim}, we compute the anomalous scaling dimension of the fermion field and the renormalized fermion propagator at two-loop, in an arbitrary gauge 
in the general case of RQED$_{4,d_e}$. The special case of RQED$_{4,3}$ is then explicitly considered as well as, for completeness, the cases of QED$_4$ and QED$_3$.
In Sec.~\ref{Sec:Conclusion} we summarize our results and conclude. Finally, App.~\ref{app-Gfuncs} contains the expansion of some master integrals
entering the two-loop self-energy and App.~\ref{App:G(alpha,beta)} presents
general formulas, valid beyond IBP relations, for diagrams with two non-integer indices.
In the following, we work in units where $\hbar=c=1$.

\section{Massless RQED}
\label{Sec:RQED}

\subsection{The model}

We consider a general model of massless RQED$_{d_\gamma,d_e}$ described by [\onlinecite{GorbarGM01,TeberRQED12}]:
\bea
S_{\text{RQED}} =&& \int \D^{d_\gamma} x \, \Lag_{\text{RQED}},
\label{rqed} \\
\Lag_{\text{RQED}}\,=&&\,\Bpsi(x) i \gamma^{\mu_e}D_{\mu_e} \psi(x)\,\delta^{(d_\gamma-d_e)}(x) -
 \nonum \\
&&-\frac{1}{4}\,F_{\mu_\gamma \nu_\gamma}F^{\mu_\gamma \nu_\gamma} - \frac{1}{2a}\left(\partial_{\mu_\gamma}A^{\mu_\gamma}\right)^2,
\nonum
\eea
where $\DDmu = \Dmu + ie \Amu$ is the covariant derivative, $\Flmn = \Dmu \Anu - \Dnu \Amu$ is the field stress tensor of the gauge potential $A^\mu$ 
and $a$ is a gauge fixing parameter.
In Eq.~(\ref{rqed}), on the one-hand, the index $\mu_e$ runs over the $d_e$-dimensional space-time in which fermions are localized: 
$\mu_e\,=\,0,\,1,\,...,\,d_e-1$, where zero index corresponds to time. On the other hand, the index $\mu_\gamma$ 
runs over the $d_\gamma$-dimensional space-time of the gauge field: $\mu_\gamma \,=\, 0,\,1,\,...,\,d_e-1,\,d_e,\,...,\,d_\gamma-1$, where we assume that $d_\gamma \geq d_e$.
The minimal coupling of the gauge field to the fermion current, $j_\mu A^\mu$, is restricted to the reduced matter space from which we deduce the expression of the fermion current:
\be
j^\mu(x) =
\left\{
          \begin{array}{ll}
                \E\, \Bpsi(x) \gamma^\mu \psi(x) \delta^{(d_\gamma-d_e)}(x) & \,\,\, \mu=0,\cdots,d_e-1, \\
                0 & \,\,\, \mu = d_e,\cdots,d_\gamma-1. \\
          \end{array}
\right.
\label{fc}
\ee
Simple dimensional analysis then shows that two epsilon parameters are naturally associated with the two dimensions $d_\gamma$ and $d_e$. 
Indeed, for a general RQED$_{d_\gamma,d_e}$, the dimensions of the fields and electric charge may be written as:
\be
[A_\mu] = 1 - \varepsilon_\gamma,
\quad [\psi]= \frac{3}{2} - \varepsilon_e - \varepsilon_\gamma,
\quad [e] = \varepsilon_\gamma,
\label{dims}
\ee
where the two epsilon parameters read:
\be
\varepsilon_\gamma = \frac{4-d_\gamma}{2},
\qquad \varepsilon_e = \frac{d_\gamma - d_e}{2},
\label{params}
\ee
and will be shown below to play a crucial role in setting up a dimensional regularization scheme for RQED.
Conversely, Eq.~(\ref{params}) yields the relations:
\be
d_\gamma = 4 - 2\veps_\gamma, \qquad d_e = 4-2\veps_e - 2\veps_\gamma.
\label{d-params}
\ee
As can be seen from Eq.~(\ref{dims}) the dimension of the coupling constant is entirely determined by the
space-time dimension of the gauge field, $d_\gamma$, or, equivalently, $\varepsilon_\gamma$; for a four-dimensional gauge field ($d_\gamma=4$) the coupling is dimensionless
so that, at least at a classical level, the theory is scale invariant. This suggests that RQED$_{4,d_e}$ are renormalizable QFTs.
Following Ref.~[\onlinecite{TeberRQED12}], let's recall that the superficial degree of divergence (SDD)
of a general RQED$_{d_\gamma,d_e}$ diagram reads:~\footnote{ In the case of RQED$_{4,3}$, Eq.~(\ref{sdd}) yields: $D = 3-N_\gamma-N_e$. The latter exactly corresponds to the result
of Ref.~[\onlinecite{AppelquistBKW86}], see discussion and equations below Eq.~(2.20) in this paper, for QED$_3$ in the large-$N_F$ limit.}
\be
D = d_e + \frac{d_\gamma - 4}{2} V - \frac{d_\gamma-2}{2} N_\gamma -\frac{d_e-1}{2}N_e,
\label{sdd}
\ee
where $V$ is the number of vertices, $N_\gamma$ the number of external gauge lines and $N_e$ the number of external
fermion lines. From Eq.~(\ref{sdd}) we see that for $d_\gamma=4$ the SDD does not depend on the number of vertices whatever
value $d_e$ takes; this confirms the renormalizability of the theory.  
A peculiar fact of reduced theories is that, while the fermion self-energy and fermion-photon vertex have the same degree of divergence
as in usual QED$_4$ (they effectively diverge logarithmically), the photon self-energies of RQED$_{4,3}$ and RQED$_{4,2}$ are finite.
This is summarized in Tab.~\ref{tab:sdd} which displays the degrees of divergence (superficial and effective) of the three most divergent amplitudes in QED$_4$, RQED$_{4,3}$ and
RQED$_{4,2}$. As a consequence, while the fermion field acquires an anomalous dimension, the coupling constant does not renormalize. 

\begin{center}
\renewcommand{\tabcolsep}{0.25cm}
\renewcommand{\arraystretch}{1.5}
\begin{table}
    \begin{tabular}{| l || c | c | c |}
      \hline
      \quad & QED$_4$ & RQED$_{4,3}$ & RQED$_{4,2}$ \\
      \hline \hline
      Photon self-energy & 2 (0) & 1 (-1) & 0 (-2)  \\
      \hline
      Fermion self-energy & 1 (0) & 1 (0) & 1 (0) \\
      \hline
      Fermion-photon vertex & 0 (0) & 0 (0) & 0 (0) \\
      \hline
    \end{tabular}
    \caption{Superficial degree of divergence (and effective degree of divergence) of the three most divergent amplitudes in RQED$_{4,d_e}$ for $d_e=4$ (QED$_4$), $d_e=3$ and $d_e=2$.}
    \label{tab:sdd}
\end{table}
\end{center}

\subsection{Perturbation theory and renormalization}

The above dimensional analysis can be made quantitative by setting up a perturbative approach to RQED.
As can be seen from Eq.~(\ref{rqed}) the free massless fermion propagator is the usual one and reads:
\be
S_0(p_e) = \frac{i\Sp}{p^2 + i0^+}\, ,
\label{fermion-prop0}
\ee
where $p=p_0,\,...,\,p_{d_e-1}$ lies in the reduced matter space and $i0^+$ is a convergence factor that will often be omitted in the following in order to simplify notations.
On the other hand, because we are interested in the properties of the reduced system, we may integrate over the $d_\gamma - d_e$ bulk gauge degrees of freedom. 
This yields an effective free gauge field propagator reading:
\be
\tilde{D}_0^{\mu \nu}(q) = \frac{i}{(4\pi)^{\varepsilon_e}}\frac{\Gamma(1-\varepsilon_e)}{(-q^2)^{1-\varepsilon_e}}\,\left( g^{\mu \nu} - \tilde{\xi}\,\frac{q^{\mu} q^{\nu}}{q^2}\right),
\label{gauge-field-prop0}
\ee
where, now, $q=q_0,\,...,\,q_{d_e-1}$ lies in the reduced matter space and $\varepsilon_e$ was defined in Eq.~(\ref{params}).
Moreover, we see that the gauge fixing parameter is affected by the integration of the bulk gauge modes: $\tilde{\xi} = \xi (1-\varepsilon_e)$
where $\xi = 1-a$. The two commonly used relativistic gauges are the Feynman and Landau gauges which are defined as:
\begin{subequations}
\label{gauges}
\bea
{\mathrm{Feynman~gauge:}}\quad \xi=0 \quad &{\mathrm{or}}& \quad a=1,
\label{Feynmangauge} \\
{\mathrm{Landau~gauge:}}\quad \xi=1 \quad &{\mathrm{or}}& \quad a=0.
\label{Landaugauge}
\eea
\end{subequations}
In the following we shall mainly work in an arbitrary gauge. 

In the case of RQED$_{4,3}$ for which $\varepsilon_e=1/2$, the effective photon propagator of Eq.~(\ref{gauge-field-prop0}) has a square root branch cut
\be
\tilde{D}_0^{\mu \nu}(q) = \frac{1}{2}\,\frac{i}{\sqrt{-q^2}}\,\left( g^{\mu \nu} - \frac{\xi}{2}\,\frac{q^{\mu} q^{\nu}}{q^2}\right).
\label{gauge-field-prop0-RQED43}
\ee
This momentum dependence is responsible for the appearance of Feynman diagrams with non-integer indices and is
a major source of technical difficulty that we shall discuss in details in the next sections. 
As already noticed in Ref.~[\onlinecite{KotikovT13}], a similar momentum dependence can be found for QED$_3$ in the large-$N_F$ limit where $N_F$ is the number of fermion species. 
The reason is the ``infrared softening'' of the photon propagator of QED$_{3}$ for large $N_F$~\cite{Appelquistetal:IRsoftening,AppelquistBKW86}.
Indeed, using a non-local gauge fixing term, $\eta$, as done in Ref.~[\onlinecite{Nash89}] 
and justified in the general case in Ref.~[\onlinecite{Shirkov90}], the dressed photon propagator together with the one-loop polarization operator of massless QED$_3$, read:
\be
D^{\mu \nu}(q) = -i\,\frac{g^{\mu \nu} - \eta \, q^{\mu} q^{\nu} / q^2}{q^2\,\left[1-\Pi(q^2)\right]},\quad \Pi_1(q^2) = -\frac{e^2\,N_F}{8\sqrt{-q^2}},
\label{gauge-field-prop-QED3}
\ee
where, for $\eta=1$, the usual Landau gauge propagator is recovered, the lower index $1$ on $\Pi$ refers to the order of perturbation theory 
and $e^2$ has dimension of mass in super-renormalizable QED$_3$.
In the $1/N_F$ approximation ($N_F \ra \infty$ and $e^2 \ra 0$ with $e^2 N_F$ fixed) and focusing on the infrared limit, 
$\sqrt{-q^2} \ll e^2 N_F$, the dressed photon propagator of Eq.~(\ref{gauge-field-prop-QED3}) becomes:
\be
D^{\mu \nu}(q) = \frac{8}{e^2 N_F}\frac{i}{\sqrt{-q^2}}\,\left( g^{\mu \nu} - \eta\,\frac{q^{\mu} q^{\nu}}{q^2}\right),
\label{gauge-field-prop-QED3-2}
\ee
and, similarly to Eq.~(\ref{gauge-field-prop0-RQED43}), behaves like $1/\sqrt{-q^2}$ instead of the usual $1/q^2$.

Finally, the effective free photon propagator of RQED$_{4,2}$, which has an exponent $\varepsilon_e=1$, has even a softer momentum dependence.
The latter is similar to the momentum dependence of the dressed photon propagator of QED$_2$ at large $N_F$.  
In the case of RQED$_{4,2}$, however, the logarithmic divergence of the effective photon propagator translates
into the appearance of a pole in the Gamma function of Eq.~(\ref{gauge-field-prop0}). 
This case requires some further regularization such as, {\it e.g.}, giving a finite width to the 1-brane~\cite{GorbarGM01}. In the following, we
shall mainly focus on the 2-brane case described by RQED$_{4,3}$.
The similarity between the reduced versions of QED and its large $N_F$ limit will be a subject of our future investigations.

\begin{figure}
    \includegraphics{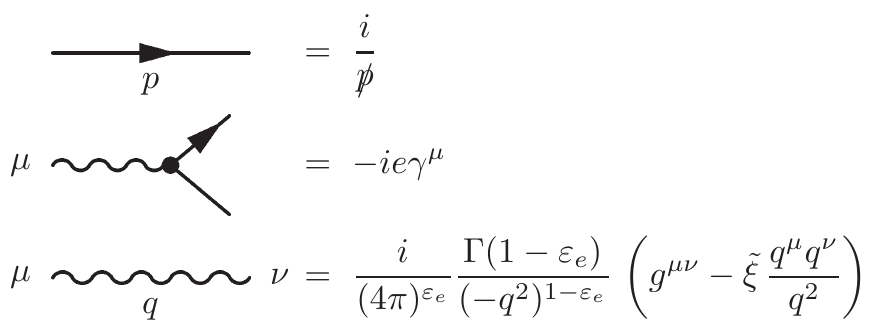}
    \caption{\label{fig:FeynmanRules}
    Feynman rules for massless RQED.}
\end{figure}

With the Feynman rules of massless RQED$_{d_\gamma,d_e}$ summarized on Fig.~\ref{fig:FeynmanRules}, 
perturbation theory is then implemented in the usual way by solving Dyson equations for the fermion and photon propagators and computing the various 
self-energies. Focusing for simplicity on the fermion propagator, which is central to the present work, the corresponding Dyson equation reads:
\be
S(p) = S_0(p) + S_0(p) \left( -i \Sigma(p) \right) S(p),
\ee
where $\Sigma$ is the fermion self-energy. The solution to this equation in the case of massless fermions can be written as:
\be
-i \Sp\,S(p) = \frac{1}{1-\Sigma_V(p^2)}, \quad \Sigma(p) = \Sp \Sigma_V(p^2)\, .
\label{fermion-prop+sigma}
\ee
The effective divergence appearing in parentheses in Tab.~\ref{tab:sdd} actually corresponds to the divergence 
of $\Sigma_V(p^2)$. Such a self-energy is divergent in all RQEDs as indicated in the table. The divergence is related to loop
integrals depending on $d_e$. Following Refs.~[\onlinecite{TeberRQED12,KotikovT13}] we use dimensional regularization 
and work with an arbitrary $d_e$. The latter may be expressed as a function of the two epsilon parameters which naturally appeared at the level of Eq.~(\ref{params}). 
The parameter $\veps_e$ takes into account the brane-like nature of the system. For fixed $\veps_e$ 
self-energies and related propagators take the form of Laurent series in $\varepsilon_\gamma$.
Working in an improved minimal subtraction scheme ($\overline{\rm{MS}}$) we may then absorb the singular 
part of the series (in $\veps_\gamma$) in renormalization constants which relate the bare fields and parameters
of the theory to renormalized ones:
\be
\psi = Z_{\psi}^{1/2}\psi_r, \quad A = Z_A^{1/2} A_r, \quad a=Z_A a_r, \quad e=Z_\al^{1/2}e_r,
\label{ren_constants}
\ee
where $Z_\psi$, $Z_A$ and $Z_\al$ are the dimensionless renormalization constants.
Moreover, in the $\overline{\rm{MS}}$ scheme we also define a dimensionless renormalized coupling constant $\al$ via the equation:
\be
\frac{\al(\mu)}{4\pi} = \mu^{-2\varepsilon_\gamma}\,\frac{e^2}{(4\pi)^{d_\gamma/2}}\,Z_\al^{-1}(\al(\mu))\,e^{-\gamma_E \varepsilon_\gamma},
\label{ren_coupling}
\ee
where $\gamma_E$ is Euler's constant and the $\mu^{-2\varepsilon_\gamma}$ factor compensates for the dimension of $e^2$.
Equivalently, from Eq.~(\ref{ren_coupling}), the bare coupling constant $e$ can be expressed via the renormalized coupling constant
$\al(\mu)$.

From the above arguments, the relation between the bare and renormalized fermion propagators reads:
\be
S(p) = Z_\psi(\al(\mu),a_r(\mu)) S_r(p\,;\mu),
\label{mpropr}
\ee
where all singularities are in $Z_\psi$ and $S_r$, the renormalized fermion propagator, is finite.
Similarly, the relation between the bare and renormalized effective photon propagators reads:
\be
\tilde{D}^{\mu \nu}(q) = Z_A(\al(\mu)) \tilde{D}_r^{\mu \nu}(q\,;\mu),
\ee
where all singularities are in $Z_A$ and $\tilde{D}_r$, the renormalized effective photon propagator, is finite.
Finally, one may introduce another constant, $Z_\Gamma$, for the vertex renormalization: $\Gamma^\mu = Z_\Gamma \Gamma^\mu_r$
where $\Gamma^\mu_r$ is finite.
Because both the renormalized coupling and vertex are finite there is a constraint among
the constants:
\be
Z_\al = (Z_\Gamma Z_\psi)^{-2} Z_A^{-1}.
\label{Zs-constraint}
\ee
Furthermore, as in usual QEDs, the Ward identity:
\be
Z_\Gamma Z_\psi = 1,
\label{ward}
\ee
is satisfied in a general RQED implying that coupling renormalization is entirely due to gauge field renormalization: 
$Z_\al = Z_A^{-1}$.~\footnote{The proof of the Ward identity, Eq.~(\ref{ward}), in RQED follows the same steps as for usual QEDs,
see, {\it e.g.}, the textbooks Refs.~[\onlinecite{ItzyksonZ05,PeskinS95,Grozin07}]. The reason is that the textbook proof
does not depend on the index of the gauge-field propagator which may then be arbitrary.}
The beta function can then be expressed as a function of the anomalous scaling dimension of the gauge field as follows:
\bea
&&\beta(\al(\mu)) = \frac{d \log \al(\mu)}{d \log \mu} = -2\varepsilon_\gamma + \gamma_A(\al(\mu)), 
\nonum \\
&&\gamma_A(\al(\mu)) = \frac{d \log Z_A(\al(\mu))}{d \log \mu}.
\label{beta+gammaA}
\eea
These functions together with the renormalized photon propagator of RQED were considered in Refs.~[\onlinecite{TeberRQED12,KotikovT13}] where it was shown, in accordance
with dimensional analysis, that: $\beta(\al) = \gamma_A(\al) = 0$ in RQED$_{4,d_e}$ ($d_e<4)$.

In the following we will compute the renormalization constant $Z_\psi$ and deduce the anomalous scaling dimension of the fermion field:
\be
\gamma_\psi(\al(\mu),a_r(\mu)) = \frac{d \log Z_\psi(\al(\mu),a_r(\mu))}{d\log \mu}.
\label{asd}
\ee
Differentiating Eq.~(\ref{mpropr}) with respect to $\mu$ and taking into account Eq.~(\ref{asd}) as well as the fact that the bare propagator does not depend on the renormalization scale
yields
\be
\frac{\D \, \log S_r(p\,;\mu)}{\D\,\log \mu} = -\gamma_\psi(\al(\mu),a_r(\mu)),
\label{ren-eq}
\ee
where $S_r$ depends on $\mu$ explicitly as well as through $\al$ and $a_r$.
As will be shown later, the explicit $\mu$ dependence enters through the combination $-p^2 / \mu^2$ and the renormalized propagator may be written:
\be
S_r(p\,;\mu) = \frac{i}{\Sp}\,s_r\left(\frac{-p^2}{\mu^2};\, \al;\, a_r \right).
\label{Sr-sr}
\ee
In partial differential form the renormalization group equation, Eq.~(\ref{ren-eq}), may then be written as:
\begin{widetext}
\be
\left ( \frac{\partial}{\partial\,\log\mu} +\beta(\al)\,\frac{\partial}{\partial\,\log\al} - \gamma_A(\al)\,\frac{\partial}{\partial\,\log a_r} + \gamma_\psi(\al,a_r) \right)\,S_r(p\,;\mu) = 0.
\label{ren-eq-part-mu}
\ee
Equivalently, we may use Eq.~(\ref{Sr-sr}) in order to trade the partial derivative with respect to $\mu$ in Eq.~(\ref{ren-eq-part-mu}) for a derivative with respect to momentum:
\be
\left ( 2\,\frac{\partial}{\partial\,\log \frac{-p^2}{\mu^2}} -\beta(\al)\,\frac{\partial}{\partial\,\log\al} 
+ \gamma_A(\al)\,\frac{\partial}{\partial\,\log a_r} - \gamma_\psi(\al,a_r) \right)\,s_r\left(\frac{-p^2}{\mu^2};\, \al;\, a_r \right) = 0.
\label{ren-eq-part-p}
\ee
We shall come back to the solution of this equation in a later section.

\subsection{One-loop results}

\begin{figure}
    \includegraphics{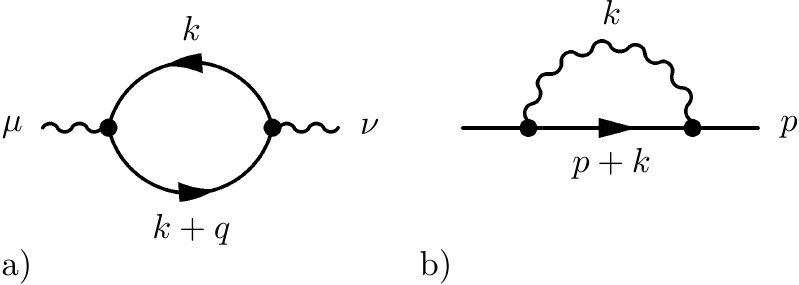}
    \caption{\label{fig:one-loop}
    One-loop diagrams: a) photon self-energy and b) fermion self-energy.}
\end{figure}

Having set the basic notations and goals, we summarize here the one-loop results obtained in Ref.~[\onlinecite{TeberRQED12}].
As will be seen below, all diagrams can be compactly expressed as a function of the one-loop massless propagator diagram defined as:
\be
\int \frac{\D^{d_e} k}{[-(k+p)^2]^{\al} [-k^2]^{\beta}} = \frac{i\,\pi^{d_e/2}}{(-p^2)^{\al+\beta-d_e/2}}\,G(\al,\beta)\, ,
\label{1loopmassless}
\ee
where $\al$ and $\beta$ are arbitrary indices labeling the diagram, the power of $-p^2$ follows from dimensional analysis
and $G(\al,\beta)$ is the (dimensionless) coefficient function associated with the diagram. The latter is known exactly
in the one-loop case:
\be
G(\al,\beta) = \frac{a(\al)a(\beta)}{a(\al+\beta-d_e/2)}\, \quad a(\al) = \frac{\Gamma(d_e/2-\al)}{\Gamma(\al)}.
\label{1loopG}
\ee

The one-loop diagrams, see Fig.~\ref{fig:one-loop}, are defined by:
\begin{subequations}
\label{1loopdiagrams}
\bea
&&i\Pi_1^{\mu \nu}(q) = - \int \frac{\D^{d_e}k}{(2\pi)^{d_e}}\, \Tr \left[ (-ie\gamma^\mu) \frac{i(\Sk +\Sq)}{(k+q)^2}(-ie\gamma^\nu)\frac{i\Sk}{k^2} \right],
\label{pimunu-1l}
\\
&&-i\Sigma_1(p) = \int \frac{\D^{d_e}k}{(2\pi)^{d_e}}\, (-ie\gamma^\mu) \frac{i(\Sp +\Sk)}{(p+k)^2}(-ie\gamma^\nu)
\frac{i}{(4\pi)^{\varepsilon_e}}\frac{\Gamma(1-\varepsilon_e)}{(-k^2)^{1-\varepsilon_e}}
\left( g_{\mu \nu} - \tilde{\xi}\,\frac{k_\mu k_\nu}{k^2} \right) \, ,
\label{sigma-1l}
\eea
\end{subequations}
where the lower index $1$ on $\Pi$ and $\Sigma$ refers to one-loop, $\Sk = \gamma^\mu k_\mu$, $k^2 = k^\mu k_\mu$, $g^{\mu \nu} = {\rm diag}(1,-1,-1,\cdots,-1)$ is the metric tensor in the 
$d_e$-dimensional space-time and $\Tr$ denotes the trace in gamma-matrix space.
The computation of the above integrals first involves gamma-matrix algebra at the level of the numerators in Eqs.~(\ref{1loopdiagrams}). 
The following trace and contraction identities summarize the numerator algebra of RQED:
\bea
&&\{\gamma^\mu,\gamma^\nu \} = 2g^{\mu \nu},\quad \Tr \left[ \mathbb{1} \right] = d, \quad \Tr \left[ \mathrm{odd~number~of~}\gamma{\mathrm{'s}} \right] = 0,\quad
\Tr \left[ \gamma^\mu \gamma^\nu \right] = dg^{\mu \nu},
\quad\Tr \left[ \gamma^\mu \gamma^\al \gamma_\mu \gamma^\beta \right] = -d(d_e-2) g^{\al \beta},
\nonum \\
&&\Tr \left[ \gamma_\mu \gamma^\al \gamma_\nu \gamma^\beta \gamma^\mu \gamma^\delta \gamma^\nu \gamma^\theta \right] = dg^{\al \beta} g^{\delta \theta} (4-d_e)(d_e-2)
-d g^{\al \delta} g^{\beta \theta} (8 - d_e)(d_e-2) +d g^{\al \theta}g^{\delta \beta}(4-d_e)(d_e-2),
\nonum \\
&&\gamma^\mu \gamma_\mu = \tensor{g}{^\mu_\mu} = d_e, \quad \gamma^\mu \gamma^\al \gamma_\mu = -(d_e-2) \gamma^\al, \quad
\gamma^\mu \gamma^\al \gamma^\beta \gamma_\mu = 2\gamma^\beta \gamma^\al +(d_e-2)\gamma^\al \gamma^\beta,
\nonum \\
&&\gamma^\mu \gamma^\al \gamma^\beta \gamma^\delta \gamma_\mu = -2\gamma^\delta \gamma^\beta \gamma^\al - (d_e-4) \gamma^\al \gamma^\beta \gamma^\delta \, ,
\label{num-algebra}
\eea
where we have used $d$-dimensional fermion spinors in such a way that the gamma matrices 
have dimension $d \times d$. Alternatively, we may use the number of massless fermion fields, $N_F$, which is related to $d$ via:
\be
N_F = \frac{d}{4},
\label{Nf}
\ee
in such a way that a single fermion specie is represented by a four-dimensional spinor such as in usual QED$_4$. In the following, we shall work 
either with an arbitrary $d$ for the sake of generality or, equivalently, with $N_F$ to relate our formulas to well known expressions
in QED$_4$ and QED$_3$ with $N_F$ fermion species. 
A further simplification of Eq.~(\ref{sigma-1l}) comes from combining the expression of $\Sigma_V$ in 
Eq.~(\ref{fermion-prop+sigma}) with the trace identities of Eq.~(\ref{num-algebra}) in such a way that:
\be
\Sigma_{V}(p^2) = \frac{-1}{d (-p^2)}\,\Tr \left[ \Sp \Sigma(p) \right].
\label{sigmaV}
\ee
We may proceed in a similar way for the polarization operator in Eq.~(\ref{pimunu-1l}). Because of current conservation the latter is constrained to have the form:
\be
\Pi^{\mu \nu}(q) = (g^{\mu \nu}q^2 - q^\mu q^\nu)\,\Pi(q^2), \quad \Pi(q^2) = \frac{- \tensor{\Pi}{^\mu_\mu}(q)}{(d_e-1)(-q^2)}\, .
\label{pimunu+piq2}
\ee
and the effective divergence appearing in parentheses in Tab.~\ref{tab:sdd} actually corresponds to the divergence
of $\Pi(q^2)$. 

With the above conventions and identities in hand, Eqs.~(\ref{1loopdiagrams}) can be straightforwardly computed in an arbitrary gauge. The final result reads:
\begin{subequations}
\label{1loopdiagrams-s}
\bea
&&
\Pi_1(q^2) = - d\,\frac{e^2}{(4\pi)^{d_e/2} (-q^2 )^{\varepsilon_\gamma+\varepsilon_e}  }\frac{d_e-2}{2(d_e-1)}\,G(1,1),
\label{pimunu-1l-s}
\\
&&
\Sigma_{V1}(p^2) = -\frac{e^2 \,\Gamma(1-\varepsilon_e)}{(4 \pi)^{d_\gamma/2}(-p^2)^{\varepsilon_\gamma}}\,\frac{d_e-2}{2}\,
\left[ \frac{2(d_e-2)}{d_\gamma+d_e-4} - \xi \right]\, G(1,1-\varepsilon_e)\, .
\label{sigma-1l-s}
\eea
\end{subequations}
As can be seen from Eqs.~(\ref{1loopdiagrams-s}), the one-loop results are expressed in terms of two master integrals:
$G(1,1)$ for the photon self-energy and $G(1,1-\varepsilon_e)$ for the fermion self-energy 
where $G$ is given by Eq.~(\ref{1loopG}). These equations have been discussed in Ref.~[\onlinecite{TeberRQED12}] and we refer the interested reader to this reference for more details.
These equations will be used in the next sections concerning the computation of the two-loop fermion self-energy and renormalization.

\section{Two-loop fermion self-energy}
\label{Sec:Sigma}

\subsection{The two-loop massless propagator diagram}
\label{SubSec:twoloop}

\begin{figure}
   \includegraphics{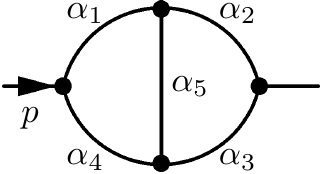}
   \caption{\label{fig:J}
   Two-loop massless propagator diagram.}
\end{figure}
%

In the one-loop case, after working out the numerator algebra, the considered diagrams could be expressed in terms of two simple scalar integrals $G(1,1)$ and $G(1,1-\varepsilon_e)$, see
Eqs.~(\ref{1loopdiagrams-s}). These one-loop integrals are known exactly from Eq.~(\ref{1loopG}). 
At the two-loop level the situation is more complicated. The two-loop massless propagator diagram is actually one of the basic building blocks of multi-loop calculations
and has a long history, see the review Ref.~[\onlinecite{Grozin12}]. For arbitrary indices $\al_i$ and external momentum $p$ 
in a Minkowski space-time of dimensionality $d_e$ the two-loop massless propagator diagram, see Fig.~\ref{fig:J}, reads:
\bea
\int \int \frac{\D^{d_e} k_1 \, \D^{d_e} k_2}{[-(k_1+p)^2]^{\al_1}\,[-(k_2+p)^2]^{\al_2}\,[-k_2^2]^{\al_3}\,[-k_1^2]^{\al_4}\,[-(k_2-k_1)^2]^{\al_5}} 
= - \frac{\pi^{d_e}}{(-p^2)^{\sum_{i=1}^5 \al_i - d_e}}\,G(\al_1,\al_2,\al_3,\al_4,\al_5)\, ,
\label{def:2loopmassless}
\eea
where, on the right-hand side, the power of $p$ follows from dimensionality and the function $G(\{\al_i\})$ is the 2-loop (dimensionless) coefficient
function of the diagram. One of the main goals of multi-loop calculations is an exact evaluation of $G(\{\al_i\})$.

When all indices are integers this function can be computed exactly using IBP identities.~\cite{VasilievPK81,ChetyrkinT81} 
The diagram can then be expressed in terms of recursively one-loop integrals as can be seen from the following well-known example:~\footnote{
The diagram $G(1,1,1,1,1)$ was actually first computed exactly in dimensional regularization with the help of the $x$-space Gegenbauer polynomial technique
in Ref.~[\onlinecite{ChetyrkinKT80}]. Subsequently, it was computed exactly with the help of IBP in Refs.~[\onlinecite{VasilievPK81,ChetyrkinT81}].}
\be
G(1,1,1,1,1) = 
C_D\left[ \quad \parbox{16mm}{
    \begin{fmfgraph*}(16,14)
      \fmfleft{i}
      \fmfright{o}
      \fmfleft{ve}
      \fmfright{vo}
      \fmftop{vn}
      \fmftop{vs}
      \fmffreeze
      \fmfforce{(-0.1w,0.5h)}{i}
      \fmfforce{(1.1w,0.5h)}{o}
      \fmfforce{(0w,0.5h)}{ve}
      \fmfforce{(1.0w,0.5h)}{vo}
      \fmfforce{(.5w,0.95h)}{vn}
      \fmfforce{(.5w,0.05h)}{vs}
      \fmffreeze
      \fmf{plain}{i,ve}
      \fmf{plain,left=0.8}{ve,vo}
      \fmf{phantom,left=0.5,label=$1$,l.d=-0.01w}{ve,vn}
      \fmf{phantom,right=0.5,label=$1$,l.d=-0.01w}{vo,vn}
      \fmf{plain,left=0.8}{vo,ve}
      \fmf{phantom,left=0.5,label=$1$,l.d=-0.01w}{vo,vs}
      \fmf{phantom,right=0.5,label=$1$,l.d=-0.01w}{ve,vs}
      \fmf{plain,label=$1$,l.d=0.05w}{vs,vn}
      \fmf{plain}{vo,o}
      \fmffreeze
      \fmfdot{ve,vn,vo,vs}
    \end{fmfgraph*}
}
\quad \right] 
=
\frac{2}{d_e-4}\, \Bigg[ (3-d_e)\,G^2(1,1) + \left( 9(d_e-2) + \frac{8}{d_e-4} \right)\,G(1,1)G(1,\varepsilon_e+\varepsilon_\gamma) \Bigg],
\label{G11111}
\ee
where $C_D$ denotes the coefficient function of the diagram.

For arbitrary indices its evaluation is however highly nontrivial: it can be represented~\cite{BierenbaumW03} as a combination
of twofold series. In some simpler cases explicit analytical results can be
obtained.~\cite{VasilievPK81,Kazakov85,Gracey92,KivelSV93+94,Kotikov96,BroadhurstGK97,BroadhurstK96,KotikovT13}
As an example, the diagram $G(1,1,1,1,\al)$ can be expressed in a compact way for a special value of
the non-integer index on the central line:~\cite{VasilievPK81,KivelSV93+94,KotikovT13}
\be
G(1,1,1,1,\lambda_e) =
C_D\left[ \quad \parbox{16mm}{
    \begin{fmfgraph*}(16,14)
      \fmfleft{i}
      \fmfright{o}
      \fmfleft{ve}
      \fmfright{vo}
      \fmftop{vn}
      \fmftop{vs}
      \fmffreeze
      \fmfforce{(-0.1w,0.5h)}{i}
      \fmfforce{(1.1w,0.5h)}{o}
      \fmfforce{(0w,0.5h)}{ve}
      \fmfforce{(1.0w,0.5h)}{vo}
      \fmfforce{(.5w,0.95h)}{vn}
      \fmfforce{(.5w,0.05h)}{vs}
      \fmffreeze
      \fmf{plain}{i,ve}
      \fmf{plain,left=0.8}{ve,vo}
      \fmf{phantom,left=0.5,label=$1$,l.d=-0.01w}{ve,vn}
      \fmf{phantom,right=0.5,label=$1$,l.d=-0.01w}{vo,vn}
      \fmf{plain,left=0.8}{vo,ve}
      \fmf{phantom,left=0.5,label=$1$,l.d=-0.01w}{vo,vs}
      \fmf{phantom,right=0.5,label=$1$,l.d=-0.01w}{ve,vs}
      \fmf{plain,label=$\lambda_e$,l.d=0.05w}{vs,vn}
      \fmf{plain}{vo,o}
      \fmffreeze
      \fmfdot{ve,vn,vo,vs}
    \end{fmfgraph*}
}
\quad \right]
=
3\,\frac{\Gamma(\lambda_e)\Gamma(1-\lambda_e)}{\Gamma(2\lambda_e)} \,\Big[ \Psi_2(\lambda_e) - \Psi_2(1) \Big]\,,\quad (\lambda_e=d_e/2-1)
\label{Ilambda}
\ee
where $\Psi_2(x) = \Psi_1'(x)$ is the trigamma function and $\Psi_1(x)$ is the digamma function. 
The result of Eq.~(\ref{Ilambda}) elegantly covers the cases of RQED$_{4,d_e}$'s where $\varepsilon_\gamma \ra 0$ and in particular RQED$_{4,3}$
where the gauge field does not renormalize.
In the general case of RQED$_{d_\gamma,d_e}$ such a diagram appears in the calculation of the 
two-loop photon self-energy but with a more general index: $\al=1-\varepsilon_e = \lambda_e +\varepsilon_\gamma$ where $\veps_\gamma$ can be non-zero. 
In this case, the result is more complicated and reads:~\cite{Kotikov96}
\bea
G(1,1,1,1,\al) &=& 
C_D\left[ \quad \parbox{16mm}{
    \begin{fmfgraph*}(16,14)
      \fmfleft{i}
      \fmfright{o}
      \fmfleft{ve}
      \fmfright{vo}
      \fmftop{vn}
      \fmftop{vs}
      \fmffreeze
      \fmfforce{(-0.1w,0.5h)}{i}
      \fmfforce{(1.1w,0.5h)}{o}
      \fmfforce{(0w,0.5h)}{ve}
      \fmfforce{(1.0w,0.5h)}{vo}
      \fmfforce{(.5w,0.95h)}{vn}
      \fmfforce{(.5w,0.05h)}{vs}
      \fmffreeze
      \fmf{plain}{i,ve}
      \fmf{plain,left=0.8}{ve,vo}
      \fmf{phantom,left=0.5,label=$1$,l.d=-0.01w}{ve,vn}
      \fmf{phantom,right=0.5,label=$1$,l.d=-0.01w}{vo,vn}
      \fmf{plain,left=0.8}{vo,ve}
      \fmf{phantom,left=0.5,label=$1$,l.d=-0.01w}{vo,vs}
      \fmf{phantom,right=0.5,label=$1$,l.d=-0.01w}{ve,vs}
      \fmf{plain,label=$\al$,l.d=0.05w}{vs,vn}
      \fmf{plain}{vo,o}
      \fmffreeze
      \fmfdot{ve,vn,vo,vs}
    \end{fmfgraph*}
}
\quad \right] \quad = \quad 
-2\, \Gamma(\lambda_e)\Gamma(\lambda_e-\al) \Gamma(1-2\lambda_e+\al) \times
\label{KotikovG}\\
&\times& \left [ \frac{\Gamma(\lambda_e)}{\Gamma(2\lambda_e)\Gamma(3\lambda_e-\al-1)}\,
\sum_{n=0}^{\infty}\,\frac{\Gamma(n+2\lambda_e)\Gamma(n+1)}{n!\,\Gamma(n+1+\al)}\,\frac{1}{n+1-\lambda_e+\al}
+\frac{\pi \cot \pi (2\lambda_e-\al)}{\Gamma(2\lambda_e)} \right ]\, ,
\nonum
\eea
where the one-fold series corresponds to a generalized hypergeometric function, ${}_3F_2$, of argument $1$.
The diagram evaluated in Eq.~(\ref{KotikovG}) is one of the simplest among a class of complicated diagrams 
that have been considered in Ref.~[\onlinecite{Kotikov96}] on the basis of a new development of the Gegenbauer polynomial technique.
This class includes diagrams with two adjacent lines having integer indices while three other lines have arbitrary indices; the corresponding coefficient
function is given by: $G(\al,1,\beta,\gamma,1)$. As proved in Ref.~[\onlinecite{Kotikov96}] all of these diagrams may be expressed in terms of 
generalized hypergeometric functions ${}_3F_2$ of argument $1$.
For this class of diagrams, similar results have been found in Ref.~[\onlinecite{BroadhurstGK97}] using an ansatz to solve the
recurrence relations for the two-loop diagram. However, beyond the simplest case with a single non-integer index represented by Eq.~(\ref{KotikovG}),
explicit expressions for other diagrams of this class are unknown.

As will be seen below, it turns out that the computation of the two-loop fermion self-energy for a general RQED$_{d_\gamma,d_e}$ requires
the knowledge of a two-loop massless propagator diagram with two non-integer indices on non-adjacent lines:
\be
G(\al,1,\beta,1,1) =
C_D\left[ \quad \parbox{16mm}{
    \begin{fmfgraph*}(16,14)
      \fmfleft{i}
      \fmfright{o}
      \fmfleft{ve}
      \fmfright{vo}
      \fmftop{vn}
      \fmftop{vs}
      \fmffreeze
      \fmfforce{(-0.1w,0.5h)}{i}
      \fmfforce{(1.1w,0.5h)}{o}
      \fmfforce{(0w,0.5h)}{ve}
      \fmfforce{(1.0w,0.5h)}{vo}
      \fmfforce{(.5w,0.95h)}{vn}
      \fmfforce{(.5w,0.05h)}{vs}
      \fmffreeze
      \fmf{plain}{i,ve}
      \fmf{plain,left=0.8}{ve,vo}
      \fmf{phantom,left=0.5,label=$\al$,l.d=-0.01w}{ve,vn}
      \fmf{phantom,right=0.5,label=$1$,l.d=-0.01w}{vo,vn}
      \fmf{plain,left=0.8}{vo,ve}
      \fmf{phantom,left=0.5,label=$\beta$,l.d=-0.01w}{vo,vs}
      \fmf{phantom,right=0.5,label=$1$,l.d=-0.01w}{ve,vs}
      \fmf{plain,label=$1$,l.d=0.05w}{vs,vn}
      \fmf{plain}{vo,o}
      \fmffreeze
      \fmfdot{ve,vn,vo,vs}
    \end{fmfgraph*}
}
\quad \right]\, .
\label{complicatedG}
\ee
Using the results of Ref.~[\onlinecite{Kotikov96}] an exact explicit expression for this diagram can be derived, see App.~\ref{App:G(alpha,beta)} for general formulas.
As can be seen from the latter, the result is of higher complexity than the one of Eq.~(\ref{KotikovG}). For the sake of simplicity we shall not consider the general solution here.
Instead, and in the spirit of Ref.~[\onlinecite{KotikovT13}], we shall focus on the less general but important case of RQED$_{4,d_e}$'s 
for which $\varepsilon_\gamma \ra 0$. As will be shown below, in this case
a simpler solution can be found using integration by parts.

\begin{figure}
    \includegraphics{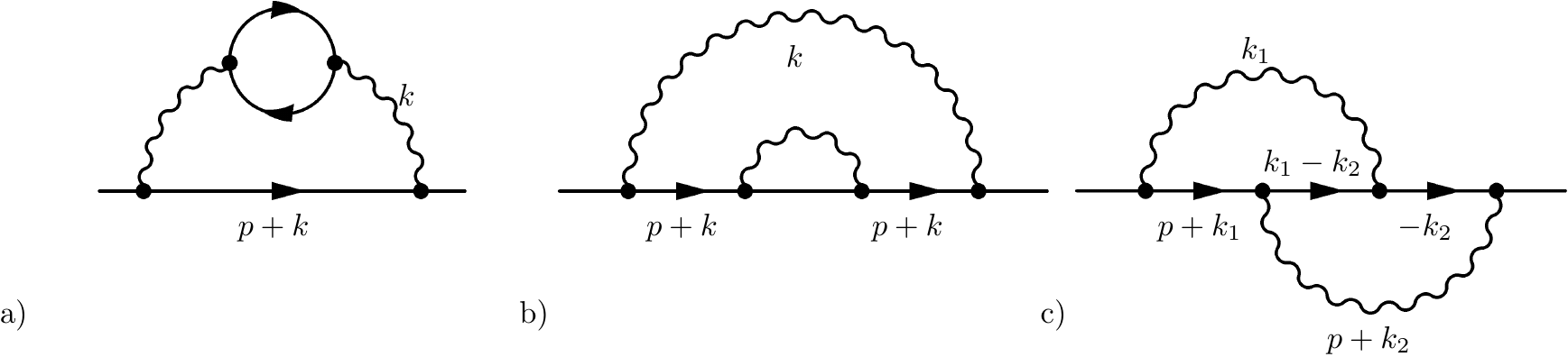}
    \caption{\label{fig:rqed-2loop-self-energy}
     2-loop fermion self-energy diagrams: a) bubble diagram, b) rainbow diagram and c) crossed photon diagram.}
\end{figure}
\FloatBarrier

\subsection{Reduction to master integrals}

Having set all the perturbative framework in Sec.~\ref{Sec:RQED} and presented the two-loop massless propagator diagram in Sec.~\ref{SubSec:twoloop},
we now proceed on computing the two-loop fermion self-energy. The latter consists of three diagrams that are displayed on Fig.~\ref{fig:rqed-2loop-self-energy}.

The first diagram, Fig.~\ref{fig:rqed-2loop-self-energy}a,
is the so-called bubble diagram which is defined as:
\bea
&&-i\Sigma_{2a}(p) =
\label{sigma-a}
\\
&&\int \frac{\D^{d_e}k}{(2\pi)^{d_e}}\, (-ie\gamma^\al) \, \frac{i(\Sp +\Sk)}{(p+k)^2} \, (-ie\gamma^\beta) \,
\frac{i}{(4\pi)^{\varepsilon_e}}\frac{\Gamma(1-\varepsilon_e)}{(-k^2)^{1-\varepsilon_e}}
\left( g_{\al \mu} - \tilde{\xi}\,\frac{k_\al k_\mu}{k^2} \right)\,i\Pi_1^{\mu \nu}(k)\,\frac{i}{(4\pi)^{\varepsilon_e}}\frac{\Gamma(1-\varepsilon_e)}{(-k^2)^{1-\varepsilon_e}}
\left( g_{\nu \beta} - \tilde{\xi}\,\frac{k_\nu k_\beta}{k^2} \right)\,,
\nonum
\eea
where the lower index $2$ in $\Sigma_{2a}$ refers to 2-loop and $\Pi_1^{\mu \nu}(k)$ is a one-loop photon self-energy insertion corresponding to Eq.~(\ref{pimunu-1l-s}). 
This diagram is obviously gauge invariant. This follows from current conservation which implies that
$k_\mu\Pi^{\mu \nu}=0$ so that all terms proportional to the gauge fixing term vanish. Using Eq.~(\ref{sigmaV}) and
performing all the numerator algebra with the help of Eqs.~(\ref{num-algebra}) yields: 
\bea
\Sigma_{V2a}(p^2) = d\,\frac{e^4 \Gamma^2(1-\varepsilon_e)}{(4\pi)^{d_\gamma} (-p^2)^{2\varepsilon_\gamma}}\,\frac{(d_e-2)^2}{2(2d_\gamma-d_e-6)}\,G(1,1)G(1,\varepsilon_\gamma-\varepsilon_e),
\label{sigma-a-s}
\eea
where the factor of $d$ is due to the fermion loop and the master integral, $G(1,1)G(1,\varepsilon_\gamma-\varepsilon_e)$, 
is a simple product of two one-loop integrals which can be straightforwardly evaluated with the help of Eq.~(\ref{1loopG}).

The second diagram, Fig.~\ref{fig:rqed-2loop-self-energy}b, is the so-called rainbow diagram which is defined as:
\bea
-i\Sigma_{2b}(p) = \int \frac{\D^{d_e}k}{(2\pi)^{d_e}}\, (-ie\gamma^\mu) \frac{i(\Sp +\Sk)}{(p+k)^2} \, \left(-i\Sigma_1(p+k)\right) \, \frac{i(\Sp +\Sk)}{(p+k)^2} (-ie \gamma^\nu)\,
\frac{i}{(4\pi)^{\varepsilon_e}}\frac{\Gamma(1-\varepsilon_e)}{(-k^2)^{1-\varepsilon_e}}
\left( g_{\mu \nu} - \tilde{\xi}\,\frac{k_\mu k_\nu}{k^2} \right),
\label{sigma-b}
\eea
where $\Sigma_1(k) = \Sk \Sigma_{V1}(k^2)$ is a one-loop fermion self-energy insertion corresponding to Eq.~(\ref{sigma-1l-s}).
This diagram is not gauge invariant.
Proceeding along the same lines as for the bubble diagram yields:
\bea
\Sigma_{V2b}(p^2) &=& \frac{e^4 \Gamma^2(1-\varepsilon_e)}{(4\pi)^{d_\gamma} (-p^2)^{2\varepsilon_\gamma}}\,\frac{(d_e-2)(d_\gamma-3)(d_\gamma+d_e-4)}{2(d_\gamma-4)}\, 
\left( \frac{2(d_e-2)}{d_\gamma+d_e-4} - \xi \right)^2\,G(1,1-\varepsilon_e)G(1-\varepsilon_e,\varepsilon_\gamma),
\label{sigma-b-s}
\eea
where the master integral, $G(1,1-\varepsilon_e)G(1-\varepsilon_e,\varepsilon_\gamma)$,
is also a simple product of two one-loop integrals which can be straightforwardly evaluated with the help of Eq.~(\ref{1loopG}).

Finally, the third diagram, Fig.~\ref{fig:rqed-2loop-self-energy}c, which is the so-called crossed photon diagram 
is a truly two-loop diagram. It is defined as:
\bea
-i\Sigma_{2c}(p) =&& \int \frac{\D^{d_e}k_1\,\D^{d_e}k_2}{(2\pi)^{2d_e}}\, (-ie\gamma^\mu) \frac{-i\Sk_2}{k_2^2}
\frac{i}{(4\pi)^{\varepsilon_e}}\frac{\Gamma(1-\varepsilon_e)}{(-(p+k_2)^2)^{1-\varepsilon_e}}
\left( g_{\mu \beta} - \tilde{\xi}\,\frac{(p+k_2)_\mu (p+k_2)_\beta}{(p+k_2)^2} \right) \times 
\label{sigma-c}
\\
&&\times (-ie\gamma^\al)\,\frac{i(\Sk_1 -\Sk_2)}{(k_1-k_2)^2} \,
\frac{i}{(4\pi)^{\varepsilon_e}}\frac{\Gamma(1-\varepsilon_e)}{(-k_1^2)^{1-\varepsilon_e}}
\left( g_{\al \nu} - \tilde{\xi}\,\frac{k_{1\,\al} k_{1\,\nu}}{k_1^2} \right)\,(-ie\gamma^\beta)\,\frac{i(\Sp + \Sk_1)}{(p+k_1)^2}\,(-ie\gamma^\nu)\,.
\nonum
\eea
Proceeding along the same lines as for the two previous diagrams yields:
\bea
\Sigma_{V2c}(p^2) &=& - \frac{e^4 \Gamma^2(1-\varepsilon_e)}{(4\pi)^{d_\gamma} (-p^2)^{2\varepsilon_\gamma}}\,\frac{d_e-2}{2}\,
\label{sigma-c-s} \\
&\times& \Bigg \{ \left( d_e-4-4\,\frac{d_\gamma-d_e}{d_\gamma+d_e-4} -2\frac{(d_e-2)^2}{d_\gamma+d_e-4}\,\xi +\frac{d_e-2}{2}\,\xi^2 \right)\,G^2(1,1-\varepsilon_e) + \Bigg .
\nonum \\
&& \Bigg . + \left( d_e + 4 + 8\,\frac{d_e-2}{d_\gamma+d_e-4}+\frac{16}{d_\gamma-4} +4 \frac{(d_e-2)(d_\gamma-3)}{d_\gamma+d_e-4}\,\xi -(d_\gamma-3)\,\xi^2 \right)\,
G(1,1-\varepsilon_e)G(1-\varepsilon_e,\varepsilon_\gamma) + \Bigg .
\nonum \\
&& \Bigg . +4G(1-\varepsilon_e,1,-1-\varepsilon_e,1,1) - 4 G(1-\varepsilon_e,1,-\varepsilon_e,1,1)+(8-d_e)G(-\varepsilon_e,1,-\varepsilon_e,1,1) \Bigg \}\, ,
\nonum
\eea
where, besides the primitively one-loop master integrals, $G^2(1,1-\varepsilon_e)$ and $G(1,1-\varepsilon_e)G(1-\varepsilon_e,\varepsilon_\gamma)$,
complicated two-loop propagator diagrams of the type of Eq.~(\ref{complicatedG}) enter the expression of the self-energy. It is interesting to notice that
none of the complicated terms depend on the gauge parameter, $\xi$. That is, working in the Feynman gauge, $\xi=0$, simplifies the calculations but does
not reduce the complexity of the diagrams. 

\subsection{Integration by parts and self-energy of the crossed photon diagram}
\label{Subsec:IBP}

In order to further reduce the complicated two-loop diagrams appearing in Eq.~(\ref{sigma-c-s}) we use an IBP identity which follows from the 
homogeneity of the two-loop propagator diagram, Eq.~(\ref{def:2loopmassless}), in $p$, see Ref.~[\onlinecite{ChetyrkinT81}] and Ref.~[\onlinecite{Grozin07}]
for a review. In graphical form, this IBP reads:
\bea
\left( \frac{d_e}{2}+\al_4-\al_1-\al_2-\al_5 \right) \quad
\parbox{16mm}{
    \begin{fmfgraph*}(16,14)
      \fmfleft{i}
      \fmfright{o}
      \fmfleft{ve}
      \fmfright{vo}
      \fmftop{vn}
      \fmftop{vs}
      \fmffreeze
      \fmfforce{(-0.1w,0.5h)}{i}
      \fmfforce{(1.1w,0.5h)}{o}
      \fmfforce{(0w,0.5h)}{ve}
      \fmfforce{(1.0w,0.5h)}{vo}
      \fmfforce{(.5w,0.95h)}{vn}
      \fmfforce{(.5w,0.05h)}{vs}
      \fmffreeze
      \fmf{plain}{i,ve}
      \fmf{plain,left=0.8}{ve,vo}
      \fmf{phantom,left=0.5,label=$\al_1$,l.d=-0.01w}{ve,vn}
      \fmf{phantom,right=0.5,label=$\al_2$,l.d=-0.01w}{vo,vn}
      \fmf{plain,left=0.8}{vo,ve}
      \fmf{phantom,left=0.5,label=$\al_3$,l.d=-0.01w}{vo,vs}
      \fmf{phantom,right=0.5,label=$\al_4$,l.d=-0.01w}{ve,vs}
      \fmf{plain,label=$\al_5$,l.d=0.05w}{vs,vn}
      \fmf{plain}{vo,o}
      \fmffreeze
      \fmfdot{ve,vn,vo,vs}
    \end{fmfgraph*}
}
& & \quad  = \quad
\al_3 \left( \quad
\parbox{16mm}{
    \begin{fmfgraph*}(16,14)
      \fmfleft{i}
      \fmfright{o}
      \fmfleft{ve}
      \fmfright{vo}
      \fmftop{vn}
      \fmftop{vs}
      \fmffreeze
      \fmfforce{(-0.1w,0.5h)}{i}
      \fmfforce{(1.1w,0.5h)}{o}
      \fmfforce{(0w,0.5h)}{ve}
      \fmfforce{(1.0w,0.5h)}{vo}
      \fmfforce{(.5w,0.95h)}{vn}
      \fmfforce{(.5w,0.05h)}{vs}
      \fmffreeze
      \fmf{plain}{i,ve}
      \fmf{plain,left=0.8}{ve,vo}
      \fmf{phantom,left=0.5}{ve,vn}
      \fmf{phantom,right=0.4}{vo,vn}
      \fmf{plain,left=0.8}{vo,ve}
      \fmf{phantom,left=0.5,label=$+$,l.d=-0.01w}{vo,vs}
      \fmf{phantom,right=0.5}{ve,vs}
      \fmf{plain,label=$-$,l.d=0.05w}{vs,vn}
      \fmf{plain}{vo,o}
      \fmffreeze
      \fmfdot{ve,vn,vo,vs}
    \end{fmfgraph*}
} \qquad - \qquad
\parbox{16mm}{
    \begin{fmfgraph*}(16,14)
      \fmfleft{i}
      \fmfright{o}
      \fmfleft{ve}
      \fmfright{vo}
      \fmftop{vn}
      \fmftop{vs}
      \fmffreeze
      \fmfforce{(-0.1w,0.5h)}{i}
      \fmfforce{(1.1w,0.5h)}{o}
      \fmfforce{(0w,0.5h)}{ve}
      \fmfforce{(1.0w,0.5h)}{vo}
      \fmfforce{(.5w,0.95h)}{vn}
      \fmfforce{(.5w,0.05h)}{vs}
      \fmffreeze
      \fmf{plain}{i,ve}
      \fmf{plain,left=0.8}{ve,vo}
      \fmf{phantom,left=0.4}{ve,vn}
      \fmf{phantom,right=0.4}{vo,vn}
      \fmf{plain,left=0.8}{vo,ve}
      \fmf{phantom,left=0.5,label=$+$,l.d=-0.01w}{vo,vs}
      \fmf{phantom,right=0.5,label=$-$,l.d=-0.01w}{ve,vs}
      \fmf{plain}{vs,vn}
      \fmf{plain}{vo,o}
      \fmffreeze
      \fmfdot{ve,vn,vo,vs}
    \end{fmfgraph*}
}
\quad \right)
\label{IBP:2loop-homog}\\
& & \quad
\nonum \\
& & + \left(\frac{3d_e}{2} - \sum_{i=1}^5\al_i \right) \,(-p^2)^{-1}\, \left( \quad
\parbox{16mm}{
    \begin{fmfgraph*}(16,14)
      \fmfleft{i}
      \fmfright{o}
      \fmfleft{ve}
      \fmfright{vo}
      \fmftop{vn}
      \fmftop{vs}
      \fmffreeze
      \fmfforce{(-0.1w,0.5h)}{i}
      \fmfforce{(1.1w,0.5h)}{o}
      \fmfforce{(0w,0.5h)}{ve}
      \fmfforce{(1.0w,0.5h)}{vo}
      \fmfforce{(.5w,0.95h)}{vn}
      \fmfforce{(.5w,0.05h)}{vs}
      \fmffreeze
      \fmf{plain}{i,ve}
      \fmf{plain,left=0.8}{ve,vo}
      \fmf{phantom,left=0.5,label=$-$,l.d=-0.01w}{ve,vn}
      \fmf{phantom,right=0.4}{vo,vn}
      \fmf{plain,left=0.8}{vo,ve}
      \fmf{phantom,left=0.5}{vo,vs}
      \fmf{phantom,right=0.5}{ve,vs}
      \fmf{plain}{vs,vn}
      \fmf{plain}{vo,o}
      \fmffreeze
      \fmfdot{ve,vn,vo,vs}
    \end{fmfgraph*}
} \qquad - \qquad
\parbox{16mm}{
    \begin{fmfgraph*}(16,14)
      \fmfleft{i}
      \fmfright{o}
      \fmfleft{ve}
      \fmfright{vo}
      \fmftop{vn}
      \fmftop{vs}
      \fmffreeze
      \fmfforce{(-0.1w,0.5h)}{i}
      \fmfforce{(1.1w,0.5h)}{o}
      \fmfforce{(0w,0.5h)}{ve}
      \fmfforce{(1.0w,0.5h)}{vo}
      \fmfforce{(.5w,0.95h)}{vn}
      \fmfforce{(.5w,0.05h)}{vs}
      \fmffreeze
      \fmf{plain}{i,ve}
      \fmf{plain,left=0.8}{ve,vo}
      \fmf{phantom,left=0.4}{ve,vn}
      \fmf{phantom,right=0.4}{vo,vn}
      \fmf{plain,left=0.8}{vo,ve}
      \fmf{phantom,left=0.5}{vo,vs}
      \fmf{phantom,right=0.5,label=$-$,l.d=-0.01w}{ve,vs}
      \fmf{plain}{vs,vn}
      \fmf{plain}{vo,o}
      \fmffreeze
      \fmfdot{ve,vn,vo,vs}
    \end{fmfgraph*}
} \quad \right),
\nonum \\
& & \quad
\nonum 
\eea
where $\pm$ on the right-hand side of the equation denotes the increase or decrease of a line index by $1$ with respect to its value on the left-hand side.
This expression slightly simplifies in the cases we are interested in: 
\bea
\left( 1 - \varepsilon_e - \varepsilon_\gamma -\al \right) &\quad&
\parbox{16mm}{
    \begin{fmfgraph*}(16,14)
      \fmfleft{i}
      \fmfright{o}
      \fmfleft{ve}
      \fmfright{vo}
      \fmftop{vn}
      \fmftop{vs}
      \fmffreeze
      \fmfforce{(-0.1w,0.5h)}{i}
      \fmfforce{(1.1w,0.5h)}{o}
      \fmfforce{(0w,0.5h)}{ve}
      \fmfforce{(1.0w,0.5h)}{vo}
      \fmfforce{(.5w,0.95h)}{vn}
      \fmfforce{(.5w,0.05h)}{vs}
      \fmffreeze
      \fmf{plain}{i,ve}
      \fmf{plain,left=0.8}{ve,vo}
      \fmf{phantom,left=0.5,label=$\al$,l.d=-0.01w}{ve,vn}
      \fmf{phantom,right=0.5,label=$1$,l.d=-0.01w}{vo,vn}
      \fmf{plain,left=0.8}{vo,ve}
      \fmf{phantom,left=0.5,label=$\beta$,l.d=-0.01w}{vo,vs}
      \fmf{phantom,right=0.5,label=$1$,l.d=-0.01w}{ve,vs}
      \fmf{plain,label=$1$,l.d=0.05w}{vs,vn}
      \fmf{plain}{vo,o}
      \fmffreeze
      \fmfdot{ve,vn,vo,vs}
    \end{fmfgraph*}
} \quad  = \quad
\beta \left( \quad
\parbox{24mm}{
  \begin{fmfgraph*}(24,14)
    \fmfleft{i}
    \fmfright{o}
    \fmfleft{ve}
    \fmfright{vo}
    \fmftop{v}
    \fmffreeze
    \fmfforce{(-0.1w,0.5h)}{i}
    \fmfforce{(1.1w,0.5h)}{o}
    \fmfforce{(0w,0.5h)}{ve}
    \fmfforce{(1.0w,0.5h)}{vo}
    \fmfforce{(.5w,0.5h)}{v}
    \fmffreeze
    \fmf{plain}{i,ve}
    \fmf{plain,left=0.8,label=$\al$,l.s=left,l.d=0.03w}{ve,v}
    \fmf{plain,left=0.8,label=$1$,l.s=left,l.d=0.03w}{v,ve}
    \fmf{plain,left=0.8,label=$1$,l.s=left,l.d=0.03w}{v,vo}
    \fmf{plain,left=0.8,label=$\beta+1$,l.d=0.03w}{vo,v}
    \fmf{plain}{vo,o}
    \fmffreeze
    \fmfdot{ve,v,vo}
  \end{fmfgraph*}
} \qquad - \qquad
\parbox{16mm}{
  \begin{fmfgraph*}(16,16)
    \fmfleft{i}
    \fmfright{o}
    \fmfleft{ve}
    \fmfright{vo}
    \fmftop{v}
    \fmffreeze
    \fmfforce{(-0.1w,0.5h)}{i}
    \fmfforce{(1.1w,0.5h)}{o}
    \fmfforce{(0w,0.5h)}{vl}
    \fmfforce{(1.0w,0.5h)}{vr}
    \fmfforce{(.5w,0.9h)}{vt}
    \fmfforce{(.5w,0.1h)}{vb}
    \fmffreeze
    \fmf{plain}{i,vl}
    \fmf{plain,left=0.8}{vl,vr}
    \fmf{plain,left=0.8}{vr,vl}
    \fmf{phantom,left=0.1,label=$\al$,l.s=left}{vl,vt}
    \fmf{plain,left=0.5,label=$1$,l.d=0.05w}{vt,vl}
    \fmf{phantom,right=0.1,label=$1$,l.s=right}{vr,vt}
    \fmf{phantom,left=0.1,label=$\beta+1$,l.s=left,l.d=0.4w}{vr,vl}
    \fmf{plain}{vr,o}
    \fmffreeze
    \fmfdot{vl,vt,vr}
  \end{fmfgraph*}
}
\quad \right)
\label{IBP:2loop-b}\\
& & \quad
\nonum \\
& & + \left( 3- 3\varepsilon_e - 3\varepsilon_\gamma - \al -\beta \right) \,(-p^2)^{-1}\, \left( \qquad
\parbox{16mm}{
    \begin{fmfgraph*}(16,14)
      \fmfleft{i}
      \fmfright{o}
      \fmfleft{ve}
      \fmfright{vo}
      \fmftop{vn}
      \fmftop{vs}
      \fmffreeze
      \fmfforce{(-0.1w,0.5h)}{i}
      \fmfforce{(1.1w,0.5h)}{o}
      \fmfforce{(0w,0.5h)}{ve}
      \fmfforce{(1.0w,0.5h)}{vo}
      \fmfforce{(.5w,0.95h)}{vn}
      \fmfforce{(.5w,0.05h)}{vs}
      \fmffreeze
      \fmf{plain}{i,ve}
      \fmf{plain,left=0.8}{ve,vo}
      \fmf{phantom,left=0.5,label=$\al-1$,l.d=-0.01w}{ve,vn}
      \fmf{phantom,right=0.5,label=$1$,l.d=-0.01w}{vo,vn}
      \fmf{plain,left=0.8}{vo,ve}
      \fmf{phantom,left=0.5,label=$\beta$,l.d=-0.01w}{vo,vs}
      \fmf{phantom,right=0.5,label=$1$,l.d=-0.01w}{ve,vs}
      \fmf{plain,label=$1$,l.d=0.05w}{vs,vn}
      \fmf{plain}{vo,o}
      \fmffreeze
      \fmfdot{ve,vn,vo,vs}
    \end{fmfgraph*}
}
\qquad - \qquad
\parbox{16mm}{
  \begin{fmfgraph*}(16,16)
    \fmfleft{i}
    \fmfright{o}
    \fmfleft{ve}
    \fmfright{vo}
    \fmftop{v}
    \fmffreeze
    \fmfforce{(-0.1w,0.5h)}{i}
    \fmfforce{(1.1w,0.5h)}{o}
    \fmfforce{(0w,0.5h)}{vl}
    \fmfforce{(1.0w,0.5h)}{vr}
    \fmfforce{(.5w,0.9h)}{vt}
    \fmfforce{(.5w,0.1h)}{vb}
    \fmffreeze
    \fmf{plain}{i,vl}
    \fmf{plain,left=0.8}{vl,vr}
    \fmf{plain,left=0.8}{vr,vl}
    \fmf{phantom,left=0.1,label=$\al$,l.s=left}{vl,vt}
    \fmf{plain,left=0.5,label=$1$,l.d=0.05w}{vt,vl}
    \fmf{phantom,right=0.1,label=$1$,l.s=right}{vr,vt}
    \fmf{phantom,left=0.1,label=$\beta$,l.s=left,l.d=0.4w}{vr,vl}
    \fmf{plain}{vr,o}
    \fmffreeze
    \fmfdot{vl,vt,vr}
  \end{fmfgraph*}
} \quad \right)\, ,
\nonum \\
& & \quad
\nonum 
\eea
where we have used Eq.~(\ref{d-params}) and the fact that, in $p$-space, a line with zero index shrinks to a point.
As a consequence, on the right-hand side of Eq.~(\ref{IBP:2loop-b}) all diagrams except the third one are recursively one-loop diagram.

In order to see how Eq.~(\ref{IBP:2loop-b}) can be usefully implemented, let's consider the case where $\al = \beta = 1-\varepsilon_e$.
In this case Eq.~(\ref{IBP:2loop-b}) reduces to:
\bea
- \varepsilon_\gamma \quad
\parbox{16mm}{
    \begin{fmfgraph*}(16,14)
      \fmfleft{i}
      \fmfright{o}
      \fmfleft{ve}
      \fmfright{vo}
      \fmftop{vn}
      \fmftop{vs}
      \fmffreeze
      \fmfforce{(-0.1w,0.5h)}{i}
      \fmfforce{(1.1w,0.5h)}{o}
      \fmfforce{(0w,0.5h)}{ve}
      \fmfforce{(1.0w,0.5h)}{vo}
      \fmfforce{(.5w,0.95h)}{vn}
      \fmfforce{(.5w,0.05h)}{vs}
      \fmffreeze
      \fmf{plain}{i,ve}
      \fmf{plain,left=0.8}{ve,vo}
      \fmf{phantom,left=0.5,label=$1-\varepsilon_e$,l.d=-0.01w}{ve,vn}
      \fmf{phantom,right=0.5,label=$1$,l.d=-0.01w}{vo,vn}
      \fmf{plain,left=0.8}{vo,ve}
      \fmf{phantom,left=0.5,label=$1-\varepsilon_e$,l.d=-0.01w}{vo,vs}
      \fmf{phantom,right=0.5,label=$1$,l.d=-0.01w}{ve,vs}
      \fmf{plain,label=$1$,l.d=0.05w}{vs,vn}
      \fmf{plain}{vo,o}
      \fmffreeze
      \fmfdot{ve,vn,vo,vs}
    \end{fmfgraph*}
} \quad  &=& \quad
(1-\varepsilon_e)\, \left( \quad
\parbox{24mm}{
  \begin{fmfgraph*}(24,14)
    \fmfleft{i}
    \fmfright{o}
    \fmfleft{ve}
    \fmfright{vo}
    \fmftop{v}
    \fmffreeze
    \fmfforce{(-0.1w,0.5h)}{i}
    \fmfforce{(1.1w,0.5h)}{o}
    \fmfforce{(0w,0.5h)}{ve}
    \fmfforce{(1.0w,0.5h)}{vo}
    \fmfforce{(.5w,0.5h)}{v}
    \fmffreeze
    \fmf{plain}{i,ve}
    \fmf{plain,left=0.8,label=$1-\varepsilon_e$,l.s=left,l.d=0.03w}{ve,v}
    \fmf{plain,left=0.8,label=$1$,l.s=left,l.d=0.03w}{v,ve}
    \fmf{plain,left=0.8,label=$1$,l.s=left,l.d=0.03w}{v,vo}
    \fmf{plain,left=0.8,label=$2-\varepsilon_e$,l.d=0.03w}{vo,v}
    \fmf{plain}{vo,o}
    \fmffreeze
    \fmfdot{ve,v,vo}
  \end{fmfgraph*}
} \qquad - \qquad
\parbox{16mm}{
  \begin{fmfgraph*}(16,16)
    \fmfleft{i}
    \fmfright{o}
    \fmfleft{ve}
    \fmfright{vo}
    \fmftop{v}
    \fmffreeze
    \fmfforce{(-0.1w,0.5h)}{i}
    \fmfforce{(1.1w,0.5h)}{o}
    \fmfforce{(0w,0.5h)}{vl}
    \fmfforce{(1.0w,0.5h)}{vr}
    \fmfforce{(.5w,0.9h)}{vt}
    \fmfforce{(.5w,0.1h)}{vb}
    \fmffreeze
    \fmf{plain}{i,vl}
    \fmf{plain,left=0.8}{vl,vr}
    \fmf{plain,left=0.8}{vr,vl}
    \fmf{phantom,left=0.1,label=$1-\varepsilon_e$,l.s=left}{vl,vt}
    \fmf{plain,left=0.5,label=$1$,l.d=0.05w}{vt,vl}
    \fmf{phantom,right=0.1,label=$1$,l.s=right}{vr,vt}
    \fmf{phantom,left=0.1,label=$2-\varepsilon_e$,l.s=left,l.d=0.4w}{vr,vl}
    \fmf{plain}{vr,o}
    \fmffreeze
    \fmfdot{vl,vt,vr}
  \end{fmfgraph*}
}
\quad \right)
\label{IBP:2loop-c}\\
& & \quad
\nonum \\
& & + \left( 1 - \varepsilon_e - 3\varepsilon_\gamma \right) \,(-p^2)^{-1}\, \left( \qquad
\parbox{16mm}{
    \begin{fmfgraph*}(16,14)
      \fmfleft{i}
      \fmfright{o}
      \fmfleft{ve}
      \fmfright{vo}
      \fmftop{vn}
      \fmftop{vs}
      \fmffreeze
      \fmfforce{(-0.1w,0.5h)}{i}
      \fmfforce{(1.1w,0.5h)}{o}
      \fmfforce{(0w,0.5h)}{ve}
      \fmfforce{(1.0w,0.5h)}{vo}
      \fmfforce{(.5w,0.95h)}{vn}
      \fmfforce{(.5w,0.05h)}{vs}
      \fmffreeze
      \fmf{plain}{i,ve}
      \fmf{plain,left=0.8}{ve,vo}
      \fmf{phantom,left=0.5,label=$1-\varepsilon_e$,l.d=-0.01w}{ve,vn}
      \fmf{phantom,right=0.5,label=$1$,l.d=-0.01w}{vo,vn}
      \fmf{plain,left=0.8}{vo,ve}
      \fmf{phantom,left=0.5,label=$-\varepsilon_e$,l.d=-0.01w}{vo,vs}
      \fmf{phantom,right=0.5,label=$1$,l.d=-0.01w}{ve,vs}
      \fmf{plain,label=$1$,l.d=0.05w}{vs,vn}
      \fmf{plain}{vo,o}
      \fmffreeze
      \fmfdot{ve,vn,vo,vs}
    \end{fmfgraph*}
}
\qquad - \qquad
\parbox{16mm}{
  \begin{fmfgraph*}(16,16)
    \fmfleft{i}
    \fmfright{o}
    \fmfleft{ve}
    \fmfright{vo}
    \fmftop{v}
    \fmffreeze
    \fmfforce{(-0.1w,0.5h)}{i}
    \fmfforce{(1.1w,0.5h)}{o}
    \fmfforce{(0w,0.5h)}{vl}
    \fmfforce{(1.0w,0.5h)}{vr}
    \fmfforce{(.5w,0.9h)}{vt}
    \fmfforce{(.5w,0.1h)}{vb}
    \fmffreeze
    \fmf{plain}{i,vl}
    \fmf{plain,left=0.8}{vl,vr}
    \fmf{plain,left=0.8}{vr,vl}
    \fmf{phantom,left=0.1,label=$1-\varepsilon_e$,l.s=left}{vl,vt}
    \fmf{plain,left=0.5,label=$1$,l.d=0.05w}{vt,vl}
    \fmf{phantom,right=0.1,label=$1$,l.s=right}{vr,vt}
    \fmf{phantom,left=0.1,label=$1-\varepsilon_e$,l.s=left,l.d=0.4w}{vr,vl}
    \fmf{plain}{vr,o}
    \fmffreeze
    \fmfdot{vl,vt,vr}
  \end{fmfgraph*}
} \quad \right)\, ,
\nonum \\
& & \quad
\nonum 
\eea
where we have used the symmetry of the two-loop diagram: $G(\al,1,\beta,1,1) = G(\beta,1,\al,1,1)$.
We then see that the truly two-loop diagram appearing on the right-hand side of Eq.~(\ref{IBP:2loop-c}) 
has a coefficient function $G(1-\varepsilon_e,1,-\varepsilon_e,1,1)$. It is therefore one of the complicated diagrams appearing in Eq.~(\ref{sigma-c-s}) that
we would like to compute.
At this point, it is important to notice that the function $G(1-\varepsilon_e,1,-\varepsilon_e,1,1)$ is associated
with an ultra-violet singular two-loop diagram in the limit $\varepsilon_\gamma \ra 0$. 
Indeed, power counting shows that this diagram is proportional to: $(-p^2)^{-2\varepsilon_\gamma}$, {\it i.e.}, diverges logarithmically.
On the other-hand, the two-loop diagram appearing on the left-hand side of Eq.~(\ref{IBP:2loop-c}), the coefficient function of which
is given by: $G(1-\varepsilon_e,1,1-\varepsilon_e,1,1)$, is proportional to $(-p^2)^{-1-2\varepsilon_\gamma}$; it is therefore ultra-violet convergent
in the limit $\varepsilon_\gamma \ra 0$. As a consequence, the Laurent series associated to its $\varepsilon_\gamma$-expansion reduces to the regular part:
\be
G(1-\varepsilon_e,1,1-\varepsilon_e,1,1) = \sum_{n=0}^\infty\,c_n(\varepsilon_e)\,\varepsilon_\gamma^n.
\ee
For $\varepsilon_e=0$, which applies to QED$_4$ and QED$_3$, this diagram corresponds to the well-known Eq.~(\ref{G11111}). On the other hand, for an arbitrary non-integer 
$\varepsilon_e$, for example $\varepsilon_e=1/2$ in the case of RQED$_{4,3}$, it belongs to the class of complicated diagrams, see Eq.~(\ref{complicatedG}) and related discussion
around this equation, whose explicit analytic expression, and hence the coefficients $c_n$, is presently unknown. 
However, the fact that this diagram is convergent together with its coefficient $\varepsilon_\gamma$, implies that the
left-hand side of Eq.~(\ref{IBP:2loop-c}) vanishes in the limit $\varepsilon_\gamma \ra 0$. Hence, from the right-hand side
of Eq.~(\ref{IBP:2loop-c}) we find that:
\bea
\parbox{16mm}{
    \begin{fmfgraph*}(16,14)
      \fmfleft{i}
      \fmfright{o}
      \fmfleft{ve}
      \fmfright{vo}
      \fmftop{vn}
      \fmftop{vs}
      \fmffreeze
      \fmfforce{(-0.1w,0.5h)}{i}
      \fmfforce{(1.1w,0.5h)}{o}
      \fmfforce{(0w,0.5h)}{ve}
      \fmfforce{(1.0w,0.5h)}{vo}
      \fmfforce{(.5w,0.95h)}{vn}
      \fmfforce{(.5w,0.05h)}{vs}
      \fmffreeze
      \fmf{plain}{i,ve}
      \fmf{plain,left=0.8}{ve,vo}
      \fmf{phantom,left=0.5,label=$1-\varepsilon_e$,l.d=-0.01w}{ve,vn}
      \fmf{phantom,right=0.5,label=$1$,l.d=-0.01w}{vo,vn}
      \fmf{plain,left=0.8}{vo,ve}
      \fmf{phantom,left=0.5,label=$-\varepsilon_e$,l.d=-0.01w}{vo,vs}
      \fmf{phantom,right=0.5,label=$1$,l.d=-0.01w}{ve,vs}
      \fmf{plain,label=$1$,l.d=0.05w}{vs,vn}
      \fmf{plain}{vo,o}
      \fmffreeze
      \fmfdot{ve,vn,vo,vs}
    \end{fmfgraph*}
}
\qquad & = & \qquad
\parbox{16mm}{
  \begin{fmfgraph*}(16,16)
    \fmfleft{i}
    \fmfright{o}
    \fmfleft{ve}
    \fmfright{vo}
    \fmftop{v}
    \fmffreeze
    \fmfforce{(-0.1w,0.5h)}{i}
    \fmfforce{(1.1w,0.5h)}{o}
    \fmfforce{(0w,0.5h)}{vl}
    \fmfforce{(1.0w,0.5h)}{vr}
    \fmfforce{(.5w,0.9h)}{vt}
    \fmfforce{(.5w,0.1h)}{vb}
    \fmffreeze
    \fmf{plain}{i,vl}
    \fmf{plain,left=0.8}{vl,vr}
    \fmf{plain,left=0.8}{vr,vl}
    \fmf{phantom,left=0.1,label=$1-\varepsilon_e$,l.s=left}{vl,vt}
    \fmf{plain,left=0.5,label=$1$,l.d=0.05w}{vt,vl}
    \fmf{phantom,right=0.1,label=$1$,l.s=right}{vr,vt}
    \fmf{phantom,left=0.1,label=$1-\varepsilon_e$,l.s=left,l.d=0.4w}{vr,vl}
    \fmf{plain}{vr,o}
    \fmffreeze
    \fmfdot{vl,vt,vr}
  \end{fmfgraph*}
} \quad - \quad
\frac{(1-\varepsilon_e)(-p^2)}{1 - \varepsilon_e - 3\varepsilon_\gamma } 
\left( \quad
\parbox{24mm}{
  \begin{fmfgraph*}(24,14)
    \fmfleft{i}
    \fmfright{o}
    \fmfleft{ve}
    \fmfright{vo}
    \fmftop{v}
    \fmffreeze
    \fmfforce{(-0.1w,0.5h)}{i}
    \fmfforce{(1.1w,0.5h)}{o}
    \fmfforce{(0w,0.5h)}{ve}
    \fmfforce{(1.0w,0.5h)}{vo}
    \fmfforce{(.5w,0.5h)}{v}
    \fmffreeze
    \fmf{plain}{i,ve}
    \fmf{plain,left=0.8,label=$1-\varepsilon_e$,l.s=left,l.d=0.03w}{ve,v}
    \fmf{plain,left=0.8,label=$1$,l.s=left,l.d=0.03w}{v,ve}
    \fmf{plain,left=0.8,label=$1$,l.s=left,l.d=0.03w}{v,vo}
    \fmf{plain,left=0.8,label=$2-\varepsilon_e$,l.d=0.03w}{vo,v}
    \fmf{plain}{vo,o}
    \fmffreeze
    \fmfdot{ve,v,vo}
  \end{fmfgraph*}
} \qquad - \qquad
\parbox{16mm}{
  \begin{fmfgraph*}(16,16)
    \fmfleft{i}
    \fmfright{o}
    \fmfleft{ve}
    \fmfright{vo}
    \fmftop{v}
    \fmffreeze
    \fmfforce{(-0.1w,0.5h)}{i}
    \fmfforce{(1.1w,0.5h)}{o}
    \fmfforce{(0w,0.5h)}{vl}
    \fmfforce{(1.0w,0.5h)}{vr}
    \fmfforce{(.5w,0.9h)}{vt}
    \fmfforce{(.5w,0.1h)}{vb}
    \fmffreeze
    \fmf{plain}{i,vl}
    \fmf{plain,left=0.8}{vl,vr}
    \fmf{plain,left=0.8}{vr,vl}
    \fmf{phantom,left=0.1,label=$1-\varepsilon_e$,l.s=left}{vl,vt}
    \fmf{plain,left=0.5,label=$1$,l.d=0.05w}{vt,vl}
    \fmf{phantom,right=0.1,label=$1$,l.s=right}{vr,vt}
    \fmf{phantom,left=0.1,label=$2-\varepsilon_e$,l.s=left,l.d=0.4w}{vr,vl}
    \fmf{plain}{vr,o}
    \fmffreeze
    \fmfdot{vl,vt,vr}
  \end{fmfgraph*}
}
\quad \right)
\nonum \\
& & \quad
\nonum \\
& & \quad - \quad \frac{\varepsilon_\gamma \, (-p^2)}{1 - \varepsilon_e - 3\varepsilon_\gamma }\,
\quad
\parbox{16mm}{
    \begin{fmfgraph*}(16,14)
      \fmfleft{i}
      \fmfright{o}
      \fmfleft{ve}
      \fmfright{vo}
      \fmftop{vn}
      \fmftop{vs}
      \fmffreeze
      \fmfforce{(-0.1w,0.5h)}{i}
      \fmfforce{(1.1w,0.5h)}{o}
      \fmfforce{(0w,0.5h)}{ve}
      \fmfforce{(1.0w,0.5h)}{vo}
      \fmfforce{(.5w,0.95h)}{vn}
      \fmfforce{(.5w,0.05h)}{vs}
      \fmffreeze
      \fmf{plain}{i,ve}
      \fmf{plain,left=0.8}{ve,vo}
      \fmf{phantom,left=0.5,label=$1-\varepsilon_e$,l.d=-0.01w}{ve,vn}
      \fmf{phantom,right=0.5,label=$1$,l.d=-0.01w}{vo,vn}
      \fmf{plain,left=0.8}{vo,ve}
      \fmf{phantom,left=0.5,label=$1-\varepsilon_e$,l.d=-0.01w}{vo,vs}
      \fmf{phantom,right=0.5,label=$1$,l.d=-0.01w}{ve,vs}
      \fmf{plain,label=$1$,l.d=0.05w}{vs,vn}
      \fmf{plain}{vo,o}
      \fmffreeze
      \fmfdot{ve,vn,vo,vs}
    \end{fmfgraph*}
}\, . 
\label{IBP:2loop-d} \\
& & \quad
\nonum
\eea
After some algebra, the recursively one-loop diagrams appearing in the left-hand side of Eq.~(\ref{IBP:2loop-d}) can be expressed in terms of the one-loop master integrals
entering the expression of the fermion self-energy Eq.~(\ref{sigma-c-s}). The result reads:
\bea
G(1-\varepsilon_e,1,-\varepsilon_e,1,1) &=& 
-2\,\frac{(d_\gamma-3)(2d_\gamma+d_e-8)}{(d_\gamma-4)(d_\gamma+d_e-6)}\,G(1,1-\varepsilon_e)G(1-\varepsilon_e,\varepsilon_\gamma) 
+ \frac{d_\gamma+d_e-6}{2d_\gamma+d_e-10}\,G^2(1,1-\varepsilon_e) 
\label{G2} \\
&+& \frac{d_\gamma-4}{2d_\gamma+d_e-10}\,G(1-\varepsilon_e,1,1-\varepsilon_e,1,1)\, ,
\nonum
\eea
where the last term does not contribute in the limit $d_\gamma \ra 4$. Hence, for RQED$_{4,d_e}$ and in particular
in the case of RQED$_{4,3}$ where the last diagram in Eq.~(\ref{G2}) is non-trivial, 
the function $G(1-\varepsilon_e,1,-\varepsilon_e,1,1)$ is known for arbitrary $\varepsilon_e$ to $\Ord(\varepsilon_\gamma)$.
The knowledge of these lowest order terms is enough, {\it e.g.}, to compute the anomalous scaling dimension of the fermion field at two-loop (see below for more).

We may proceed in a similar way for the two other complicated diagrams appearing in Eq.~(\ref{sigma-c-s}).
The corresponding coefficient functions, $G(1-\varepsilon_e,1,-1-\varepsilon_e,1,1)$ and $G(-\varepsilon_e,1,-\varepsilon_e,1,1)$,
are proportional to $(-p^2)^{1-2\varepsilon_\gamma}$ and are strongly UV divergent. They can both be expressed
in terms of $G(1-\varepsilon_e,1,-\varepsilon_e,1,1)$. In order to see this, we consider the case where $\al = -\varepsilon_e$ and $\beta = 1-\varepsilon_e$
for which Eq.~(\ref{IBP:2loop-b}) reduces to:
\bea
(1- \varepsilon_\gamma) \quad
\parbox{16mm}{
    \begin{fmfgraph*}(16,14)
      \fmfleft{i}
      \fmfright{o}
      \fmfleft{ve}
      \fmfright{vo}
      \fmftop{vn}
      \fmftop{vs}
      \fmffreeze
      \fmfforce{(-0.1w,0.5h)}{i}
      \fmfforce{(1.1w,0.5h)}{o}
      \fmfforce{(0w,0.5h)}{ve}
      \fmfforce{(1.0w,0.5h)}{vo}
      \fmfforce{(.5w,0.95h)}{vn}
      \fmfforce{(.5w,0.05h)}{vs}
      \fmffreeze
      \fmf{plain}{i,ve}
      \fmf{plain,left=0.8}{ve,vo}
      \fmf{phantom,left=0.5,label=$-\varepsilon_e$,l.d=-0.01w}{ve,vn}
      \fmf{phantom,right=0.5,label=$1$,l.d=-0.01w}{vo,vn}
      \fmf{plain,left=0.8}{vo,ve}
      \fmf{phantom,left=0.5,label=$1-\varepsilon_e$,l.d=-0.01w}{vo,vs}
      \fmf{phantom,right=0.5,label=$1$,l.d=-0.01w}{ve,vs}
      \fmf{plain,label=$1$,l.d=0.05w}{vs,vn}
      \fmf{plain}{vo,o}
      \fmffreeze
      \fmfdot{ve,vn,vo,vs}
    \end{fmfgraph*}
} \quad  &=& \quad
(1-\varepsilon_e)\, \left( \quad
\parbox{24mm}{
  \begin{fmfgraph*}(24,14)
    \fmfleft{i}
    \fmfright{o}
    \fmfleft{ve}
    \fmfright{vo}
    \fmftop{v}
    \fmffreeze
    \fmfforce{(-0.1w,0.5h)}{i}
    \fmfforce{(1.1w,0.5h)}{o}
    \fmfforce{(0w,0.5h)}{ve}
    \fmfforce{(1.0w,0.5h)}{vo}
    \fmfforce{(.5w,0.5h)}{v}
    \fmffreeze
    \fmf{plain}{i,ve}
    \fmf{plain,left=0.8,label=$-\varepsilon_e$,l.s=left,l.d=0.03w}{ve,v}
    \fmf{plain,left=0.8,label=$1$,l.s=left,l.d=0.03w}{v,ve}
    \fmf{plain,left=0.8,label=$1$,l.s=left,l.d=0.03w}{v,vo}
    \fmf{plain,left=0.8,label=$2-\varepsilon_e$,l.d=0.03w}{vo,v}
    \fmf{plain}{vo,o}
    \fmffreeze
    \fmfdot{ve,v,vo}
  \end{fmfgraph*}
} \qquad - \qquad
\parbox{16mm}{
  \begin{fmfgraph*}(16,16)
    \fmfleft{i}
    \fmfright{o}
    \fmfleft{ve}
    \fmfright{vo}
    \fmftop{v}
    \fmffreeze
    \fmfforce{(-0.1w,0.5h)}{i}
    \fmfforce{(1.1w,0.5h)}{o}
    \fmfforce{(0w,0.5h)}{vl}
    \fmfforce{(1.0w,0.5h)}{vr}
    \fmfforce{(.5w,0.9h)}{vt}
    \fmfforce{(.5w,0.1h)}{vb}
    \fmffreeze
    \fmf{plain}{i,vl}
    \fmf{plain,left=0.8}{vl,vr}
    \fmf{plain,left=0.8}{vr,vl}
    \fmf{phantom,left=0.1,label=$-\varepsilon_e$,l.s=left}{vl,vt}
    \fmf{plain,left=0.5,label=$1$,l.d=0.05w}{vt,vl}
    \fmf{phantom,right=0.1,label=$1$,l.s=right}{vr,vt}
    \fmf{phantom,left=0.1,label=$2-\varepsilon_e$,l.s=left,l.d=0.4w}{vr,vl}
    \fmf{plain}{vr,o}
    \fmffreeze
    \fmfdot{vl,vt,vr}
  \end{fmfgraph*}
}
\quad \right)
\label{IBP:2loop-e}\\
& & \quad
\nonum \\
& & + \left( 2 - \varepsilon_e - 3\varepsilon_\gamma \right) \,(-p^2)^{-1}\, \left( \qquad \qquad
\parbox{16mm}{
    \begin{fmfgraph*}(16,14)
      \fmfleft{i}
      \fmfright{o}
      \fmfleft{ve}
      \fmfright{vo}
      \fmftop{vn}
      \fmftop{vs}
      \fmffreeze
      \fmfforce{(-0.1w,0.5h)}{i}
      \fmfforce{(1.1w,0.5h)}{o}
      \fmfforce{(0w,0.5h)}{ve}
      \fmfforce{(1.0w,0.5h)}{vo}
      \fmfforce{(.5w,0.95h)}{vn}
      \fmfforce{(.5w,0.05h)}{vs}
      \fmffreeze
      \fmf{plain}{i,ve}
      \fmf{plain,left=0.8}{ve,vo}
      \fmf{phantom,left=0.5,label=$-1-\varepsilon_e$,l.d=-0.01w}{ve,vn}
      \fmf{phantom,right=0.5,label=$1$,l.d=-0.01w}{vo,vn}
      \fmf{plain,left=0.8}{vo,ve}
      \fmf{phantom,left=0.5,label=$1-\varepsilon_e$,l.d=-0.01w}{vo,vs}
      \fmf{phantom,right=0.5,label=$1$,l.d=-0.01w}{ve,vs}
      \fmf{plain,label=$1$,l.d=0.05w}{vs,vn}
      \fmf{plain}{vo,o}
      \fmffreeze
      \fmfdot{ve,vn,vo,vs}
    \end{fmfgraph*}
}
\qquad - \qquad
\parbox{16mm}{
  \begin{fmfgraph*}(16,16)
    \fmfleft{i}
    \fmfright{o}
    \fmfleft{ve}
    \fmfright{vo}
    \fmftop{v}
    \fmffreeze
    \fmfforce{(-0.1w,0.5h)}{i}
    \fmfforce{(1.1w,0.5h)}{o}
    \fmfforce{(0w,0.5h)}{vl}
    \fmfforce{(1.0w,0.5h)}{vr}
    \fmfforce{(.5w,0.9h)}{vt}
    \fmfforce{(.5w,0.1h)}{vb}
    \fmffreeze
    \fmf{plain}{i,vl}
    \fmf{plain,left=0.8}{vl,vr}
    \fmf{plain,left=0.8}{vr,vl}
    \fmf{phantom,left=0.1,label=$-\varepsilon_e$,l.s=left}{vl,vt}
    \fmf{plain,left=0.5,label=$1$,l.d=0.05w}{vt,vl}
    \fmf{phantom,right=0.1,label=$1$,l.s=right}{vr,vt}
    \fmf{phantom,left=0.1,label=$1-\varepsilon_e$,l.s=left,l.d=0.4w}{vr,vl}
    \fmf{plain}{vr,o}
    \fmffreeze
    \fmfdot{vl,vt,vr}
  \end{fmfgraph*}
} \quad \right)\, .
\nonum \\
& & \quad
\nonum 
\eea
Eq.~(\ref{IBP:2loop-e}) shows that there is a simple relation, {\it i.e.}, involving recursively one-loop diagrams, between
two of the complicated diagrams, $G(1-\varepsilon_e,1,-1-\varepsilon_e,1,1)$ and $G(1-\varepsilon_e,1,-\varepsilon_e,1,1)$, entering the fermion self-energy, 
Eq.~(\ref{sigma-c-s}). Expressing the, still unknown, coefficient function $G(1-\varepsilon_e,1,-1-\varepsilon_e,1,1)$ in terms of
$G(1-\varepsilon_e,1,-\varepsilon_e,1,1)$ that we have computed above, see Eq.~(\ref{G2}), yields, in graphical form:
\bea
\parbox{16mm}{
    \begin{fmfgraph*}(16,14)
      \fmfleft{i}
      \fmfright{o}
      \fmfleft{ve}
      \fmfright{vo}
      \fmftop{vn}
      \fmftop{vs}
      \fmffreeze
      \fmfforce{(-0.1w,0.5h)}{i}
      \fmfforce{(1.1w,0.5h)}{o}
      \fmfforce{(0w,0.5h)}{ve}
      \fmfforce{(1.0w,0.5h)}{vo}
      \fmfforce{(.5w,0.95h)}{vn}
      \fmfforce{(.5w,0.05h)}{vs}
      \fmffreeze
      \fmf{plain}{i,ve}
      \fmf{plain,left=0.8}{ve,vo}
      \fmf{phantom,left=0.5,label=$1-\varepsilon_e$,l.d=-0.01w}{ve,vn}
      \fmf{phantom,right=0.5,label=$1$,l.d=-0.01w}{vo,vn}
      \fmf{plain,left=0.8}{vo,ve}
      \fmf{phantom,left=0.5,label=$-1-\varepsilon_e$,l.d=-0.01w}{vo,vs}
      \fmf{phantom,right=0.5,label=$1$,l.d=-0.01w}{ve,vs}
      \fmf{plain,label=$1$,l.d=0.05w}{vs,vn}
      \fmf{plain}{vo,o}
      \fmffreeze
      \fmfdot{ve,vn,vo,vs}
    \end{fmfgraph*}
}
\qquad & = & \qquad
\parbox{16mm}{
  \begin{fmfgraph*}(16,16)
    \fmfleft{i}
    \fmfright{o}
    \fmfleft{ve}
    \fmfright{vo}
    \fmftop{v}
    \fmffreeze
    \fmfforce{(-0.1w,0.5h)}{i}
    \fmfforce{(1.1w,0.5h)}{o}
    \fmfforce{(0w,0.5h)}{vl}
    \fmfforce{(1.0w,0.5h)}{vr}
    \fmfforce{(.5w,0.9h)}{vt}
    \fmfforce{(.5w,0.1h)}{vb}
    \fmffreeze
    \fmf{plain}{i,vl}
    \fmf{plain,left=0.8}{vl,vr}
    \fmf{plain,left=0.8}{vr,vl}
    \fmf{phantom,left=0.1,label=$-\varepsilon_e$,l.s=left}{vl,vt}
    \fmf{plain,left=0.5,label=$1$,l.d=0.05w}{vt,vl}
    \fmf{phantom,right=0.1,label=$1$,l.s=right}{vr,vt}
    \fmf{phantom,left=0.1,label=$1-\varepsilon_e$,l.s=left,l.d=0.4w}{vr,vl}
    \fmf{plain}{vr,o}
    \fmffreeze
    \fmfdot{vl,vt,vr}
  \end{fmfgraph*}
}\quad  - \quad
\frac{(1-\varepsilon_e)(-p^2)}{2 - \varepsilon_e - 3\varepsilon_\gamma}\, \left( \quad
\parbox{24mm}{
  \begin{fmfgraph*}(24,14)
    \fmfleft{i}
    \fmfright{o}
    \fmfleft{ve}
    \fmfright{vo}
    \fmftop{v}
    \fmffreeze
    \fmfforce{(-0.1w,0.5h)}{i}
    \fmfforce{(1.1w,0.5h)}{o}
    \fmfforce{(0w,0.5h)}{ve}
    \fmfforce{(1.0w,0.5h)}{vo}
    \fmfforce{(.5w,0.5h)}{v}
    \fmffreeze
    \fmf{plain}{i,ve}
    \fmf{plain,left=0.8,label=$-\varepsilon_e$,l.s=left,l.d=0.03w}{ve,v}
    \fmf{plain,left=0.8,label=$1$,l.s=left,l.d=0.03w}{v,ve}
    \fmf{plain,left=0.8,label=$1$,l.s=left,l.d=0.03w}{v,vo}
    \fmf{plain,left=0.8,label=$2-\varepsilon_e$,l.d=0.03w}{vo,v}
    \fmf{plain}{vo,o}
    \fmffreeze
    \fmfdot{ve,v,vo}
  \end{fmfgraph*}
} \qquad - \qquad
\parbox{16mm}{
  \begin{fmfgraph*}(16,16)
    \fmfleft{i}
    \fmfright{o}
    \fmfleft{ve}
    \fmfright{vo}
    \fmftop{v}
    \fmffreeze
    \fmfforce{(-0.1w,0.5h)}{i}
    \fmfforce{(1.1w,0.5h)}{o}
    \fmfforce{(0w,0.5h)}{vl}
    \fmfforce{(1.0w,0.5h)}{vr}
    \fmfforce{(.5w,0.9h)}{vt}
    \fmfforce{(.5w,0.1h)}{vb}
    \fmffreeze
    \fmf{plain}{i,vl}
    \fmf{plain,left=0.8}{vl,vr}
    \fmf{plain,left=0.8}{vr,vl}
    \fmf{phantom,left=0.1,label=$-\varepsilon_e$,l.s=left}{vl,vt}
    \fmf{plain,left=0.5,label=$1$,l.d=0.05w}{vt,vl}
    \fmf{phantom,right=0.1,label=$1$,l.s=right}{vr,vt}
    \fmf{phantom,left=0.1,label=$2-\varepsilon_e$,l.s=left,l.d=0.4w}{vr,vl}
    \fmf{plain}{vr,o}
    \fmffreeze
    \fmfdot{vl,vt,vr}
  \end{fmfgraph*}
}
\quad \right) 
\nonum \\
& & \quad
\nonum \\
& & \qquad + \quad \frac{(1- \varepsilon_\gamma)(-p^2)}{2 - \varepsilon_e - 3\varepsilon_\gamma}\, \quad
\parbox{16mm}{
    \begin{fmfgraph*}(16,14)
      \fmfleft{i}
      \fmfright{o}
      \fmfleft{ve}
      \fmfright{vo}
      \fmftop{vn}
      \fmftop{vs}
      \fmffreeze
      \fmfforce{(-0.1w,0.5h)}{i}
      \fmfforce{(1.1w,0.5h)}{o}
      \fmfforce{(0w,0.5h)}{ve}
      \fmfforce{(1.0w,0.5h)}{vo}
      \fmfforce{(.5w,0.95h)}{vn}
      \fmfforce{(.5w,0.05h)}{vs}
      \fmffreeze
      \fmf{plain}{i,ve}
      \fmf{plain,left=0.8}{ve,vo}
      \fmf{phantom,left=0.5,label=$-\varepsilon_e$,l.d=-0.01w}{ve,vn}
      \fmf{phantom,right=0.5,label=$1$,l.d=-0.01w}{vo,vn}
      \fmf{plain,left=0.8}{vo,ve}
      \fmf{phantom,left=0.5,label=$1-\varepsilon_e$,l.d=-0.01w}{vo,vs}
      \fmf{phantom,right=0.5,label=$1$,l.d=-0.01w}{ve,vs}
      \fmf{plain,label=$1$,l.d=0.05w}{vs,vn}
      \fmf{plain}{vo,o}
      \fmffreeze
      \fmfdot{ve,vn,vo,vs}
    \end{fmfgraph*}
}\, .
\label{IBP:2loop-f}
\eea
After some algebra, the recursively one-loop diagrams appearing in the right-hand side of 
Eq.~(\ref{IBP:2loop-f}) can also be expressed in terms of the one-loop master integrals
that enter the expression of the fermion self-energy Eq.~(\ref{sigma-c-s}). 
Together with Eq.~(\ref{G2}), the final result reads:
\bea
G(1-\varepsilon_e,1,-1-\varepsilon_e,1,1) &=& 
-\frac{d_\gamma-2}{d_\gamma-4}\,\left( \frac{2(d_\gamma-3)}{d_\gamma+d_e-6}+\frac{d_\gamma-d_e}{d_\gamma+d_e-4}\right)\,G(1,1-\varepsilon_e)G(1-\varepsilon_e,\varepsilon_\gamma) 
\label{G1} \\
&+& \frac{d_\gamma+d_e-6}{2d_\gamma+d_e-8}\,\left( \frac{d_\gamma-2}{2d_\gamma+d_e-10}+\frac{d_\gamma-d_e}{d_\gamma+d_e-4}\right)\,G^2(1,1-\varepsilon_e) 
\nonum \\
&+& \frac{(d_\gamma-2)(d_\gamma-4)}{(2d_\gamma+d_e-10)(2d_\gamma+d_e-8)}\,G(1-\varepsilon_e,1,1-\varepsilon_e,1,1)\, .
\nonum
\eea

Finally, let's consider the case where $\al = 1-\varepsilon_e$ and $\beta = -\varepsilon_e$
for which Eq.~(\ref{IBP:2loop-b}) reduces to:
\bea
- \varepsilon_\gamma \quad
\parbox{16mm}{
    \begin{fmfgraph*}(16,14)
      \fmfleft{i}
      \fmfright{o}
      \fmfleft{ve}
      \fmfright{vo}
      \fmftop{vn}
      \fmftop{vs}
      \fmffreeze
      \fmfforce{(-0.1w,0.5h)}{i}
      \fmfforce{(1.1w,0.5h)}{o}
      \fmfforce{(0w,0.5h)}{ve}
      \fmfforce{(1.0w,0.5h)}{vo}
      \fmfforce{(.5w,0.95h)}{vn}
      \fmfforce{(.5w,0.05h)}{vs}
      \fmffreeze
      \fmf{plain}{i,ve}
      \fmf{plain,left=0.8}{ve,vo}
      \fmf{phantom,left=0.5,label=$1-\varepsilon_e$,l.d=-0.01w}{ve,vn}
      \fmf{phantom,right=0.5,label=$1$,l.d=-0.01w}{vo,vn}
      \fmf{plain,left=0.8}{vo,ve}
      \fmf{phantom,left=0.5,label=$-\varepsilon_e$,l.d=-0.01w}{vo,vs}
      \fmf{phantom,right=0.5,label=$1$,l.d=-0.01w}{ve,vs}
      \fmf{plain,label=$1$,l.d=0.05w}{vs,vn}
      \fmf{plain}{vo,o}
      \fmffreeze
      \fmfdot{ve,vn,vo,vs}
    \end{fmfgraph*}
} \quad  &=& \quad
-\varepsilon_e\, \left( \quad
\parbox{24mm}{
  \begin{fmfgraph*}(24,14)
    \fmfleft{i}
    \fmfright{o}
    \fmfleft{ve}
    \fmfright{vo}
    \fmftop{v}
    \fmffreeze
    \fmfforce{(-0.1w,0.5h)}{i}
    \fmfforce{(1.1w,0.5h)}{o}
    \fmfforce{(0w,0.5h)}{ve}
    \fmfforce{(1.0w,0.5h)}{vo}
    \fmfforce{(.5w,0.5h)}{v}
    \fmffreeze
    \fmf{plain}{i,ve}
    \fmf{plain,left=0.8,label=$1-\varepsilon_e$,l.s=left,l.d=0.03w}{ve,v}
    \fmf{plain,left=0.8,label=$1$,l.s=left,l.d=0.03w}{v,ve}
    \fmf{plain,left=0.8,label=$1$,l.s=left,l.d=0.03w}{v,vo}
    \fmf{plain,left=0.8,label=$1-\varepsilon_e$,l.d=0.03w}{vo,v}
    \fmf{plain}{vo,o}
    \fmffreeze
    \fmfdot{ve,v,vo}
  \end{fmfgraph*}
} \qquad - \qquad
\parbox{16mm}{
  \begin{fmfgraph*}(16,16)
    \fmfleft{i}
    \fmfright{o}
    \fmfleft{ve}
    \fmfright{vo}
    \fmftop{v}
    \fmffreeze
    \fmfforce{(-0.1w,0.5h)}{i}
    \fmfforce{(1.1w,0.5h)}{o}
    \fmfforce{(0w,0.5h)}{vl}
    \fmfforce{(1.0w,0.5h)}{vr}
    \fmfforce{(.5w,0.9h)}{vt}
    \fmfforce{(.5w,0.1h)}{vb}
    \fmffreeze
    \fmf{plain}{i,vl}
    \fmf{plain,left=0.8}{vl,vr}
    \fmf{plain,left=0.8}{vr,vl}
    \fmf{phantom,left=0.1,label=$1-\varepsilon_e$,l.s=left}{vl,vt}
    \fmf{plain,left=0.5,label=$1$,l.d=0.05w}{vt,vl}
    \fmf{phantom,right=0.1,label=$1$,l.s=right}{vr,vt}
    \fmf{phantom,left=0.1,label=$1-\varepsilon_e$,l.s=left,l.d=0.4w}{vr,vl}
    \fmf{plain}{vr,o}
    \fmffreeze
    \fmfdot{vl,vt,vr}
  \end{fmfgraph*}
}
\quad \right)
\label{IBP:2loop-g}\\
& & \quad
\nonum \\
& & + \left( 2 - \varepsilon_e - 3\varepsilon_\gamma \right) \,(-p^2)^{-1}\, \left( \quad
\parbox{16mm}{
    \begin{fmfgraph*}(16,14)
      \fmfleft{i}
      \fmfright{o}
      \fmfleft{ve}
      \fmfright{vo}
      \fmftop{vn}
      \fmftop{vs}
      \fmffreeze
      \fmfforce{(-0.1w,0.5h)}{i}
      \fmfforce{(1.1w,0.5h)}{o}
      \fmfforce{(0w,0.5h)}{ve}
      \fmfforce{(1.0w,0.5h)}{vo}
      \fmfforce{(.5w,0.95h)}{vn}
      \fmfforce{(.5w,0.05h)}{vs}
      \fmffreeze
      \fmf{plain}{i,ve}
      \fmf{plain,left=0.8}{ve,vo}
      \fmf{phantom,left=0.5,label=$-\varepsilon_e$,l.d=-0.01w}{ve,vn}
      \fmf{phantom,right=0.5,label=$1$,l.d=-0.01w}{vo,vn}
      \fmf{plain,left=0.8}{vo,ve}
      \fmf{phantom,left=0.5,label=$-\varepsilon_e$,l.d=-0.01w}{vo,vs}
      \fmf{phantom,right=0.5,label=$1$,l.d=-0.01w}{ve,vs}
      \fmf{plain,label=$1$,l.d=0.05w}{vs,vn}
      \fmf{plain}{vo,o}
      \fmffreeze
      \fmfdot{ve,vn,vo,vs}
    \end{fmfgraph*}
}
\qquad - \qquad
\parbox{16mm}{
  \begin{fmfgraph*}(16,16)
    \fmfleft{i}
    \fmfright{o}
    \fmfleft{ve}
    \fmfright{vo}
    \fmftop{v}
    \fmffreeze
    \fmfforce{(-0.1w,0.5h)}{i}
    \fmfforce{(1.1w,0.5h)}{o}
    \fmfforce{(0w,0.5h)}{vl}
    \fmfforce{(1.0w,0.5h)}{vr}
    \fmfforce{(.5w,0.9h)}{vt}
    \fmfforce{(.5w,0.1h)}{vb}
    \fmffreeze
    \fmf{plain}{i,vl}
    \fmf{plain,left=0.8}{vl,vr}
    \fmf{plain,left=0.8}{vr,vl}
    \fmf{phantom,left=0.1,label=$1-\varepsilon_e$,l.s=left}{vl,vt}
    \fmf{plain,left=0.5,label=$1$,l.d=0.05w}{vt,vl}
    \fmf{phantom,right=0.1,label=$1$,l.s=right}{vr,vt}
    \fmf{phantom,left=0.1,label=$-\varepsilon_e$,l.s=left,l.d=0.4w}{vr,vl}
    \fmf{plain}{vr,o}
    \fmffreeze
    \fmfdot{vl,vt,vr}
  \end{fmfgraph*}
} \quad \right)\, .
\nonum \\
& & \quad
\nonum 
\eea
From Eq.~(\ref{IBP:2loop-g}) we now have a relation between $G(-\varepsilon_e,1,-\varepsilon_e,1,1)$ and $G(1-\varepsilon_e,1,-\varepsilon_e,1,1)$. 
Expressing the former as a function of the later, yields, in graphical form:
\bea
\parbox{16mm}{
    \begin{fmfgraph*}(16,14)
      \fmfleft{i}
      \fmfright{o}
      \fmfleft{ve}
      \fmfright{vo}
      \fmftop{vn}
      \fmftop{vs}
      \fmffreeze
      \fmfforce{(-0.1w,0.5h)}{i}
      \fmfforce{(1.1w,0.5h)}{o}
      \fmfforce{(0w,0.5h)}{ve}
      \fmfforce{(1.0w,0.5h)}{vo}
      \fmfforce{(.5w,0.95h)}{vn}
      \fmfforce{(.5w,0.05h)}{vs}
      \fmffreeze
      \fmf{plain}{i,ve}
      \fmf{plain,left=0.8}{ve,vo}
      \fmf{phantom,left=0.5,label=$-\varepsilon_e$,l.d=-0.01w}{ve,vn}
      \fmf{phantom,right=0.5,label=$1$,l.d=-0.01w}{vo,vn}
      \fmf{plain,left=0.8}{vo,ve}
      \fmf{phantom,left=0.5,label=$-\varepsilon_e$,l.d=-0.01w}{vo,vs}
      \fmf{phantom,right=0.5,label=$1$,l.d=-0.01w}{ve,vs}
      \fmf{plain,label=$1$,l.d=0.05w}{vs,vn}
      \fmf{plain}{vo,o}
      \fmffreeze
      \fmfdot{ve,vn,vo,vs}
    \end{fmfgraph*}
}
\qquad & = & \qquad
\parbox{16mm}{
  \begin{fmfgraph*}(16,16)
    \fmfleft{i}
    \fmfright{o}
    \fmfleft{ve}
    \fmfright{vo}
    \fmftop{v}
    \fmffreeze
    \fmfforce{(-0.1w,0.5h)}{i}
    \fmfforce{(1.1w,0.5h)}{o}
    \fmfforce{(0w,0.5h)}{vl}
    \fmfforce{(1.0w,0.5h)}{vr}
    \fmfforce{(.5w,0.9h)}{vt}
    \fmfforce{(.5w,0.1h)}{vb}
    \fmffreeze
    \fmf{plain}{i,vl}
    \fmf{plain,left=0.8}{vl,vr}
    \fmf{plain,left=0.8}{vr,vl}
    \fmf{phantom,left=0.1,label=$1-\varepsilon_e$,l.s=left}{vl,vt}
    \fmf{plain,left=0.5,label=$1$,l.d=0.05w}{vt,vl}
    \fmf{phantom,right=0.1,label=$1$,l.s=right}{vr,vt}
    \fmf{phantom,left=0.1,label=$-\varepsilon_e$,l.s=left,l.d=0.4w}{vr,vl}
    \fmf{plain}{vr,o}
    \fmffreeze
    \fmfdot{vl,vt,vr}
  \end{fmfgraph*}
}\quad  + \quad
\frac{\varepsilon_e\,(-p^2)}{2 - \varepsilon_e - 3\varepsilon_\gamma}\, \left( \quad
\parbox{24mm}{
  \begin{fmfgraph*}(24,14)
    \fmfleft{i}
    \fmfright{o}
    \fmfleft{ve}
    \fmfright{vo}
    \fmftop{v}
    \fmffreeze
    \fmfforce{(-0.1w,0.5h)}{i}
    \fmfforce{(1.1w,0.5h)}{o}
    \fmfforce{(0w,0.5h)}{ve}
    \fmfforce{(1.0w,0.5h)}{vo}
    \fmfforce{(.5w,0.5h)}{v}
    \fmffreeze
    \fmf{plain}{i,ve}
    \fmf{plain,left=0.8,label=$1-\varepsilon_e$,l.s=left,l.d=0.03w}{ve,v}
    \fmf{plain,left=0.8,label=$1$,l.s=left,l.d=0.03w}{v,ve}
    \fmf{plain,left=0.8,label=$1$,l.s=left,l.d=0.03w}{v,vo}
    \fmf{plain,left=0.8,label=$1-\varepsilon_e$,l.d=0.03w}{vo,v}
    \fmf{plain}{vo,o}
    \fmffreeze
    \fmfdot{ve,v,vo}
  \end{fmfgraph*}
} \qquad - \qquad
\parbox{16mm}{
  \begin{fmfgraph*}(16,16)
    \fmfleft{i}
    \fmfright{o}
    \fmfleft{ve}
    \fmfright{vo}
    \fmftop{v}
    \fmffreeze
    \fmfforce{(-0.1w,0.5h)}{i}
    \fmfforce{(1.1w,0.5h)}{o}
    \fmfforce{(0w,0.5h)}{vl}
    \fmfforce{(1.0w,0.5h)}{vr}
    \fmfforce{(.5w,0.9h)}{vt}
    \fmfforce{(.5w,0.1h)}{vb}
    \fmffreeze
    \fmf{plain}{i,vl}
    \fmf{plain,left=0.8}{vl,vr}
    \fmf{plain,left=0.8}{vr,vl}
    \fmf{phantom,left=0.1,label=$1-\varepsilon_e$,l.s=left}{vl,vt}
    \fmf{plain,left=0.5,label=$1$,l.d=0.05w}{vt,vl}
    \fmf{phantom,right=0.1,label=$1$,l.s=right}{vr,vt}
    \fmf{phantom,left=0.1,label=$1-\varepsilon_e$,l.s=left,l.d=0.4w}{vr,vl}
    \fmf{plain}{vr,o}
    \fmffreeze
    \fmfdot{vl,vt,vr}
  \end{fmfgraph*}
}
\quad \right) 
\nonum \\
& & \quad
\nonum \\
& & \qquad - \quad \frac{\varepsilon_\gamma \,(-p^2)}{2 - \varepsilon_e - 3\varepsilon_\gamma}\, \qquad
\parbox{16mm}{
    \begin{fmfgraph*}(16,14)
      \fmfleft{i}
      \fmfright{o}
      \fmfleft{ve}
      \fmfright{vo}
      \fmftop{vn}
      \fmftop{vs}
      \fmffreeze
      \fmfforce{(-0.1w,0.5h)}{i}
      \fmfforce{(1.1w,0.5h)}{o}
      \fmfforce{(0w,0.5h)}{ve}
      \fmfforce{(1.0w,0.5h)}{vo}
      \fmfforce{(.5w,0.95h)}{vn}
      \fmfforce{(.5w,0.05h)}{vs}
      \fmffreeze
      \fmf{plain}{i,ve}
      \fmf{plain,left=0.8}{ve,vo}
      \fmf{phantom,left=0.5,label=$1-\varepsilon_e$,l.d=-0.01w}{ve,vn}
      \fmf{phantom,right=0.5,label=$1$,l.d=-0.01w}{vo,vn}
      \fmf{plain,left=0.8}{vo,ve}
      \fmf{phantom,left=0.5,label=$-\varepsilon_e$,l.d=-0.01w}{vo,vs}
      \fmf{phantom,right=0.5,label=$1$,l.d=-0.01w}{ve,vs}
      \fmf{plain,label=$1$,l.d=0.05w}{vs,vn}
      \fmf{plain}{vo,o}
      \fmffreeze
      \fmfdot{ve,vn,vo,vs}
    \end{fmfgraph*}
}\, .
\label{IBP:2loop-h}
\eea
Similarly to the previous cases, the recursively one-loop diagrams appearing in the right-hand side of
Eq.~(\ref{IBP:2loop-h}) can be expressed in terms of the one-loop master integrals
entering the expression of the fermion self-energy Eq.~(\ref{sigma-c-s}).
Together with Eq.~(\ref{G2}), the final result reads:
\bea
G(-\varepsilon_e,1,-\varepsilon_e,1,1) &=& 
-\frac{3d_\gamma-d_e-6}{d_\gamma+d_e-6}\,G(1,1-\varepsilon_e)G(1-\varepsilon_e,\varepsilon_\gamma) 
\label{G3} \\
&+& \frac{1}{2d_\gamma+d_e-8}\,\left( \frac{(d_\gamma-4)(d_\gamma+d_e-6)}{2d_\gamma+d_e-10} + d_\gamma - d_e \right)\,G^2(1,1-\varepsilon_e) 
\nonum \\
&+& \frac{(d_\gamma-4)^2}{(2d_\gamma+d_e-10)(2d_\gamma+d_e-8)}\,G(1-\varepsilon_e,1,1-\varepsilon_e,1,1)\, .
\nonum
\eea
Eqs.~(\ref{G2}), (\ref{G1}) and (\ref{G3}) constitute the central results of this paragraph. 
They show that all three complicated diagrams appearing in Eq.~(\ref{sigma-c-s}) are related: they can all be expressed
in terms of the ultra-violet convergent diagram $G(1-\varepsilon_e,1,1-\varepsilon_e,1,1)$ with a factor proportional to $\varepsilon_\gamma$.

With the help of Eqs.~(\ref{G2}), (\ref{G1}) and (\ref{G3}) the two-loop self-energy associated with the crossed photon diagram, Eq.~(\ref{sigma-c-s}), reads:
\bea
&&\Sigma_{V2c}(p^2) = - \frac{e^4 \Gamma^2(1-\varepsilon_e)}{(4\pi)^{d_\gamma} (-p^2)^{2\varepsilon_\gamma}}\,\frac{d_e-2}{2}\,
\times \Bigg \{ \left[ d_e-4 -2\xi\,\frac{(d_e-2)^2}{d_\gamma+d_e-4} +\xi^2\,\frac{d_e-2}{2} 
\right . \Bigg .
\label{sigma-c-s2} \\
&& \left . -\frac{(d_\gamma+d_e-6)( d_\gamma(d_e-4)+8)}{(2d_\gamma+d_e-10)(2d_\gamma+d_e-8)} 
-\frac{4(d_\gamma-d_e)}{d_\gamma+d_e-4} - \frac{d_\gamma-d_e}{2d_\gamma+d_e-8}\, \left( d_e-8 - 4 \, \frac{d_\gamma+d_e-6}{d_\gamma+d_e-4} \right)\right]\,G^2(1,1-\varepsilon_e)
\nonum \\
&& \Bigg . + \left[ 2d_e - 4 + 8\,\frac{d_e-2}{d_\gamma+d_e-4}+\frac{8(d_\gamma-1)}{d_\gamma-4} + 4\xi\,\frac{(d_e-2)(d_\gamma-3)}{d_\gamma+d_e-4} -\xi^2 (d_\gamma-3) \right .
\nonum \\
&\quad& \quad \left . + \frac{2(d_e-8)(d_\gamma-d_e)}{d_\gamma+d_e-6}-\frac{4(d_\gamma-2)(d_\gamma-d_e)}{(d_\gamma-4)(d_\gamma+d_e-4)} \right] 
G(1,1-\varepsilon_e)G(1-\varepsilon_e,\varepsilon_\gamma)  \Bigg .
\nonum \\
&& \Bigg . - \frac{(d_\gamma-4)(d_\gamma(d_e-4)+8)}{(2d_\gamma+d_e-8)(2d_\gamma+d_e-10)}\,G(1-\varepsilon_e,1,1-\varepsilon_e,1,1) \Bigg \}\, .
\nonum
\eea

\subsection{Total two-loop fermion self-energy (explicit expression)}

The sum of all the self-energy contributions, Eqs.~(\ref{sigma-a-s}), (\ref{sigma-b-s}) and Eq.~(\ref{sigma-c-s2}), yields the total two-loop fermion self-energy in an arbitrary gauge:
\bea
\Sigma_{V2}(p^2) &=& \frac{e^4 \Gamma^2(1-\varepsilon_e)}{(4\pi)^{d_\gamma} (-p^2)^{2\varepsilon_\gamma}}\,\frac{d_e-2}{2}\,
\Bigg \{ d\,\frac{d_e-2}{2d_\gamma-d_e-6}\,G(1,1)G(1,\varepsilon_\gamma-\varepsilon_e) \Bigg . 
\label{sigma-2l-s} \\
\nonum \\
&\quad& + \left[ 4-d_e +2\xi\frac{(d_e-2)^2}{d_\gamma+d_e-4} -\xi^2\,\frac{d_e-2}{2} +\frac{(d_\gamma+d_e-6)( d_\gamma(d_e-4)+8)}{(2d_\gamma+d_e-10)(2d_\gamma+d_e-8)} \right .
\nonum \\
&\quad& \qquad \left . +\frac{4(d_\gamma-d_e)}{d_\gamma+d_e-4} + \frac{d_\gamma-d_e}{2d_\gamma+d_e-8}\, \left( d_e-8 - 4 \, \frac{d_\gamma+d_e-6}{d_\gamma+d_e-4} \right)\right]\,G^2(1,1-\varepsilon_e)
\nonum \\
&& \Bigg . + \left[ 4-2d_e - \frac{8(d_e-2)}{d_\gamma+d_e-4} - \frac{8(d_\gamma-1)}{d_\gamma-4} -4\xi\,\frac{(d_e-2)(d_\gamma-3)}{d_\gamma+d_e-4} +\xi^2 (d_\gamma-3) \right .
\nonum \\
&\quad& \qquad \left . + \frac{(d_\gamma-3)(d_\gamma+d_e-4)}{d_\gamma-4}\,\left( \frac{2(d_e-2)}{d_\gamma+d_e-4}-\xi \right)^2 \right .
\nonum \\
&\quad& \qquad \left . - \frac{2(d_e-8)(d_\gamma-d_e)}{d_\gamma+d_e-6}+\frac{4(d_\gamma-2)(d_\gamma-d_e)}{(d_\gamma-4)(d_\gamma+d_e-4)} \right]
G(1,1-\varepsilon_e)G(1-\varepsilon_e,\varepsilon_\gamma) + \Bigg .
\nonum \\
&& \Bigg . + \frac{(d_\gamma-4)(d_\gamma(d_e-4)+8)}{(2d_\gamma+d_e-8)(2d_\gamma+d_e-10)}\,G(1-\varepsilon_e,1,1-\varepsilon_e,1,1) \Bigg \}\, .
\nonum
\eea
As explained in the previous paragraph, the last line of Eq.~(\ref{sigma-2l-s}) contains the UV-convergent diagram $G(1-\varepsilon_e,1,1-\varepsilon_e,1,1)$ 
with a prefactor proportional to $d_\gamma -4$. It therefore vanishes
for all RQED$_{4,d_e}$ and in particular in RQED$_{4,3}$ which enables us to avoid computing the complicated diagram $G(1/2,1,1/2,1,1)$ 
and obtain an expansion valid to $\Ord(\varepsilon_\gamma)$ for the two-loop fermion self-energy. 

Eq.~(\ref{sigma-2l-s}) simplifies for usual QEDs where $d_\gamma=d_e$ corresponding to $\varepsilon_e=0$. In this case, the complicated diagram reduces to the well known $G(1,1,1,1,1)$ which
is given by Eq.~(\ref{G11111}) and Eq.~(\ref{sigma-2l-s}) becomes:
\bea
\Sigma_{V2}(p^2) = \frac{e^4 (-p^2)^{-2\varepsilon_\gamma} }{(4\pi)^{d_\gamma}}\,(d_\gamma-2)\,
\Bigg \{ && 2N_F\,\frac{d_\gamma-2}{d_\gamma-6}\,G(1,1)G(1,\varepsilon_\gamma) -\frac{1}{4}\,\bigg(d_\gamma-6+(d_\gamma-2)a^2\bigg)G^2(1,1) \Bigg .
\label{sigma-2l-s-QEDd} \\
&& + \quad \Bigg . \frac{1}{2}\,\frac{d_\gamma-3}{d_\gamma-4}\,\bigg(a^2(3d_\gamma-8) - d_\gamma-4\bigg)\,G(1,1)G(1,\varepsilon_\gamma)\quad \Bigg \}\, ,
\nonum 
\eea
where Eq.~(\ref{Nf}) has been used. Eq.~(\ref{sigma-2l-s-QEDd}) agrees with well known results in the literature, see {\it e.g.}, Ref.~[\onlinecite{Grozin07}].
 
\section{Anomalous dimension of the fermion field and renormalized fermion propagator up to two loops}
\label{Sec:FermiDim}

We now compute the renormalization constant associated to the fermion field $Z_\psi$, the related anomalous dimension $\gamma_\psi$ and renormalized propagator 
$S_r(p;\mu)$ up to two loops. To achieve this task we use for RQED well known methods used for usual QEDs, see, {\it e.g.}, Ref.~[\onlinecite{Grozin07}].

\subsection{Renormalization}

In the following, we shall expand the one-loop and two-loop fermion self-energies, for fixed $\varepsilon_e$, in Laurent series in $\varepsilon_\gamma$.
For this purpose, we introduce two functions, $\sigma_1$ and $\sigma_2$, which
are defined with the help of the expressions of the one-loop (Eq.~(\ref{sigma-1l-s})) and two-loop (Eq.~(\ref{sigma-2l-s})) fermion self-energies as follows:
\begin{subequations}
\label{ren-sigmas}
\bea
\Sigma_{V1}(p^2) &=& \frac{e^2 \,(-p^2)^{-\varepsilon_\gamma}}{(4 \pi)^{d_\gamma/2}}\,\sigma_1(\varepsilon_e,\varepsilon_\gamma,a),
\label{Ren:def:sigma1}\\
\Sigma_{V2}(p^2) &=& \frac{e^4 (-p^2)^{-2\varepsilon_\gamma} }{(4\pi)^{d_\gamma}}\,\sigma_2(\varepsilon_e,\varepsilon_\gamma,a).
\label{Ren:def:sigma2}
\eea
\end{subequations}
As will be seen shortly, it is important to single out in the function $\sigma_1$ the part which depends on the (bare) gauge fixing parameter $a$. 
From Eq.~(\ref{sigma-1l-s}), this yields
\begin{subequations}
\label{ren-sigma1s}
\bea
&&\sigma_1(\varepsilon_e,\varepsilon_\gamma,a) = \sigma_1'(\varepsilon_e,\varepsilon_\gamma) + a\, \sigma_1''(\varepsilon_e,\varepsilon_\gamma),
\label{sigma1-sep}\\
&&\sigma_1'(\varepsilon_e,\varepsilon_\gamma) = \Gamma(1-\varepsilon_e)\,\frac{\varepsilon_e\,(1-\varepsilon_e-\varepsilon_\gamma)}{2-\varepsilon_e-2\varepsilon_\gamma}\,G(1,1-\varepsilon_e)\, ,
\label{Ren:sigma1}\\
&&\sigma_1''(\varepsilon_e,\varepsilon_\gamma) = -\Gamma(1-\varepsilon_e)\,(1-\varepsilon_e-\varepsilon_\gamma)\,G(1,1-\varepsilon_e)\, .
\eea
\end{subequations}
On the other hand, in the 2-loop function $\sigma_2$ we may single out the three contributions corresponding to the three diagrams of 
Fig.~\ref{fig:rqed-2loop-self-energy} whose expressions are given by Eqs.~(\ref{sigma-a-s}), (\ref{sigma-b-s}) and (\ref{sigma-c-s2}). This yields:
\begin{subequations}
\label{ren-sigma2s}
\bea
&&\sigma_2(\varepsilon_e,\varepsilon_\gamma,a) 
= \sigma^{(2)}_a(\varepsilon_e,\varepsilon_\gamma)+\sigma^{(2)}_b(\varepsilon_e,\varepsilon_\gamma,a)+\sigma^{(2)}_c(\varepsilon_e,\varepsilon_\gamma,a)\, ,
\\
&&\sigma^{(2)}_a(\varepsilon_e,\varepsilon_\gamma) 
= -4N_F\,\Gamma^2(1-\varepsilon_e)\,\frac{(1-\varepsilon_e-\varepsilon_\gamma)^2}{1-\varepsilon_e+\varepsilon_\gamma}\,G(1,1)G(1,\varepsilon_\gamma-\varepsilon_e)\, ,
\\
&&\sigma^{(2)}_b(\varepsilon_e,\varepsilon_\gamma,a) = 
-\Gamma^2(1-\varepsilon_e)\,\frac{(1-2\varepsilon_\gamma)(1-\varepsilon_e-\varepsilon_\gamma)\left(\varepsilon_e-a(2-\varepsilon_e-2\varepsilon_\gamma)\right)^2}
{\varepsilon_\gamma (2-\varepsilon_e-2\varepsilon_\gamma)}\,G(1,1-\varepsilon_e)G(1-\varepsilon_e,\varepsilon_\gamma)\, ,
\\
%
&&\sigma^{(2)}_c(\varepsilon_e,\varepsilon_\gamma,a) = \Gamma^2(1-\varepsilon_e)\, (1-\veps_e-\veps_\gamma)\,\times
\\
&&\times \Bigg \{ G^2(1,1-\veps_e)\, \left[ 
 2(\veps_e + \veps_\gamma) +4(1-a)\frac{(1-\veps_e-\veps_\gamma)^2}{2-\veps_e-2\veps_\gamma}
-(1-a)^2\,(1-\veps_e-\veps_\gamma) \right . \Bigg .
\nonumber \\
&&\left . + 2\,\frac{(1-\veps_e-2\veps_\gamma)\,[2-(2-\veps_\gamma)(\veps_e+\veps_\gamma)]}{(1-\veps_e-3\veps_\gamma)(2-\veps_e-3\veps_\gamma)} +
\frac{4\veps_e}{2-\veps_e-2\veps_\gamma}-\frac{2\veps_e}{2-\veps_e-3\veps_\gamma}\,\left(2+\veps_e+\veps_\gamma+2\,\frac{1-\veps_e-2\veps_\gamma}{2-\veps_e-2\veps_\gamma} \right) \right]
\nonum \\
&&- G(1,1-\veps_e)\,G(1-\veps_e,\veps_\gamma)\, \left[ 4 - 4\veps_e - 4\veps_\gamma + \frac{8(1-\veps_e-\veps_\gamma)}{2-\veps_e-2\veps_\gamma}-\frac{4(3-2\veps_\gamma)}{\veps_\gamma} +
4(1-a)\,\frac{(1-\veps_e-\veps_\gamma)(1-2\veps_\gamma)}{2-\veps_e-2\veps_\gamma} \right .
\nonum \\
&&\left . -(1-a)^2\,(1-2\veps_\gamma) -4\,\frac{\veps_e(2+\veps_e+\veps_\gamma)}{1-\veps_e-2\veps_\gamma}+4\,\frac{\veps_e(1-\veps_\gamma)}{\veps_\gamma(2-\veps_e-2\veps_\gamma)}\right]
\nonum \\
&& \Bigg .- \frac{2\veps_\gamma\,[2 - (2-\veps_\gamma)(\veps_e+\veps_\gamma)]}{(2-\veps_e-3\veps_\gamma)(1-\veps_e-3\veps_\gamma)}\,G(1-\veps_e,1,1-\veps_e,1,1) \Bigg \}\, .
\nonum
\eea
\end{subequations}
These equations are no more than a re-writing of the two-loop self-energies, Eqs.~(\ref{sigma-a-s}), (\ref{sigma-b-s}) and (\ref{sigma-c-s2}), 
in terms of the $\veps$-parameters instead of the dimensions.
 
In order to implement the renormalization procedure all bare quantities (charge and gauge fixing parameter) appearing in the self-energies Eqs.~(\ref{ren-sigmas})
have to be expressed in terms of renormalized ones:
\be
\frac{e^2}{(4\pi)^{d_\gamma/2}} = Z_\al(\al(\mu))\,\frac{\al(\mu)}{4\pi}\,\mu^{2\varepsilon_\gamma}\,e^{\gamma_E \varepsilon_\gamma}, \qquad a = Z_A(\al(\mu)) \,a_r(\mu), 
\ee
where Eqs.~(\ref{ren_coupling}) and (\ref{ren_constants}) were used.
For an expansion to $\Ord(\al^2)$ it is enough to know the coupling and gauge field renormalizations at one-loop.
From Ref.~[\onlinecite{TeberRQED12}] they read:
\be
Z_\al(\al) = Z_A^{-1}(\al) = 1+z_\al\,\frac{\al}{4\pi\varepsilon_\gamma} + \Ord(\al^2), \quad \gamma_A(\al) = 2z_\al\,\frac{\al}{4\pi}, \quad \beta(\al) = -2\varepsilon_\gamma + 2z_\al\,\frac{\al}{4\pi},
\label{1loopren-gauge}
\ee
where the coefficient $z_\al$ depends of the theory under consideration:
\be
z_\al = \frac{4N_F}{3}\,\,\,(\mathrm{QED}_4),\qquad z_\al = 0\,\,\,(\mathrm{RQED}_{4,3}).
\label{1loopren-gauge-zal}
\ee
Summing the one-loop and two-loop contributions, the total self-energy up to two loops can then be written as:
\be
\Sigma_V(p^2) = \frac{\al}{4\pi}\,e^{(\gamma_E-L_p)\varepsilon_\gamma}\,\sigma_1(a_r) + 
\left( \frac{\al}{4\pi} \right)^2\, \left( \frac{z_\al}{\veps_\gamma}\,e^{(\gamma_E-L_p)\varepsilon_\gamma}\,\sigma_1'+e^{2(\gamma_E-L_p)\varepsilon_\gamma}\,\sigma_2(a_r) \right) +\Ord(\al^3)\, ,
\label{self-energy-exp}
\ee
where $L_p=\log(-p^2/\mu^2)$ and the functions $\sigma_1 = \sigma_1'+a_r\,\sigma_1''$ and $\sigma_2$ now depend on the {\it renormalized} gauge fixing parameter.
With the help of Eq.~(\ref{self-energy-exp}) the expansion of the fermion propagator Eq.~(\ref{fermion-prop+sigma}) reads:
\be
-i \Sp\,S(p) = 1 +
\frac{\al}{4\pi}\,e^{(\gamma_E-L_p)\varepsilon_\gamma}\,\sigma_1(a_r) +
\left( \frac{\al}{4\pi} \right)^2\, \left[ \frac{z_\al}{\veps_\gamma}\,
e^{(\gamma_E-L_p)\varepsilon_\gamma}\,\sigma_1'+e^{2(\gamma_E-L_p)\varepsilon_\gamma}\,\left( \sigma_1^2(a_r) + \sigma_2(a_r) \right) \right] 
+\Ord(\al^3).
\label{fermion-prop-exp1}
\ee
The Laurent series associated with the various functions entering Eq.~(\ref{fermion-prop-exp1}) can be written as:
\begin{subequations}
\label{exps}
\bea
&&e^{(\gamma_E-L_p)\varepsilon_\gamma}\,\sigma_1' = \frac{c_{10}'}{\varepsilon_\gamma} + c_{11}' + c_{12}'\varepsilon_\gamma + \cdots \, , \quad
e^{(\gamma_E-L_p)\varepsilon_\gamma}\,\sigma_1'' = \frac{c_{10}''}{\varepsilon_\gamma} + c_{11}'' + c_{12}''\varepsilon_\gamma + \cdots \, ,
\label{c1j} \\
&&e^{2(\gamma_E-L_p)\varepsilon_\gamma}\,\left( \sigma_1^2 + \sigma_2 \right) = \frac{c_{20}}{\varepsilon_\gamma^2} + \frac{c_{21}}{\varepsilon_\gamma} + c_{22} + \cdots \, ,
\label{c2j}
\eea
\end{subequations}
where the dots indicate higher order terms in $\veps_\gamma$.
In Eq.~(\ref{c2j}) the two-loop contribution has been expanded up to $\Ord(1)$ in $\veps_\gamma$ which is all we need for two-loop renormalization and is within the accuracy of our 
loop calculations in RQED$_{4,d_e}$. With these notations Eq.~(\ref{fermion-prop-exp1}) can then be re-written as:
\bea
-i \Sp\,S(p) = 1 + 
\frac{\al}{4\pi}\,\left( \frac{c_{10}}{\varepsilon_\gamma} + c_{11} + c_{12}\varepsilon_\gamma + \cdots \right)  +
\left( \frac{\al}{4\pi} \right)^2\, \left( \frac{c_{20} + z_\al c_{10}'}{\varepsilon_\gamma^2} + \frac{c_{21}+z_\al c_{11}'}{\varepsilon_\gamma} + c_{22} + z_\al c_{12}' + \cdots \right)
+ \Ord(\al^3)\, ,
\label{def:series:S}
\eea
where the one-loop coefficients $c_{1j} = c_{1j}' + a_r\, c_{1j}''$ are defined by Eqs.~(\ref{c1j}) and the coefficients $c_{2j}$, which also depend on $a_r$, are defined by Eq.~(\ref{c2j}).

In the $\overline{\rm{MS}}$ scheme all singular terms are factorized in the renormalization constant $Z_\psi$ while 
the regular terms contribute to the renormalized propagator $S_r$, yielding general expressions of the form:
\begin{subequations}
\label{ren-Z+S}
\bea
Z_\psi(\al,a_r) &=& 1 + \frac{\al}{4\pi}\,\frac{z_1}{\varepsilon_\gamma} 
+ \left( \frac{\al}{4\pi} \right)^2\, \left(\frac{z_{20}}{\varepsilon_\gamma^2} + \frac{z_{21}}{\varepsilon_\gamma}\right) + \Ord(\al^3),
\label{def:series:Z}\\
-i \Sp\,S_r(p;\mu) &=& 1 + \frac{\al}{4\pi}\,\left(r_1 + r_{11}\varepsilon_\gamma +\cdots \right) 
+ \left( \frac{\al}{4\pi} \right)^2\, \left( r_2 + \cdots \right) + \Ord(\al^3),
\label{def:series:Sr}
\eea
\end{subequations}
where the coefficients $z_{ij}$ and $r_{ij}$ are to be determined. 
This can be done with the help of Eq.~(\ref{mpropr}) by identifying, at a given loop order, the coefficients of equal powers of $\varepsilon_\gamma$.
The result reads:
\begin{subequations}
\label{ren-z+r}
\bea
&&z_1 = c_{10}, \qquad z_{20} = c_{20} + z_\al\,c_{10}', \qquad z_{21} = c_{21} + z_\al\,c_{11}'- c_{10} c_{11},
\label{def:zcoeff}\\
&&r_1 = c_{11}, \qquad r_{11} = c_{12}, \qquad \qquad \qquad r_2 = c_{22}+z_\al\,c_{12}'- c_{10} c_{12}.
\label{def:rcoeff}
\eea
\end{subequations}
We are now in a position to present a general expression for the anomalous scaling dimension of the fermion field. In partial derivative form the latter reads
\be
\gamma_\psi(\al(\mu),a_r(\mu)) = \frac{d \log Z_\psi(\al(\mu),a_r(\mu))}{d \log \mu} = 
\beta(\al(\mu))\,\frac{\partial \log Z_\psi(\al,a_r)}{\partial \log \al} - \gamma_A(\al)\,\frac{\partial \log Z_\psi(\al,a_r)}{\partial \log a_r},
\label{gamma_psi_2l}
\ee
where the one-loop beta-function and anomalous scaling dimension of the gauge field are given by Eqs.~(\ref{1loopren-gauge}).
From Eqs.~(\ref{def:series:Z}), (\ref{def:zcoeff}) and (\ref{1loopren-gauge}), we recover the fact, well known in usual QED, that the simple poles of $Z_\psi$
determine the anomalous scaling dimension:
\be
\gamma_\psi(\al,a_r) = -2z_1\,\frac{\al}{4\pi} - 4z_{21}\,\left( \frac{\al}{4\pi} \right)^2 + \Ord(\al^3),
\label{gamma_psi-z}
\ee
and that, together with the beta function, they also determine the higher order poles:
\be
z_{20} = \frac{z_1^2 + z_\al \,c_{10}'}{2}.
\label{z-constraint}
\ee
The constraint Eq.~(\ref{z-constraint}) comes from the fact that the anomalous dimension has to be finite as $\veps_\gamma \ra 0$.

\subsection{Case of RQED$_{4,d_e}$}

In this paragraph we shall compute the anomalous dimension of the fermion field and renormalized fermion propagator in a general theory of RQED$_{4,d_e}$.
In order to do so, we expand the $\sigma$ functions of Eqs.~(\ref{ren-sigma1s}) and (\ref{ren-sigma2s}) in Laurent series in $\veps_\gamma$ for arbitrary $\veps_e$. 
The coefficients of the expansions read:
\begin{subequations}
\label{ren-sigmas-RQED4de}
\bea
&& e^{(\gamma_E-L_p)\veps_\gamma}\,\sigma_1' = \frac{\veps_e}{(2-\veps_e)\,\veps_\gamma} - \frac{\veps_e}{2-\veps_e}\,\left(\bar{L}_p
- \frac{2}{2-\veps_e} \right) ,
\label{sigma'-RQED4de}\\
&& \qquad + \frac{\veps_e}{2(2-\veps_e)}\,\Bigg( \bar{L}_p
\left( \bar{L}_p - \frac{4}{2-\veps_e} \right) +2 \zeta_2 -3\,\Psi_2(2-\veps_e) + \frac{8}{(2-\veps_e)^2} \Bigg)\,\veps_\gamma 
+ \Ord(\veps_\gamma^2)\, , 
\nonum \\
&& e^{(\gamma_E-L_p)\veps_\gamma}\,\sigma_1'' = -\frac{1}{\veps_\gamma}+\bar{L}_p
- \frac{1}{2}\,\Bigg(\bar{L}_p^2 + 2\zeta_2 - 3 \Psi_2(2-\veps_e) \Bigg)\,\veps_\gamma + \Ord(\veps_\gamma^2),
\label{sigma''-RQED4de} \\
&& e^{2(\gamma_E-L_p)\varepsilon_\gamma}\,\left( \sigma_1^2 + \sigma_2 \right) = \frac{1}{2}\,\left(a_r - \frac{\veps_e}{2-\veps_e} \right)^2\,\frac{1}{\veps_\gamma^2}
- \Bigg( \big( a_r - \frac{\veps_e}{2-\veps_e} \big)^2\,\bar{L}_p
+ 2\,a_r\,\frac{\veps_e}{(2-\veps_e)^2}  \Bigg .
\label{sigma2-RQED4de} \\
&& \qquad \Bigg .  - 2\,\frac{\veps_e^2 (10-3\veps_e)-11\veps_e+3}{(1-\veps_e)(2-\veps_e)^3}
+ 2 N_F K_1\,\frac{\veps_e}{2-\veps_e} \Bigg) \, \frac{1}{\veps_\gamma} \nonumber \\
&& \qquad +\left(a_r-\frac{\veps_e}{2-\veps_e}\right)^2\,\left( \bar{L}_p^2+\frac{3}{2}\,\zeta_2-
2\Psi_2(2-\veps_e) \right) 
+4a_r\,\frac{\veps_e}{(2-\veps_e)^2}\,\left( \bar{L}_p - \frac{1}{2-\veps_e}\right)
\nonumber \\
&& \qquad + 4\,\frac{3-2\veps_e}{(2-\veps_e)^2}\,\Bigg(\Psi_2(1-\veps_e) - \zeta_2 \Bigg)
-4\bar{L}_p\,\frac{\veps_e^2(10-3\veps_e)-11\veps_e+3}{(1-\veps_e)(2-\veps_e)^3}
+ 2\,\frac{\veps_e^2(32-10\veps_e) -23\veps_e - 7}{(1-\veps_e)(2-\veps_e)^4}
\nonumber \\
&& \qquad -2\,\frac{\veps_e^2}{(1-\veps_e)^2(2-\veps_e)^2} + 4N_F K_1\,\frac{\veps_e}{2-\veps_e}\,\left( \bar{L}_p+\frac{1}{2}\,(\overline{\Psi}_1 - \overline{\Psi}_2)
+\frac{1}{\veps_e(1-\veps_e)(2-\veps_e)}-\frac{1}{2-\veps_e} \right)\,
+ \Ord(\veps_\gamma) \, ,
\nonum
\eea
\end{subequations}
where 
\begin{subequations}
\label{ren-params-RQED4de}
\bea
&&\bar{L}_p = L_p - \Psi_1(2-\veps_e)+ \Psi_1(1), \qquad K_1 = \frac{\Gamma^3(1-\veps_e)\Gamma(\veps_e)}{\Gamma(2-2\veps_e)}\, ,
\label{ren-params-RQED4de-Lp+K1} \\
&&\overline{\Psi}_1=\Psi_1(1-\veps_e)-\Psi_1(1), \qquad
\overline{\Psi}_2=\Psi_1(\veps_e)-2\Psi_1(1-\veps_e)+2\Psi_1(2-2\veps_e) -\Psi_1(1)\, ,
\label{ren-params-RQED4de-Psis}
\eea
\end{subequations}
and $\Psi_1$ and $\Psi_2$ are the digamma and trigamma functions, respectively.
From Eqs.~(\ref{ren-sigmas-RQED4de}) the coefficients $c_{1j}$ and $c_{2j}$ can be deduced. 

Focusing for the moment on the singular contributions, which are obtained by combining Eqs.~(\ref{ren-sigmas-RQED4de}) with Eqs.~(\ref{def:zcoeff}), yields:
\begin{subequations}
\label{ren-z-RQED4de}
\bea
&&z_1 = -\left(a_r - \frac{\veps_e}{2-\veps_e}\right), \qquad z_{20} = \frac{1}{2}\,\left(a_r - \frac{\veps_e}{2-\veps_e} \right)^2 + z_\al\,\frac{\veps_e}{2-\veps_e}\, , 
\label{z1+z20-RQED4de}\\
&&z_{21} = -2N_F K_1\,\frac{\veps_e}{2-\veps_e} + 2\,\frac{(3-2\veps_e)\,[1-\veps_e\,(3-\veps_e)]}{(1-\veps_e)(2-\veps_e)^3} 
-z_\al\,\frac{\veps_e}{2-\veps_e}\,\left(\bar{L}_p
- \frac{2}{2-\veps_e} \right)\, ,
\label{z21-RQED4de}
\eea
\end{subequations}
where all gauge depend terms cancel out from $z_{21}$.
In order for the consistency relation, Eq.~(\ref{z-constraint}), to be valid we see from Eqs.~(\ref{z1+z20-RQED4de}) that we must either have $\veps_e = 0$ if $z_\al \not= 0$ which 
is the case of QED$_4$ or $z_\al=0$ if $\veps_e \not= 0$ which is the case of RQED$_{4,3}$, see Eq.~(\ref{1loopren-gauge-zal}). Taking this constraint into account 
we see that momentum dependent terms also cancel out from $z_{21}$. 
Then, the general expression of the anomalous dimension of the fermion field in RQED$_{4,d_e}$ reads
\be
\gamma_\psi(\al,a_r) = 2\left(a_r - \frac{\veps_e}{2-\veps_e} \right)\,\frac{\al}{4\pi} 
+8\,\left( N_F K_1 \frac{\veps_e}{2-\veps_e} 
- \frac{(3-2\veps_e)\,[1-\veps_e\,(3-\veps_e)]}{(1-\veps_e)(2-\veps_e)^3} \right)\,\left( \frac{\al}{4\pi} \right)^2 + \Ord(\al^3)\, .
\label{asd-RQED4de}
\ee
It is important that Eq.~(\ref{asd-RQED4de}) does not depend on the external momenta and digamma functions. 
This property is an extension of the 
rule \cite{Vladimirov80}
for the results of the anomalous dimensions in standard
QFT ($\veps_e=0$ and $\veps_\gamma \to 0$), where the corresponding
anomalous dimensions do not depend on the external momenta and 
(in $\overline{MS}$-scheme) on
$\gamma_E$ and $\zeta_2$. Note that in the framework of RQED$_{4,3}$ there is 
a contribution of $\zeta_2$ in the corresponding anomalous dimension 
(see Eq. (\ref{asd-RQED43}) below) but this
contribution appears, not from the expansion of $\Gamma$-functions in 
 $\veps_\gamma$
but from the 
factor $K_1$ in the expression $G(1,1)$, see App.~\ref{app-Gfuncs}.

Finally, the finite part can be obtained in a similar way 
by combining Eqs.~(\ref{ren-sigmas-RQED4de}) with Eqs.~(\ref{def:rcoeff}). This yields:
\begin{subequations}
\label{ren-r-RQED4de}
\bea
r_1 &=& \left(a_r-\frac{\veps_e}{2-\veps_e}\right)\,\bar{L}_p + \frac{2\veps_e}{(2-\veps_e)^2}\, ,
\label{r1-RQED4de} \\
r_{11} &=&  -\frac{1}{2}\,\left(a_r-\frac{\veps_e}{2-\veps_e}\right)\,\biggl(\bar{L}_p^2+2\zeta_2-3\Psi_2(2-\veps_e) \biggr) -
\frac{2\veps_e}{(2-\veps_e)^2}\,\left(\bar{L}_p-\frac{2}{2-\veps_e} \right)\, ,
\label{r11-RQED4de} \\
r_{2} &=& \frac{1}{2}\,\left(a_r-\frac{\veps_e}{2-\veps_e}\right)^2\,\biggl( \bar{L}_p^2+\zeta_2-\Psi_2(2-\veps_e) \biggr)
+2a_r\,\frac{\veps_e}{(2-\veps_e)^2}\,\bar{L}_p
+ 4\,\frac{3-2\veps_e}{(2-\veps_e)^2}\,\Bigg(\Psi_2(1-\veps_e) - \zeta_2 \Bigg)
\nonumber \\
&-& 2\bar{L}_p\,\frac{6-22\veps_e+19\veps_e^2-5\veps_e^3}{(1-\veps_e)(2-\veps_e)^3}
-2\,\frac{7+23\veps_e-30\veps_e^2+8\veps_e^3}{(1-\veps_e)(2-\veps_e)^4}-2\,\frac{\veps_e^2}{(1-\veps_e)^2(2-\veps_e)^2}
\nonumber \\
&+& 4N_F K_1\,\frac{\veps_e}{2-\veps_e}\,\left( \bar{L}_p+\frac{1}{2}\,(\overline{\Psi}_1 - \overline{\Psi}_2)
+\frac{1}{\veps_e(1-\veps_e)(2-\veps_e)}-\frac{1}{2-\veps_e} \right)\, .
\label{r2-RQED4de}
\eea
\end{subequations}
A few remarks are in order concerning the above results starting from Eqs.~(\ref{ren-sigmas-RQED4de}).
The Laurent series in $\veps_\gamma$ are plagued by singularities for some values of $\veps_e$. The singularity at $\veps_e=2$ (case
of a 0-brane or quantum mechanics) reflects the fact that the self-energy of a point-like particle is ill-defined already at one-loop.
The singularity at $\veps_e=1$ (case of a 1-brane or RQED$_{4,2}$) appears starting from two-loop and is related to the zero-width of the filament, {\it i.e.},
the higher powers of $\Gamma(1-\veps_e)$, coming from the effective free gauge field propagator (\ref{gauge-field-prop0}), as the order increases. 
Finally, at two-loop, a singularity appears for $\veps_e=0$ (case of QED$_4$) as can be seen
in particular from the last term in Eq.~(\ref{r2-RQED4de}).
This singularity is gauge invariant and can be traced back to the UV behaviour of the one-loop polarization operator entering the bubble diagram
which is contained in the master integral $G(1,1)$, see App.~\ref{app-Gfuncs}. The above results are therefore valid only in application to RQED$_{4,3}$.

\subsection{Application to massless RQED$_{4,3}$}

In the case of massless RQED$_{4,3}$ ($\varepsilon_e=1/2$ and $\varepsilon_\gamma \ra 0$) the $\veps_\gamma$-expansions of Eqs.~(\ref{ren-sigmas-RQED4de}) read:
\begin{subequations}
\label{ren-sigmas-RQED43}
\bea
e^{(\gamma_E-L_p)\veps_\gamma}\,\sigma_1' &=& \frac{1}{3\veps_\gamma} + \frac{4-3\bar{L}_p}{9}
+ \frac{1}{6}\,\left( \bar{L}_p \left( \bar{L}_p -\frac{8}{3} \right) -   7\zeta_2 + \frac{140}{9} \right)\,\veps_\gamma + \Ord(\veps_\gamma^2)\, ,
\label{sigma'-RQED43}\\
e^{(\gamma_E-L_p)\veps_\gamma}\,\sigma_1'' &=& -\frac{1}{\veps_\gamma}+\bar{L}_p -
\frac{\bar{L}_p^2-7\zeta_2 +12}{2}\,\veps_\gamma + \Ord(\veps_\gamma^2)\, ,
\label{sigma''-RQED4} \\
e^{2(\gamma_E-L_p)\varepsilon_\gamma}\,\left( \sigma_1^2 + \sigma_2 \right) &=& \frac{(1-3a_r)^2}{18\veps_\gamma^2}
-\frac{(1-3a_r)^2\,\bar{L}_p + 4a_r + 36 \zeta_2 N_F + 4}{9 \veps_\gamma} +
\nonum \\
&+& \frac{(1-3a_r)^2}{9}\,\left(\bar{L}_p^2 -\frac{9}{2}\,\zeta_2+8 \right)
+ \frac{8a}{9}\,\left( \bar{L}_p -\frac{2}{3} \right) +\frac{64}{9}\,\zeta_2 + \frac{8}{9}\,\bar{L}_p
\nonum \\
&+& 8 \, \zeta_2 \,N_F \,(\bar{L}_p +2 - \log 4) - \frac{824}{81} + \Ord(\veps_\gamma)\, ,
\label{sigma2-RQED43}
\eea
\end{subequations}
where $\bar{L}_p = \log(-p^2/\mu^2)+\log4-2$.
The coefficients then read:
\begin{subequations}
\label{ren-z+r-RQED43}
\bea
&&z_1 = \frac{1-3a_r}{3}\, , \qquad z_{20} = \frac{(1-3a_r)^2}{18}\, , \qquad z_{21} = -4\,\zeta_2\,N_F - \frac{16}{27}\, ,
\label{def:zcoeff-RQED43}\\
&&r_1 = \frac{4}{9}-\frac{1-3a_r}{3}\,\bar{L}_p\, , \qquad
r_{11} = \frac{1-3a_r}{6}\,\bar{L}_p^2 -\frac{4}{9}\,\bar{L}_p-7\,\zeta_2\,\frac{1-3a_r}{6}-6a_r + \frac{70}{27}\, ,
\nonum \\
&&r_2 = \frac{(1-3a_r)^2}{18}\,\left( \bar{L}_p^2 - 2\,\zeta_2 + 4 \right) +4\,\frac{(3a_r+7)\,\bar{L}_p + 48 \zeta_2}{27} -8\,\zeta_2\,N_F\,(\bar{L}_p+2-\log4) - \frac{280}{27}\, .
\label{def:rcoeff-RQED43}
\eea
\end{subequations}
The corresponding renormalization constant and renormalized fermion propagator read:
\begin{subequations}
\label{2loopren-RQED43}
\bea
&&Z_\psi(\al,a_r) = 1 +\frac{1-3a_r}{3}\,\frac{\al}{4\pi\varepsilon_\gamma}
+ \bigg[  \frac{(1-3a_r)^2}{18} - 4\left(\zeta_2 N_F + \frac{4}{27}\right)\varepsilon_\gamma + \cdots \bigg]\,\left(\frac{\al}{4\pi \varepsilon_\gamma}\right)^2 + \Ord(\al^3)\, ,
\label{2loopren-RQED43-Zpsi}\\
&&-i \Sp\,S_r(p;\mu) = 1 + \frac{\al}{4\pi}\,\biggl[ \frac{4}{9}-\frac{1-3a_r}{3}\,\bar{L}_p + 
\biggl( \frac{1-3a_r}{6}\,\bar{L}_p^2 -\frac{4}{9}\,\bar{L}_p-7\,\zeta_2\,\frac{1-3a_r}{6}-6a_r + \frac{70}{27} \biggr)\,\veps_\gamma + \cdots \biggr]
\label{2loopren-RQED43-Sr}\\
&&+\left(\frac{\al}{4\pi}\right)^2\,\biggl[ \frac{(1-3a_r)^2}{18}\,\left( \bar{L}_p^2 - 2\,\zeta_2 + 4 \right) 
+4\,\frac{(3a_r+7)\,\bar{L}_p + 48 \zeta_2}{27} -8\,\zeta_2\,N_F\,(\bar{L}_p+2-\log4) - \frac{280}{27} +\cdots \biggr] + \Ord(\al^3)\, .
\nonum
\eea
\end{subequations}
Finally, the anomalous scaling dimension of the fermion field in RQED$_{4,3}$ reads:
\be
\gamma_\psi(\al,a_r) = 2\frac{3a_r-1}{3}\,\frac{\al}{4\pi} +16\left(\zeta_2 N_F + \frac{4}{27}\right)\,\left(\frac{\al}{4\pi}\right)^2 + \Ord(\al^3)\, .
\label{asd-RQED43}
\ee
As already noticed in the general formulas, this anomalous dimension does not depend on the external momentum and
the two-loop contribution does not depend on the choice of gauge. In order to see how it affects the momentum dependence
of the fermion propagator we solve Eq.~(\ref{ren-eq-part-p}). For this purpose, we use the fact that: $\beta(\al)=\gamma_A(\al)=0$,
as the coupling and gauge field do not renormalize in RQED$_{4,3}$. The solution then reads: 
\be
S_r(p\,;\mu) = \frac{i}{\Sp}\,s_r(1;\al; a)\,\left( \frac{-p^2}{\mu^2} \right)^{\frac{1}{2}\,\gamma_\psi(\al,a)}\, ,
\label{ren-prop-RQED43}
\ee
where $\gamma_\psi$ is given by Eq.~(\ref{asd-RQED43}).

\subsection{Case of massless QED$_4$}

The case of massless QED$_4$ corresponds to: $\varepsilon_e=0$ and $\varepsilon_\gamma \ra 0$.
Starting from Eq.~(\ref{sigma-2l-s-QEDd}) the $\veps_\gamma$-expansions of the self-energies read:
\begin{subequations}
\label{ren-sigmas-QED4}
\bea
 e^{(\gamma_E-L_p)\veps_\gamma}\,\sigma_1' &=& 0,
\label{sigma'-QED4}\\
 e^{(\gamma_E-L_p)\veps_\gamma}\,\sigma_1'' &=& -\frac{1}{\veps_\gamma}+L_p-1+\frac{\zeta_2 -L_p(L_p-2) - 4}{2}\,\veps_\gamma + \Ord(\veps_\gamma^2),
\label{sigma''-QED4} \\
 e^{2(\gamma_E-L_p)\varepsilon_\gamma}\,\left( \sigma_1^2 + \sigma_2 \right) &=& \frac{a_r^2}{2\veps_\gamma^2} + \frac{a_r^2\,(1-L_p)+N_F+3/4}{\veps_\gamma} +
\nonum \\
&+& a_r^2\Big( L_p(L_p-2) - \frac{1}{2}\zeta_2+3 \Big) -2\,L_p\left(N_F+\frac{3}{4}\right) + \frac{7}{2}N_F+\frac{5}{8} + \Ord(\veps_\gamma),
\label{sigma2-QED4}
\eea
\end{subequations}
from which the coefficients $c_{1j}$ and $c_{2j}$ can be deduced; in particular, the coefficients $c_{1j}'=0$ in QED$_4$. 
Substituting these coefficients in Eqs.~(\ref{ren-z+r}), yields:
\begin{subequations}
\label{ren-z+r-QED4}
\bea
&&z_1 = -a_r, \qquad z_{20} = \frac{a_r^2}{2}, \qquad z_{21} = N_F + \frac{3}{4},
\label{def:zcoeff-QED4}\\
&&r_1 = a_r(L_p-1), \qquad 
r_{11} = \frac{a_r}{2}\,\left(\zeta_2 - L_p(L_p-2)-4\right), 
\nonum \\ 
&&r_2 = a_r^2\left(1+\frac{1}{2}L_p(L_p-2)\right) -2L_p\left(N_F+\frac{3}{4}\right) + \frac{7}{2}N_F + \frac{5}{8}.
\label{def:rcoeff-QED4}
\eea
\end{subequations}
From Eq.~(\ref{def:zcoeff-QED4}) we see that Eq.~(\ref{z-constraint}) is indeed satisfied. 
Moreover, similarly to the case of QED$_{4,3}$, while $z_1$ is gauge-variant, the gauge dependence cancels out in $z_{21}$.
Substituting the coefficients of Eqs.~(\ref{ren-z+r-QED4}) in Eqs.~(\ref{ren-Z+S}) yields the following renormalization constant and renormalized fermion propagator: 
\begin{subequations}
\label{2loopren-QED4}
\bea
Z_\psi(\al,a_r) &=& 1 -a_r\,\frac{\al}{4\pi\varepsilon_\gamma} 
+ \left[ \frac{a_r^2}{2} + (N_F + \frac{3}{4})\,\varepsilon_\gamma + \cdots \right]\,\left(\frac{\al}{4\pi \varepsilon_\gamma}\right)^2 + \Ord(\al^3),
\label{2loopren-QED4-Zpsi}\\
-i \Sp\,S_r(p;\mu) &=& 1 + \frac{\al}{4\pi}\,\bigg[a_r(L_p-1) + \frac{a_r}{2} \left(\zeta_2-L_p(L_P-2)-4\right)\,\varepsilon_\gamma + \cdots\bigg] + \\
&\quad& + \left(\frac{\al}{4\pi}\right)^2\,\bigg[a_r^2\left(1+\frac{1}{2}L_p(L_p-2)\right) -2L_p\left(N_F+\frac{3}{4}\right) + \frac{7}{2}N_F + \frac{5}{8} + \cdots\bigg] + \Ord(\al^3),
\label{2loopren-QED4-Sr}
\eea
\end{subequations}
Finally, from Eqs.~(\ref{def:zcoeff-QED4}) and (\ref{gamma_psi-z}) the anomalous scaling dimension of the fermion field in QED$_4$ is recovered:
\be
\gamma_\psi(\al,a_r) = 2a_r\,\frac{\al}{4\pi} - 4\left(N_F + \frac{3}{4}\right)\,\left(\frac{\al}{4\pi}\right)^2 + \Ord(\al^3) \, .
\label{asd-QED4}
\ee 
As in the case of RQED$_{4,3}$, see Eq.~(\ref{asd-RQED43}), the fermion anomalous dimension of QED$_4$ does not depend on the external momentum and the
two-loop contribution does not depend on the choice of gauge. Furthermore, Eq.~(\ref{asd-QED4}) is transcendentally simpler than Eq.~(\ref{asd-RQED43})
as no $\zeta_2$ appears in Eq.~(\ref{asd-QED4}).

\subsection{Case of massless QED$_3$}

Finally, we consider the case of massless QED$_3$. Such a model differs considerably from the previous ones because it is super-renormalizable.
Nevertheless we shall proceed along the same lines as the previous, renormalizable, models.
In QED$_3$ we have $\veps_e=0$ and the expansion parameter is $\delta_\gamma = \veps_\gamma -1/2$.
Eqs.~(\ref{ren-sigmas}) can then be written as:
\begin{subequations}
\label{ren-sigmas-QED3}
\bea
\Sigma_{V1}(p^2) &=& \frac{e^2 \,(-p^2)^{-\varepsilon_\gamma}}{(4 \pi)^{d_\gamma/2}}\,\sigma_1(\varepsilon_e,\varepsilon_\gamma,a)
= \frac{\tilde{\al}}{4\pi}\,e^{(\gamma_E-L_p)\delta_\gamma}\,\sigma_1(\varepsilon_e,\varepsilon_\gamma,a)\, ,
\label{Ren:def:sigma1-QED3}\\
\Sigma_{V2}(p^2) &=& \frac{e^4 (-p^2)^{-2\varepsilon_\gamma} }{(4\pi)^{d_\gamma}}\,\sigma_2(\varepsilon_e,\varepsilon_\gamma,a)
= \left( \frac{\tilde{\al}}{4\pi} \right)^2\, e^{2(\gamma_E-L_p)\delta_\gamma}\,\sigma_2(\varepsilon_e,\varepsilon_\gamma,a)\, .
\label{Ren:def:sigma2-QED3}
\eea
\end{subequations}
where $L_p = \log(-p^2/\mu^2)$, $\tilde{\al}$ is a momentum-dependent dimensionless coupling constant ($e^2$ has dimension of mass in QED$_3$) defined as:
\be
\tilde{\al} = \frac{e^2}{\sqrt{4\pi}\,\sqrt{-p^2}}\, ,
\label{coupl-QED3}
\ee
and we have used the fact that $z_\al=0$ for QED$_3$, see Eq.~(\ref{1loopren-gauge}).
With the help of Eq.~(\ref{sigma-2l-s-QEDd}), the expansions read:
\begin{subequations}
\label{exp-sigmas-QED3}
\bea
&& e^{(\gamma_E-L_p)\delta_\gamma}\,\sigma_1 = -\frac{a \pi^{3/2}}{2} + \frac{a \pi^{3/2}}{2}\,\left( L_p - \log(4)+2 \right)\,\delta_\gamma + \Ord(\delta_\gamma^2)\, ,
\label{exp-1loop-QED3}\\
&& e^{2(\gamma_E-L_p)\delta_\gamma}\,\left( \sigma_1^2 + \sigma_2 \right) = - \frac{2\pi N_F}{3\delta_\gamma} + \pi \left( a^2 + \frac{4}{9}\,(3L_p-2)\,N_F + \frac{3 \pi^2}{4} - 7 \right)
+ \Ord(\delta_\gamma)\, .
\label{exp-2loop-QED3}
\eea
\end{subequations}
From these results we see that a singularity appears only at two-loop: the $1/\delta_\gamma$ pole in Eq.~(\ref{exp-2loop-QED3}).
This singularity is gauge independent and has a coefficient proportional to $N_F$. From Eq.~(\ref{sigma-2l-s-QEDd}) it can therefore be traced back to the bubble diagram
which involves the master integrals $G(1,1)$ and $G(1,\varepsilon_\gamma)$. While $G(1,1)$ is finite in three-dimensional QED,
the one-loop master integral $G(1,\varepsilon_\gamma)$ is indeed divergent in the limit $\varepsilon_\gamma \ra 1/2$:
\be
G(1,\varepsilon_\gamma) =  \frac{\Gamma(1+\delta_{\gamma})}{\sqrt{\pi}} \left[
\frac{1}{\delta_\gamma} + 6 - 2\log 2 
+ \Ord(\delta_\gamma)\right] \, .
\ee
At this point, the singularity looks like a UV one. Indeed, coming back to the general expression for the one-loop master 
integral (\ref{1loopG}) and expressing it in terms of gamma functions yields:
\be
G(\al,\beta) = \frac{\Gamma(\al + \beta - d_e/2)\Gamma(d_e/2-\al) \Gamma(d_e/2-\beta)}{\Gamma(\al)\Gamma(\beta)\Gamma(d_e-\al-\beta)}\, ,
\label{1loopG+}
\ee
where $d_e = 4 - 2\veps_e - 2\veps_\gamma$. Dimensional analysis then shows that a pole in the first gamma function in the numerator of (\ref{1loopG+})
is associated with a UV singularity whereas a pole in either of the two other gamma functions in the numerator is associated with an IR singularity.
For $\veps_e=0$, $\al=1$ and $\beta=\veps_\gamma \ra 1/2$, we then see that the singularity is in the first gamma function of the numerator:
\be
G(1,\varepsilon_\gamma) = 
\frac{\Gamma(2\veps_\gamma - 1)\Gamma(1-\veps_\gamma) \Gamma(2-2\veps_\gamma)}{\Gamma(\veps_\gamma)\Gamma(3-3\veps_\gamma)} = 
\frac{\Gamma(2\delta_\gamma)\Gamma(1/2-\delta_\gamma) \Gamma(1-2\delta_\gamma)}{\Gamma(1/2+\delta_\gamma)\Gamma(3/2-3\delta_\gamma)} = 
\frac{1}{2\delta_\gamma}\,\frac{\Gamma(1+2\delta_\gamma)\Gamma(1/2-\delta_\gamma) \Gamma(1-2\delta_\gamma)}{\Gamma(1/2+\delta_\gamma)\Gamma(3/2-3\delta_\gamma)}\, .
\ee
It turns out, however, that the present example is one in which there is an interchange between 
UV and IR types of singularities. Indeed, in dimensional regularization, both of these singularities correspond to poles of $\Gamma$-functions. 
To see the interchange, let's consider the example in more details.
The above considered UV type of singularity was actually related to our choice of master integral: we took 
$G(1,\varepsilon_\gamma-\varepsilon_e)$ in (\ref{sigma-2l-s-QEDd}). In our calculations, however, it is the integral
$G(1,\varepsilon_\gamma-\varepsilon_e+1)$ that is involved. On the one hand, it can be related to the integral $G(1,\varepsilon_\gamma-\varepsilon_e)$ by using
the simple property
\be
G(1,\alpha+1) = - \frac{d_e-2-\alpha}{\alpha} G(1,\alpha) \, .
\ee
On the other hand, the singularity in $G(1,\varepsilon_\gamma-\varepsilon_e+1)$ is an IR one. 
Indeed:
%
\be
G(1,\varepsilon_\gamma+1) = 
\frac{\Gamma(2\veps_\gamma)\Gamma(1-\veps_\gamma) \Gamma(1-2\veps_\gamma)}{
\Gamma(\veps_\gamma+1)\Gamma(2-3\veps_\gamma)} = 
\frac{\Gamma(1+2\delta_\gamma)\Gamma(1/2-\delta_\gamma) 
\Gamma(-2\delta_\gamma)}{\Gamma(3/2+\delta_\gamma)\Gamma(1/2-3\delta_\gamma)} 
= -
\frac{1}{2\delta_\gamma}\,\frac{\Gamma(1+2\delta_\gamma)\Gamma(1/2-\delta_\gamma) \Gamma(1-2\delta_\gamma)}{\Gamma(3/2+\delta_\gamma)\Gamma(1/2-3\delta_\gamma)}\, ,
\ee
and 
\be
G(1,\varepsilon_\gamma+1) =  \frac{\Gamma(1+\delta_{\gamma})}{\sqrt{\pi}} 
\left[
-\frac{1}{\delta_\gamma} - 2 + 2\log 2 
+ \Ord(\delta_\gamma)\right] \, .
\ee
Hence, in QED$_3$ the singularity in the two-loop fermion self-energy is of IR origin, as it was shown earlier in
Refs.~[\onlinecite{QED3_IR_papers}]. This is to be contrasted with the cases of QED$_4$ and RQED$_{4,3}$ where the corresponding singularities are of the UV type.

From equations (\ref{exp-sigmas-QED3}) we find the following coefficients:
\begin{subequations}
\label{ren-z+r-QED3}
\bea
&&z_1 = 0, \qquad z_{20} = 0, \qquad z_{21} = c_{21} = - \frac{2\pi N_F}{3}\, ,
\label{zcoeff-QED3}\\
&&r_1 = -\frac{a \pi^{3/2}}{2}, \qquad r_{11} = \frac{a \pi^{3/2}}{2}\,\left( L_p - \log(4)+2 \right), \qquad
r_2 = c_{22}=  \pi \left( a^2 + \frac{4}{9}\,(3L_p-2)\,N_F + \frac{3 \pi^2}{4} - 7 \right)  \, .
\label{rcoeff-QED3}
\eea
\end{subequations}
This yields the following result for the renormalized fermion propagator:
\bea
-i \Sp\,S_r(p;\mu) &=& 1 -\frac{a \sqrt{\pi}}{8}\,\tilde{\al}\,\bigg(1 - (L_p-\log(4)+2)\,\delta_\gamma + \cdots \bigg)
+ \frac{\tilde{\al}^2}{16\pi}\,\left( a^2 + \frac{4}{9}\,(3L_p-2)\,N_F + \frac{3 \pi^2}{4} - 7   \right) + \Ord(\tilde{\al}^3) \\
&=& 1 -\frac{a e^2}{16\sqrt{-p^2}}\,\bigg(1 - (L_p-\log(4)+2)\,\delta_\gamma + \cdots \bigg)
+ \frac{e^4}{64(-p^2)}\,\left( a^2 + \frac{4}{9}\,(3L_p-2)\,N_F + \frac{3 \pi^2}{4} - 7   \right) + \Ord(e^6)\, .
\nonumber
\eea
Similarly, the renormalization constant reads:
\be
Z_\psi = 1 - \frac{2\pi N_F}{3 \delta_\gamma}\, \left( \frac{\tilde{\al}}{4\pi} \right)^2 + \Ord(\tilde{\al}^3) =
1 - \frac{N_F}{24 \pi \delta_\gamma}\, \tilde{\al}^2 + \Ord(\tilde{\al}^3) \, .
\ee
Finally, the anomalous scaling dimension reads:
\be
\gamma_\psi = \frac{8\pi N_F}{3}\, \left( \frac{\tilde{\al}}{4\pi} \right)^2 + \Ord(\tilde{\al}^3) = +\frac{N_F}{6\pi}\,\tilde{\al}^2 + \Ord(\tilde{\al}^3) \, .
\label{asd-QED3}
\ee
Interestingly, because there is no one-loop contribution, this anomalous dimension is fully gauge-invariant.
In order to see how it affects the momentum dependence of the fermion propagator, we combine Eq.~(\ref{Sr-sr}) with the general solution of Eq.~(\ref{ren-eq-part-p}).
This yields:
\be
S_r(p\,;\mu) = \frac{i}{\Sp}\,s_r(1;\tilde{\al};a)\,e^{\frac{1}{2}
\int_{0}^{\log \left(\frac{-p^2}{\mu^2}\right)}
\D \log \left(\frac{-{p'}^2}{\mu^2}\right)\,\gamma_\psi \big(\tilde{\al}(p'\,;\mu),a\big)}, 
\label{ren-prop-QED3:def}
\ee
where we have taken into account the fact that, in QED$_3$, $\gamma_A=0$ 
and the lack of running of the coupling constant $e^2$ at the considered level of accuracy.
The momentum dependence of this coupling constant implies that perturbation theory is valid, {\it i.e.}, $\tilde{\al} \ll 1$,
as long as: $e^2 \ll \sqrt{-p^2}$, {\it i.e.}, for large euclidean momenta. With the help of Eq.~(\ref{asd-QED3}), the asymptotic form of
the dressed fermion propagator defined in the left-hand side of (\ref{ren-prop-QED3:def}) reads:
\be
S_r(p\,;\mu) = \frac{i}{\Sp}\,s_r(1;\tilde{\al};a)\,e^{\frac{N_F}{48 \pi^2}\frac{e^4}{\mu^2}}\,e^{-\frac{N_F}{48 \pi^2}\frac{e^4}{-p^2}}\, .
\label{ren-prop-QED3}
\ee
Deep in the UV, the momentum dependence of the dressed propagator is essentially the one of a free fermion in accordance with the fact that QED$_3$ is asymptotically free. 

\section{Conclusion and outlook}
\label{Sec:Conclusion}

The central result of this paper is the formula, Eq.~(\ref{sigma-2l-s}), for the two-loop fermion self-energy $\Sigma_{V2}(p^2)$
of massless RQED$_{d_\gamma,d_e}$ 
in an arbitrary gauge. This formula was derived using simple IBP relations. It allowed us, in the limit $d_\gamma \ra 4$, to compute exactly 
$\Sigma_{V2}(p^2)$ 
of massless RQED$_{4,d_e}$ 
without calculation of the complicated
two-loop diagram $G(1/2,1,1/2,1,1)$ (the last term in Eq.~(\ref{sigma-2l-s})). 
From Eq.~(\ref{sigma-2l-s}), as well as the one-loop self-energy Eq.~(\ref{sigma-1l-s}), general expressions 
were derived for the fermion anomalous scaling dimension (\ref{asd-RQED4de}) and the
renormalized fermion propagator, Eqs.~(\ref{ren-r-RQED4de}) and (\ref{def:series:Sr}), in the limit $d_\gamma \ra 4$.
These results were then applied to RQED$_{4,3}$ and compared with the cases of (renormalizable) QED$_4$ and 
(super-renormalizable) QED$_3$. In all cases, the two-loop contribution to the anomalous dimension was found to be gauge invariant.
The latter was also shown to be transcendentally more complex in RQED$_{4,3}$ than in usual QEDs as witnessed by the appearance of $\zeta_2$
in Eq.~(\ref{asd-RQED43}) with respect to Eqs.~(\ref{asd-QED4}) and (\ref{asd-QED3}).

From the condensed matter physics point of view, as explained in the introduction, the massless RQED$_{4,3}$ model describes the ultra-relativistic limit of undoped graphene.
Our results are a first step towards a rigorous understanding of interaction corrections to the spectral properties of graphene in the ultra-relativistic limit.
Because of the Lorentz invariance of the present model there is no renormalization of the Fermi velocity.~\footnote{The renormalization of the Fermi velocity to order $e^4$ has been done in
the recent preprint by E.~Barnes, E.\ H.~Hwang, R.\ Throckmorton and S.\ Das~Sarma, arXiv:1401.7011 [cond-mat.mes-hall].} 
The effect of interactions manifests at the level of
the finite part of the fermion propagator as well as in the anomalous scaling dimension of the fermion field, Eq.~(\ref{asd-RQED43}).
The later indicates how radiative corrections affect the momentum dependence of the dressed fermion propagator
Eq.~(\ref{ren-prop-RQED43}). For RQED$_{4,3}$ our present results yield a gauge-variant dressed propagator coming from the one-loop contribution; the two-loop
contribution, on the other hand, is gauge-invariant and positive (this is to be contrasted with the QED$_3$ result which is fully gauge-invariant, see Eq.~(\ref{ren-prop-QED3})).
From the field theory point of view, the results obtained can be extended to an arbitrary model of massless RQED$_{d_\gamma,d_e}$
with the help of an exact evaluation of $G(1/2,1,1/2,1,1)$, see App.~\ref{App:G(alpha,beta)} for general formulas. 
We plan to return to this task, with application to the case of RQED$_{3,2}$ which requires the knowledge of the 
complicated contribution $G(1/2,1,1/2,1,1)$, in our future investigations.  


\acknowledgments

The work of A.V.K. was supported in part by the Russian Foundation for 
Basic Research (Grant No. 13-02-01005) and by the Universit\'e Pierre et Marie Curie (UPMC).

\appendix

\section{Expansion of master integrals}
\label{app-Gfuncs}

Here we give the expansion of the master integrals, or $G$-functions, contributing to the fermion self-energies
in Eqs.~(\ref{ren-sigmas})-(\ref{ren-sigma2s}) at fixed $\varepsilon_e$. The results have the following form:
\begin{subequations}
\label{exp:Gfuncs}
\bea
&&G(1,1-\veps_e)=\frac{\exp[-\gamma_E\varepsilon_\gamma]}{\varepsilon_\gamma(1-\veps_e-2\varepsilon_\gamma)}
\frac{1}{\Gamma(1-\veps_e)}
\biggl(1+\overline{\Psi}_1\varepsilon_\gamma + \frac{\varepsilon_\gamma^2}{2}
\Bigl(\overline{\Psi}_1^2+2\zeta_2 -3\Psi_2(1-\veps_e)\Bigr)\biggr), \label{G1,1-ee} 
\nonum \\
&&G(1,\varepsilon_\gamma-\veps_e)=
-\frac{\exp[-\gamma_E\varepsilon_\gamma]}{2\varepsilon_\gamma}
\frac{\varepsilon_\gamma-\veps_e}{(1-\veps_e-3\varepsilon_\gamma)(2-\veps_e-3\varepsilon_\gamma)}
\frac{1}{\Gamma(1-\veps_e)}
\biggl(1+\overline{\Psi}_1\varepsilon_\gamma + \frac{\varepsilon_\gamma^2}{2}
\Bigl(\overline{\Psi}_1^2+8\zeta_2 -9\Psi_2(1-\veps_e)\Bigr)\biggr), \label{G1,eg-ee} 
\nonum \\
&&G(1-\veps_e,\varepsilon_\gamma)=
-\frac{\exp[-\gamma_E\varepsilon_\gamma]}{2(1-2\varepsilon_\gamma)}
\frac{1-\veps_e-2\varepsilon_\gamma}{(1-\veps_e-3\varepsilon_\gamma)(2-\veps_e-3\varepsilon_\gamma)}
\frac{1}{\Gamma(1-\veps_e)}
\biggl(1+\overline{\Psi}_1\varepsilon_\gamma + \frac{\varepsilon_\gamma^2}{2}
\Bigl(\overline{\Psi}_1^2+4\zeta_2 -5\Psi_2(1-\veps_e)\Bigr)\biggr), \label{G1-ee,eg} 
\nonum \\
&&G(1,1)=\exp[-\gamma_E\varepsilon_\gamma]
\frac{K_1}{\Gamma(1-\veps_e)}
\biggl(1+\overline{\Psi}_2\varepsilon_\gamma + \frac{\varepsilon_\gamma^2}{2}
\Bigl(\overline{\Psi}_2^2+\Psi_2(\veps_e)+2\Psi_2(1-\veps_e)-4\Psi_2(2-2\veps_e)\Bigr) \biggr), 
\nonum
\eea
\end{subequations}
where $\gamma_E$ is the Euler constant and $K_1$, $\overline{\Psi}_1$ and $\overline{\Psi}_2$ were determined
in (\ref{ren-params-RQED4de}).

\section{Exact expression of the two-loop massless propagator diagram with two non-integer indices}
\label{App:G(alpha,beta)}

In this appendix, we derive general expressions for the coefficient function $G(\al,1,\beta,1,1)$, see 
Eq.~(\ref{complicatedG}). This function can be expressed in the following form:
\be
G(\al,1,\beta,1,1) = \frac{a^3(1)a(\al)a(\beta)}{a(\al+\beta+3-D)}
G(\tilde{\al},\lambda,\tilde{\beta},\lambda,\lambda),~~~ a(\al)=\frac{\Gamma(\tilde{\al})}{\Gamma(\al)},~~~ \tilde{\al}=\frac{D}{2}-\al,~~
\lambda=\frac{D}{2}-1 \, ,
\label{complicatedG1}
\ee
where
\be
G(\tilde{\al},\lambda,\tilde{\beta},\lambda,\lambda)= 
C_D\left[ \quad \parbox{16mm}{
    \begin{fmfgraph*}(16,14)
      \fmfleft{i}
      \fmfright{o}
      \fmfleft{ve}
      \fmfright{vo}
      \fmftop{vn}
      \fmftop{vs}
      \fmffreeze
      \fmfforce{(-0.1w,0.5h)}{i}
      \fmfforce{(1.1w,0.5h)}{o}
      \fmfforce{(0w,0.5h)}{ve}
      \fmfforce{(1.0w,0.5h)}{vo}
      \fmfforce{(.5w,0.95h)}{vn}
      \fmfforce{(.5w,0.05h)}{vs}
      \fmffreeze
      \fmf{plain}{i,ve}
      \fmf{plain,left=0.8}{ve,vo}
      \fmf{phantom,left=0.5,label=$\tilde{\al}$,l.d=-0.01w}{ve,vn}
      \fmf{phantom,right=0.5,label=$\lambda$,l.d=-0.01w}{vo,vn}
      \fmf{plain,left=0.8}{vo,ve}
      \fmf{phantom,left=0.5,label=$\tilde{\beta}$,l.d=-0.01w}{vo,vs}
      \fmf{phantom,right=0.5,label=$\lambda$,l.d=-0.01w}{ve,vs}
      \fmf{plain,label=$\lambda$,l.d=0.05w}{vs,vn}
      \fmf{plain}{vo,o}
      \fmffreeze
      \fmfdot{ve,vn,vo,vs}
    \end{fmfgraph*}
}
\quad \right]\, .
\label{complicatedGt}
\ee
The function $G(\tilde{\al},\lambda,\tilde{\beta},\lambda,\lambda)$ belongs to the type of Feynman integrals considered in
[\onlinecite{Kotikov96}]: it corresponds to the notation $C_f[A(\tilde{\al},\tilde{\beta},\lambda)]$ in [\onlinecite{Kotikov96}]
and can be considered as the particular case where $\gamma \to \lambda$ of the results in  [\onlinecite{Kotikov96}].
The general results contain four hypergeometric functions ${}_3F_2$ of argument $1$. 
We will show that the diagram  $G(\tilde{\al},\lambda,\tilde{\beta},\lambda,\lambda)$ contains only two hypergeometric functions ${}_3F_2$ of argument $1$.

Following [\onlinecite{Kotikov96}], $G(\tilde{\al},\lambda,\tilde{\beta},\lambda,\lambda)$ can be represented in the following
form:~\footnote{Notice that, because of our definition, Eq.~(\ref{def:2loopmassless}), the factor $1/(4\pi)^D$, which appears explicitly in 
$C_f[A]$ in [\onlinecite{Kotikov96}], has already been extracted from the expression of $G$.}
\be 
G(\tilde{\al},\lambda,\tilde{\beta},\lambda,\lambda)= 
\frac{1}{\Gamma(\lambda)} \frac{1}{\tilde{\al}-1}
\hat{I}(\tilde{\al},\tilde{\beta}),~~~ 
\hat{I}(\tilde{\al},\tilde{\beta})=\overline{I}(\tilde{\al},\tilde{\beta})-\tilde{I}(\tilde{\al},\tilde{\beta}) \, .
\label{complicatedGt1}
\ee
Taking $\gamma \to \lambda$, the results (17) in  [\onlinecite{Kotikov96}] 
becomes 
%
\bea
&&\overline{I}(\tilde{\al},\tilde{\beta}) = \frac{\Gamma(\lambda+1-\tilde{\beta})}{\Gamma(\tilde{\beta})} \biggl\{\sum_{n=0}^{\infty}
\frac{\Gamma(n+2\lambda)}{n!\Gamma(2\lambda)} \,
\biggl[\frac{1}{n+\lambda+1-\tilde{\al}} \,
\frac{\Gamma(n+2-\tilde{\al})\Gamma(\tilde{\al}+\tilde{\beta}-2)}{\Gamma(n+3+\lambda-\tilde{\al}-\tilde{\beta})\Gamma(\tilde{\al}
+\lambda-1)} \nonumber \\
&& - \frac{1}{n+\lambda+\tilde{\al}-1} \, \biggl(
\frac{\Gamma(n+1)\Gamma(\tilde{\beta}-1)}{\Gamma(n+2+\lambda-\tilde{\beta})\Gamma(\lambda)}
+ \frac{\Gamma(n+\lambda-2+\tilde{\al}+\tilde{\beta})\Gamma(2-\lambda-\tilde{\al})}{\Gamma(n-1+2\lambda+\tilde{\al})\Gamma(3-\tilde{\al}
-\tilde{\beta})} \biggl)
\biggr] \nonumber \\
&&+ \frac{\Gamma(1-\lambda)
\Gamma(1+\lambda-\tilde{\al})
}{\Gamma(2\lambda)\Gamma(2-\tilde{\beta})\Gamma(3-\tilde{\al}-\tilde{\beta})}
\frac{\pi}{\sin[\pi (\lambda-1+\tilde{\beta})]}
\biggr\} \, ,
\label{oI}
\eea
where we have used the property: $\Gamma(a)\Gamma(1-a)=\pi \sin^{-1}[\pi a]$ for any $a$.
Note that the second term in the r.h.s. of  (17) in [\onlinecite{Kotikov96}] has been summed in the product of $\Gamma$-functions, yielding the last term in Eq.~(\ref{oI}).

For the part $\tilde{I}$ there are four representations, {\it i.e.}, the equations (18)-(22) in [\onlinecite{Kotikov96}].  
We use  equation (21), 
\footnote{Notice that there is a slip in the first term of the r.h.s.\ of Eq.~(21) in [\onlinecite{Kotikov96}]: 
in the denominator, $\Gamma(2+\lambda-\al-\beta)$ should be replaced by  $\Gamma(2+\lambda-\gamma-\beta)$.}
where the last terms is zero at $\gamma = \lambda$. So, we have
%
\bea
&&\tilde{I}(\tilde{\al},\tilde{\beta})= \frac{\Gamma(1-\lambda)\Gamma(1-\tilde{\beta})\Gamma(1+\lambda-\tilde{\al})
\Gamma(1+\lambda-\tilde{\beta})}{
\Gamma(2\lambda)\Gamma(2-\tilde{\beta})\Gamma(3-\tilde{\al}-\tilde{\beta})}
\frac{\sin[\pi \tilde{\al}]}{\sin[\pi (\lambda-1+\tilde{\al})]} \nonumber \\
&&+ \sum_{n=0}^{\infty} \frac{\Gamma(n+2\lambda)}{n!\Gamma(2\lambda)} \, \frac{(-1)^n}{n+\lambda+1-\tilde{\al}} \,
\frac{\Gamma(1-\tilde{\beta})}{\Gamma(\tilde{\beta}-\lambda)}
\frac{\Gamma(\tilde{\al}+\tilde{\beta}-2)\Gamma(2-\lambda-\tilde{\al})}{\Gamma(n+3+\lambda-\tilde{\al}-\tilde{\beta})\Gamma(\tilde{\al}-1-n)}
\, .
\label{tI}
\eea
Now, we consider the difference $\overline{I}-\tilde{I}$, {\it i.e.}, $\hat{I}(\tilde{\al},\tilde{\beta})$. The first term and the 
second one in the r.h.s. of (\ref{oI}) and (\ref{tI}), respectively, combine to one term. The terms without series also combine to one term. 
So, we have
%
\bea
&&\hat{I}(\tilde{\al},\tilde{\beta}) =
-\sum_{n=0}^{\infty} \frac{\Gamma(n+2\lambda)}{n!\Gamma(2\lambda)}
\biggl[\frac{1}{n+\lambda+1-\tilde{\al}} \,
\frac{\Gamma(n+2-\tilde{\al})\Gamma(1-\tilde{\beta})\Gamma(2-\lambda-\tilde{\al})}
{\Gamma(n+3+\lambda-\tilde{\al}-\tilde{\beta})\Gamma(3-\tilde{\al}-\tilde{\beta})\Gamma(\tilde{\beta}-\lambda)}
\,\frac{\sin[\pi \lambda]}{\sin[\pi (\lambda+1-\tilde{\beta})]}
 \nonumber \\
&&+ \frac{\Gamma(\lambda+1-\tilde{\beta})}{\Gamma(\tilde{\beta})} \, \frac{1}{n+\lambda+\tilde{\al}-1} \biggl(
\frac{\Gamma(n+1)\Gamma(\tilde{\beta}-1)}{\Gamma(n+2+\lambda-\tilde{\beta})\Gamma(\lambda)}
+ \frac{\Gamma(n-2+\lambda +\tilde{\al}+\tilde{\beta})\Gamma(2-\lambda-\tilde{\al})
}{\Gamma(n-1+2\lambda+\tilde{\al})\Gamma(3-\tilde{\al}
-\tilde{\beta})} \biggl)
\biggr] \nonumber \\
&&- \frac{1}{1-\tilde{\beta}} \,
\frac{\Gamma(1+\lambda-\tilde{\al})\Gamma(1+\lambda-\tilde{\beta})
}{\Gamma(2\lambda)\Gamma(\lambda)\Gamma(3-\tilde{\al}-\tilde{\beta})}
\frac{\pi \sin[\pi(\tilde{\beta}- \tilde{\al})]}{\sin[\pi (\lambda-1+\tilde{\beta})] \sin[\pi (\lambda-1+\tilde{\al})]}
\label{oI-tI}
\eea
It is convenient to transform the first term in the r.h.s. to the new form containing the factor ${(n+\lambda+\tilde{\al}-1)}^{-1}$ in its 
denominator. It is possible to obtain from the transformation of ${}_3F_2$ hypergeometric functions of argument $1$ (see Eq.~(9) in 
[\onlinecite{Kotikov96}]: 
%
%
\be
\sum_{n=0}^{\infty} \frac{\Gamma(n+\hat{a})\Gamma(n+\hat{c})}{n!\Gamma(n+\hat{f})}
\, \frac{1}{n+\hat{b}} = \frac{
\Gamma(\hat{b})\Gamma(\hat{c}-\hat{b})}{\Gamma(\hat{f}-\hat{b})
\Gamma(1+\hat{b}-\hat{a})} 
\frac{\pi }{\sin[\pi \hat{a}]}
- \frac{\sin[\pi(\hat{f}-\hat{c})]}{\sin[\pi \hat{a}]}
\sum_{n=0}^{\infty} \frac{\Gamma(n+\hat{c}-\hat{f}+1)\Gamma(n+\hat{c})}{n!\Gamma(n+1+ \hat{c}-\hat{a})}
\, \frac{1}{n+\hat{c}-\hat{b}} \, , 
\label{transform}
\ee
with arbitrary $\hat{a}$, $\hat{b}$, $\hat{c}$ and $\hat{f}$.
Indeed, if $\hat{c}=2\lambda$, $\hat{a}=2-\tilde{\al}$, $\hat{b}=1+\lambda-\tilde{\al}$ and $\hat{f}=3+\lambda-\tilde{\al}-\tilde{\beta}$,
we have %
%
\bea
&&\sum_{n=0}^{\infty} \frac{\Gamma(n+2\lambda)}{n!\Gamma(2\lambda)}
\frac{\Gamma(n+2-\tilde{\al})}{\Gamma(n+3+\lambda-\tilde{\al}-\tilde{\beta})}
\frac{1}{n+\lambda+1-\tilde{\al}} = 
\frac{
\Gamma(1+\lambda-\tilde{\al})\Gamma(\lambda +\tilde{\al}-1)
}{\Gamma(2\lambda)\Gamma(\lambda)\Gamma(2-\tilde{\beta})} \frac{\pi }{\sin[\pi (\tilde{\al}-1)]}
\nonumber \\
&&- 
\frac{\sin[\pi (\lambda+\tilde{\al}+\tilde{\beta}-2)] }{\sin[\pi (\tilde{\al}-1)]}
\sum_{n=0}^{\infty} \frac{\Gamma(n+2\lambda)}{n!\Gamma(2\lambda)}
\frac{\Gamma(n-2+\lambda+\tilde{\al}+\tilde{\beta})}{\Gamma(n+2\lambda+\tilde{\al}-1)}
\frac{1}{n+\lambda+\tilde{\al}-1}\, .
\label{transform1}
\eea

Taking together this new results with the last two terms in Eq. (\ref{oI-tI}), we have:
\be
\hat{I}(\tilde{\al},\tilde{\beta}) = \frac{\Gamma(\lambda+1-\tilde{\beta})}{\Gamma(2\lambda)\Gamma(\lambda)} \,
\frac{1}{(1-\tilde{\beta})} \, I(\tilde{\al},\tilde{\beta}) \, , 
\label{Ht}
\ee
where
\bea
&&I(\tilde{\al},\tilde{\beta}) = 
\frac{\Gamma(1+\lambda-\tilde{\al})}{
\Gamma(3-\tilde{\al}-\tilde{\beta})}
\frac{\pi \sin[\pi(\tilde{\beta}- \tilde{\al}+\lambda)]}{\sin[\pi (\lambda-1+\tilde{\beta})]\sin[\pi \tilde{\al}]} 
+ \sum_{n=0}^{\infty} \frac{\Gamma(n+2\lambda)}{n!(n+\lambda+\tilde{\al}-1)}
\nonumber \\&& \times
\biggl(
\frac{\Gamma(n+1)}{\Gamma(n+2+\lambda-\tilde{\beta})} 
- \frac{\Gamma(n-2+\lambda +\tilde{\al}+\tilde{\beta})\Gamma(2-\tilde{\beta}) \Gamma(\lambda)
}{\Gamma(n-1+2\lambda+\tilde{\al})\Gamma(3-\tilde{\al}-\tilde{\beta})\Gamma(\lambda+\tilde{\al}-1)} 
\frac{\sin[\pi(\tilde{\beta}+\lambda-1)]}{\sin[\pi \tilde{\al}]}
\biggl)
\, .
\label{oI-tIN}
\eea
%
The results (\ref{Ht}) and  (\ref{oI-tIN}) together with (\ref{complicatedG1}) and (\ref{complicatedGt}) can be considered as the 
final result for the initial diagram $G(\al,1,\beta,1,1)$:
%
\be
G(\al,1,\beta,1,1) = 
\frac{1}{\tilde{\al}-1} \frac{1}{1-\tilde{\beta}} \,
\frac{\Gamma(\tilde{\al})\Gamma(\tilde{\beta})
\Gamma(3-\tilde{\al}-\tilde{\beta})}{\Gamma(\al)
\Gamma(\lambda-2+\tilde{\al}+\tilde{\beta})} 
\frac{\Gamma(\lambda)}{\Gamma(2\lambda)} \, I(\tilde{\al},\tilde{\beta}) \, .
\label{GN1}
\ee
However, in the case where $\al$ and $\beta$ are close to $\lambda$, it is not so 
convenient because (\ref{oI-tIN}) contains several additional singularities, which are canceled only at the end of calculations.

Another form of the final result can be obtained by application of the transformation (\ref{transform}) to the last term in 
Eq.~(\ref{oI-tIN}). Some simple algebra yields the following expression:
\bea
&&I(\tilde{\al},\tilde{\beta}) = 
\frac{\Gamma(1+\lambda - \tilde{\al})}{
\Gamma(3-\tilde{\al}-\tilde{\beta})}
\frac{\pi \sin[\pi \tilde{\al}]}{\sin[\pi (\lambda-1+\tilde{\beta})] \sin[\pi (\tilde{\al}+\tilde{\beta}+\lambda-1) ]} 
+ \sum_{n=0}^{\infty} \frac{\Gamma(n+2\lambda)}{n!
}
\biggl(\frac{1}{n+\lambda+\tilde{\al}-1}
\frac{\Gamma(n+1)}{\Gamma(n+2+\lambda-\tilde{\beta})} 
\nonumber \\&&
+ \frac{1}{n+\lambda+1-\tilde{\al}}
\frac{\Gamma(n+2-\tilde{\al})\Gamma(2-\tilde{\beta})\Gamma(\lambda)
}{\Gamma(n+3+\lambda-\tilde{\al}-\tilde{\beta})\Gamma(3-\tilde{\al}-\tilde{\beta})\Gamma(\lambda+\tilde{\al}-1)} 
\frac{\sin[\pi(\tilde{\beta}+\lambda-1)]}{\sin[\pi (\tilde{\al}+\tilde{\beta}+\lambda-1)]}
\biggl)
\, .
\label{oI-tIN1}
\eea
%
%
It seems that the result (\ref{oI-tIN1}) together with Eq.~(\ref{GN1})
is the most convenient final form of the result for the initial diagram $G(\al,1,\beta,1,1)$.

There are two other forms of the result for $I(\tilde{\al},\tilde{\beta})$,
which can be obtained by application of the transform (\ref{transform}) to the second term in Eqs.~(\ref{oI-tIN}) and (\ref{oI-tIN1}). 
That leads to
\be
 \sum_{n=0}^{\infty} 
\frac{\Gamma(n+2\lambda)}{\Gamma(n+2+\lambda-\tilde{\beta})} \, \frac{1}{n+\lambda+\tilde{\al}-1} =
\frac{\sin[\pi(\tilde{\beta}+\lambda-1)]}{\pi } \sum_{n=0}^{\infty} \frac{\Gamma(n+\tilde{\beta}+\lambda-1)}{n!
(n+\lambda+1-\tilde{\al})} \Psi_1(n+2\lambda) - \frac{\Gamma(1+\lambda-\tilde{\al})}{\Gamma(3-\tilde{\al}-\tilde{\beta})}
\Psi_1(\lambda+\tilde{\al}-1) \, .
\label{term}
\ee
%
Using this formula the expressions of $I(\tilde{\al},\tilde{\beta})$ become
more cumbersome because they contain the Euler (digamma) $\Psi_1$-functions. Nevertheless, we also present them for completeness. 
They are following
\bea
&&I(\tilde{\al},\tilde{\beta}) = 
\frac{\Gamma(1+\lambda-\tilde{\al})}{
\Gamma(3-\tilde{\al}-\tilde{\beta})}
\frac{\pi \sin[\pi(\tilde{\beta}- \tilde{\al}+\lambda)]}{\sin[\pi (\lambda-1+\tilde{\beta})]\sin[\pi \tilde{\al}]} 
- \frac{\Gamma(1+\lambda-\tilde{\al})}{\Gamma(3-\tilde{\al}-\tilde{\beta})}
\Psi_1(\lambda+\tilde{\al}-1) \nonumber \\&&
+ \sum_{n=0}^{\infty} \frac{\Gamma(n+\tilde{\beta}+\lambda-1)}{n!
(n+\lambda+1-\tilde{\al})} \Psi_1(n+2\lambda) \frac{\sin[\pi(\tilde{\beta}+\lambda-1)]}{\pi } \nonumber \\&&
- \sum_{n=0}^{\infty} \frac{\Gamma(n+2\lambda)}{n!(n+\lambda+\tilde{\al}-1)}
\frac{\Gamma(n-2+\lambda +\tilde{\al}+\tilde{\beta})\Gamma(2-\tilde{\beta})\Gamma(\lambda)
}{\Gamma(n-1+2\lambda+\tilde{\al})\Gamma(3-\tilde{\al}-\tilde{\beta})\Gamma(\lambda+\tilde{\al}-1)} 
\frac{\sin[\pi(\tilde{\beta}+\lambda-1)]}{\sin[\pi \tilde{\al}]}
\, .
\label{oI-tIN2}
\eea
%
and
\bea
&&I(\tilde{\al},\tilde{\beta}) = 
\frac{\Gamma(1+\lambda - \tilde{\al})}{
\Gamma(3-\tilde{\al}-\tilde{\beta})}
\frac{\pi \sin[\pi \tilde{\al}]}{\sin[\pi (\lambda-1+\tilde{\beta})\sin[\pi (\tilde{\al}+\tilde{\beta}+\lambda-1) ]} 
- \frac{\Gamma(1+\lambda-\tilde{\al})}{\Gamma(3-\tilde{\al}-\tilde{\beta})}
\Psi_1(\lambda+\tilde{\al}-1)
\nonumber \\&&
+ \sum_{n=0}^{\infty} \frac{\Gamma(n+2\lambda)}{n!(n+\lambda+1-\tilde{\al})} \biggl(\
\frac{\Gamma(n+2-\tilde{\al})\Gamma(2-\tilde{\beta})\Gamma(\lambda)
}{\Gamma(n+3+\lambda-\tilde{\al}-\tilde{\beta})\Gamma(3-\tilde{\al}-\tilde{\beta})\Gamma(\lambda+\tilde{\al}-1)} 
\frac{\sin[\pi(\tilde{\beta}+\lambda-1)]}{\sin[\pi (\tilde{\al}+\tilde{\beta}+\lambda-1)]} 
\nonumber \\&& 
+  \frac{\Gamma(n+\tilde{\beta}+\lambda-1)}{\Gamma(n+2\lambda)} \Psi_1(n+2\lambda) \frac{\sin[\pi(\tilde{\beta}+\lambda-1)]}{\pi }
\biggl)
\, .
\label{oI-tIN3}
\eea
Unfortunately, the case $\alpha=\beta$ ({\it i.e.}, $\tilde{\al}=\tilde{\beta}$) does not produce
any additional simplification and, in this case, we should use the combination of the equations 
(\ref{Ht}) and (\ref{oI-tIN}) together with (\ref{complicatedG1}) and (\ref{complicatedGt}).

\end{widetext}

\end{fmffile}
\end{document}